\documentstyle[12pt]{article}
\input epsf
\begin{document}

\newcommand{\mysection}[1]{\section{#1}\setcounter{equation}{0}}
\renewcommand{\theequation}{\thesection.\arabic{equation}}
\newcommand{\eq}{\begin{equation}}
\newcommand{\en}{\end{equation}}
\newcommand{\gsi}{\,\raisebox{-0.13cm}{$\stackrel{\textstyle>}
{\textstyle\sim}$}\,}
\newcommand{\lsi}{\,\raisebox{-0.13cm}{$\stackrel{\textstyle<}
{\textstyle\sim}$}\,}

\rightline{CERN-TH.6734/93}
\rightline{RU-93-11}
\baselineskip=18pt
\vskip 0.5in
\begin{center}
{\bf \LARGE Baryon Asymmetry of the Universe\\
in the Standard Model}
\end{center}
\vspace*{0.7in}
\begin{center}{\large Glennys R. Farrar}\footnote{Research supported
in part by NSF-PHY-91-21039} \\
\vspace{.05in}
{\it Department of Physics and Astronomy \\ Rutgers University,
Piscataway, NJ 08855, USA}

\vspace*{0.5in}
{\large M. E. Shaposhnikov}\footnote{On leave of absence from
Institute for Nuclear Research of Russian Academy of Sciences, Moscow
117312, Russia} \\
\vspace{.05in}
{\it CERN, TH Division \\ CH-1211, Geneva 23, Switzerland}
\end{center}
\vskip  0.5in
\noindent   Extended version\\
\noindent   November 1993

\eject
\vspace*{1.in}

{\bf Abstract:}

We study the interactions of quarks and antiquarks with the changing
Higgs field during the electroweak phase transition, including quantum
mechanical and some thermal effects, with the only source of CP
violation being the known CKM phase.  We show that the GIM
cancellation, which has been commonly thought to imply a prediction
which is at least 10 orders of magnitude too small, can be evaded in
certain kinematic regimes, for instance when the strange quark is
totally reflected but the down quark
is not.  We report on a quantitative calculation of the asymmetry in a
one-dimensional approximation based on the present understanding of
the physics of the high-temperature environment, but with some aspects
of the problem over-simplified.  The resulting
prediction for the magnitude and sign of the present baryonic
asymmetry of the universe agrees with the observed value, with
moderately optimistic assumptions about the dynamics of the phase
transition.  Both magnitude and sign of the asymmetry have an
intricate dependence on quark masses and mixings, so that quantitative
agreement between prediction and observation would be highly
non-trivial.  At present uncertainties related to the dynamics of the
ew phase transition and the oversimplifications of our treatment
are too great to decide whether or not this is the
correct explanation for the presence of remnant matter in our
universe, however the present work makes it clear that the minimal
standard model cannot be discounted as a contender for explaining this
phenomenon.

\thispagestyle{empty}
\newpage
\addtocounter{page}{-1}
\tableofcontents
\newpage

\mysection{Introduction}
\label{sec:intro}

\hspace*{2em} The non-zero
ratio of baryon number to entropy, $n_B/s \sim (4-6)~10^{-11}$, is an
important challenge to particle theory.  Many possible mechanisms
have
been advanced to explain it
\cite{sakharov,kuzmin,ikkt:bau,yosh,weinberg,dimsus}(for
reviews and references to more recent work see, e.g.
\cite{dz:rev,koltur:rev,dolgov:rev,s:nobel,s:rev,turok:rev,ckn:rev}).
The standard electroweak theory contains in principle all the
elements
necessary \cite{sakharov} for generation of the baryonic asymmetry of
the universe (BAU):
\begin{enumerate}
\item C and CP violation, in the
fundamental gauge and Higgs interactions of the quarks.
\item Anomalous electroweak baryon number violation
\cite{hooft:prl,hooft:pr}.\footnote{It was shown in ref. \cite{krs85}
(for earlier discussion see
\cite{linde:pl77,dimsus,kman}) that although anomalous B violation is
negligible at zero temperature, it is enormously enhanced at high
temperature, so that it can be large enough that the baryonic
asymmetry of the universe may have been produced during the
electroweak phase transition.}
\item A departure from thermal equilibrium, assuming
the cosmological \\ SU(2)xU(1) phase transition \cite{kir,kirlin} is
first order.
\end{enumerate}

However, conventional wisdom holds that the minimal standard model
(MSM) cannot by itself cause the observed baryonic asymmetry of the
universe. The most important reason is that the CP violation present
in the MSM to account for the observed CP violation in kaon decays is
commonly believed to be inadequate, by 10-12 orders of magnitude or
more, to explain the observed $\sim 10^{-11}-10^{-10}$ level of the
asymmetry.  The second reason is that such a large asymmetry can only
be generated in a strongly first order phase transition, and the higgs
sector of the MSM may not produce a sufficiently strong phase
transition.  In particular, the rate of sphaleron transitions after
the phase transition must be small enough not to wash out the
asymmetry, so the mass of the W just after the phase transition must
not be too small\cite{s:m^14}. Using the one-loop high temperature effective
potential together with the one-loop sphaleron rate results in the
upper bound on the Higgs mass in the MSM: $M_H^{crit} = 45$ GeV
\cite{s:sm87,bs:higgs}, which is in contradiction with the LEP
experimental lower bound: $M_H \gsi 60$ GeV.\footnote{For the current
best limit from all four experiments, see \cite{lep}.}

For these reasons, people have considered extensions of the standard
model in a search for a mechanism for electroweak baryogenesis
\cite{krs85,mcl:prl,s:sm88,tz:prl,tz:np,mstv,dhss,ckn:L1,ckn:L2,ckn:Y1,ckn:Y2}.
Variants of the electroweak theory contain more free parameters than
the MSM alone, allowing the introduction of an extra source of CP
violation and allowing the upper bound on the Higgs mass to be
relaxed\cite{bks:higgs_pl,bks:higgs,tz:higgs,myint:higgs,piet:higgs}.
The general conclusion is quite optimistic: the baryon asymmetry of
the universe can plausibly be a natural consequence of several
extended versions of the standard model, provided the electroweak
phase transition is sufficiently strongly first order\footnote{All
present models of electroweak baryogenesis depend on two essential
aspects of the strength of the phase transition which are in principle
independent, but which for simplicity we lump together when speaking
of a requirement that the phase transition be strong enough: the vev
after the transition must be large enough that sphaleron transitions
in the broken phase are turned off, and the bubbles of low temperature
phase which expand to fill the universe must not be too flimsy or slow
moving.  The first condition is the one which is better understood, so
that it is normally the one whose constraint is given quantitatively.}
and the additional source of CP violation is strong enough. In these
``scenarios'', however, the magnitude and sign of the asymmetry cannot
be predicted; instead, the observed BAU must be used to constrain the
parameters of the extended theory.

We shall demonstrate in this paper that contrary to popular belief,
known MSM physics alone may in fact be responsible for production of
the baryonic asymmetry, with no new source of CP violation required,
as long as the usual requirement of a sufficiently strongly first
order phase transition is met.\footnote{Whether or not the minimal
standard model with a single Higgs produces a sufficiently strongly
first order phase transition is a separate question from whether the
MSM CP violation in the CKM matrix is suffieciently large.  In fact,
it is still an open question, since the uncertainties in the upper
bound on the Higgs mass are rather large. Although the one loop
calculation gives $M_H^{crit} = 45$ GeV, taking into account Debye
screening effects reduces the critical mass to $M_H^{crit} = 35$ GeV
\cite{dlhll:pl,dlhll:pr}.  The two-loop corrections increase the
latter number by about $5$ GeV\cite{bd,ae}, while non-perturbative
effects may change it up to $M_H^{crit} \sim 100$ GeV
\cite{s:condensate}.}.  We make a detailed calculation
of the asymmetry which is produced when the quarks and antiquarks,
treated as quasiparticle excitations of the plasma (whose dispersion
relation we obtain using 1-loop high temperature perturbation theory)
are quantum-mechanically reflected from the barrier presented to them
by the interface between the regions of small and large higgs vev.  We
simplify the problem in several important respects, neglecting
incoherent scattering of the quasiparticles and considering only
1-dimensional scattering.  Since the results can be consistent with
observation in sign and magnitude, our simplified treatment provides
the necessary incentive to investigate MSM production of the BAU with
a more realistic treatment.  It also provides insight into the most
important and problematic aspects of the physics, and thus can be of
guidance in future work.

The bulk of this paper is devoted to elucidating the essential aspects
of the physics involved in production of the BAU in the minimal
standard model, developing necessary theoretical machinery for solving
the problem, and presenting the quantitative results of a fairly
realistic but nonetheless over-simplified calculation.

The plan of the paper is as follows.  We begin with a section
describing CP violation in the MSM and the argument leading to the
conventional wisdom on the smallness of MSM CP violation.  We discuss
CP violation in the $K^0$ system, which is known to be $O(10^{-3}$),
in order to understand how the physics of the MSM may lead to C and CP
violation at the level required for the production of the observed
BAU. This allows us to identify aspects of the physics which must be
treated adequately if we are to avoid missing the effect in
cosmology.

In section \ref{sec:overview} we give an overview of the various
dynamical mechanisms for electroweak baryogenesis which have been
discussed in the literature, and describe the
mechanism\cite{ckn:L2,ckn:Y1,s:msm} which we quantitatively
investigate in the latter portion of this paper.  In this mechanism,
the CP violating scattering of thermal quarks from the bubble wall of
the expanding Higgs vev produces a baryonic current flowing from the
unbroken to the broken phase.  In section \ref{sec:rough} we make a
rough estimate of the size of the baryonic current which might be
anticipated from this process, when the important region of phase
space is not overlooked.  The quantitative calculation of the current
is deferred to the latter portion of the paper.

Given a non-zero current of baryon number produced by some process
involving the bubble wall, we must estimate the ratio $n_B/s$ which
remains after the phase transition is complete, produced on account of
sphaleron processes which diminish the antibaryon excess in the
unbroken phase.  We cannot give a firm estimate of this ratio as a
function of the sphaleron rate in the unbroken phase, since it also
depends on the nature of the bubble wall.  However we can obtain a
conservative estimate of the magnitude of $n_B/s$ by considering the
case that the bubble wall does not disequilibrate the medium as it
passes, i.e., in quasi-static approximation.  This analysis is
presented in section \ref{sec:asym}.  Presumably improvements on the
quasi-static approximation will only increase the magnitude of
$n_B/s$.

Interactions of the fermions with the gauge and Higgs fields of the
high temperature plasma are crucial to the existence of non-trivial CP
violation, as explained in section \ref{sec:CPinSM}.  Moreover they
are quantitatively important in other regards.  Thus we present in
section \ref{sec:thermal} a discussion of the properties of quark
excitations in the thermal plasma.

While the discussion in the first part of the paper is rather general,
independent of the precise source of the baryonic current, in order to
make a quantitative prediction of the asymmetry we must adopt a
particular mechanism for its production.  The formalism which we have
developed in order to calculate the asymmetry in the reflection
probabilities of quarks and antiquarks scattering from the domain
wall, is given in section \ref{sec:scattering} and the appendices.
Using this formalism, section \ref{sec:analytic} is devoted to
obtaining analytic results under various simplifying assumptions.  The
analytic expression presented here is helpful for understanding how
the GIM mechanism can be evaded, and explaining the final dependence
on quark Yukawa couplings.  However in order to do a calculation with
sufficient accuracy to address the question of the sign of the result,
we must do an exact numerical calculation.  The results of this are
given in section \ref{sec:results}.  We explore in some detail how the
asymmetry depends on quark masses and mixings.

Having determined the asymmetry in the reflection probabilities, we
then combine it with the estimates of the sphaleron conversion
efficiency and the asymmetry in the fluxes, obtained in earlier
sections, and give in section \ref{sec:prediction} our final estimate
for $n_b/s$.  We review and elaborate on the uncertainties and
inadequacies of the present calculation.  The last section is the
Conclusion, where we summarize the present situation, describe the
problems which must be solved, and mention some consequences.

The main ideas of the paper can be understood by reading sections
\ref{sec:CPinSM}-\ref{sec:rough}, \ref{sec:results},
\ref{sec:prediction}, and the Conclusion, although section
\ref{sec:asym} is important in that it shows how the final prediction
is connected to the kinetics and dynamics near the bubble wall, and
section \ref{sec:thermal} will aid in comprehending some of the
unusual properties of the thermal excitations which are the actual
eigenstates of the scattering problem.  Many readers will also want to
study the analytic formula derived in Section \ref{sec:analytic} in
thin wall, small $p/\omega$ approximation, working to lowest
non-vanishing order in the mixing angles, in order to understand the
dependence on quark masses of the final asymmetry.  Other sections and
the appendices are intended for readers wanting to understand the
details as well as the general ideas.

A brief description this work has been published in ref. \cite{fs:1}.
The present paper was originally issued as a preprint\cite{fs:2} in
May, 1993.  This version corrects some typographical errors and has
been somewhat reorganized (e.g., moving more to the appendices) and
elaborated (especially the section on analytic results) in order to
make it more readily understandable.  In addition we include two
effects which were previously neglected: $L-R$ mixing due to QCD
sphalerons, and a diminution of the electroweak gauge and Higgs
effects in the broken phase due to mass corrections in the 1-loop
approximation to the quasi-particle propagator.  The results and
conclusion are not significantly modified; the discussion of the
various uncertainties is more complete and quantitative.

\mysection{CP Violation in the Standard Model}
\label{sec:CPinSM}

\hspace*{2em} In the minimal standard model, CP violation occurs
because of relative phases between the electroweak gauge interactions
and the Higgs interactions of the quarks.  As shown by Kobayashi and
Maskawa \cite{km}, if there are at least three generations of quarks,
there can be a physically meaningful phase which leads to observable
CP-violating effects.  The part of the MSM Lagrangian which involves
quarks is:
\eq
{\cal L} = {\cal L}_G + {\cal L}_Y.
\label{ewlagr}
\en
In the ``gauge'' basis,
\eq
{\cal L}_G = \bar{Q}_L {\not{\cal D}} Q_L + \bar{U}_R {\not{\cal D}}
U_R + \bar{D}_R {\not{\cal D}} D_R,
\en
and
\eq
{\cal L}_Y = \frac{g_W}{\sqrt{2}M_W} \{\bar{Q}_L^i K^{ij} M_d^{jj}
D_R^j\phi +
\bar{Q}_L^i M_u^{ii} U_R^i\tilde{\phi} + h.c.\},
\label{Yukawa}
\en
where ${\not{\cal D}}$ is the appropriate covariant derivative,
$Q_L^i$ are the left-handed quark doublets ($i$ is the generation
index), $U_R^i$ and $D_R^i$ are the right handed quarks with electric
charges $\frac{2}{3}$ and $-\frac{1}{3}$ respectively, $K$ is the
Cabibbo-Kobayashi-Maskawa (CKM) matrix, $M_u$ and $M_d$ are the
diagonal mass matrices of the quarks, and $\tilde{\phi}_i =
\epsilon_{ij}\phi^{\dagger}_j$.  In this basis, the Lagrangian has
been written in terms of the fields which are eigenstates of the
$SU(2)_L$ gauge interactions, and the CP violation is contained in a
phase in the matrix $K$, relating the gauge eigenstates to the mass
eigenstates.  When a specific parametrization of the CKM matrix is
required, we adopt that of the Particle Data Group \cite{PDG}.  Note
that by redefining the basis, any one of the three types of fermionic
interactions -- with charged or neutral Higgs, or gauge bosons -- can
be made purely real.  This means that in order for a process in the
minimal standard model to violate CP, it must involve in an essential
way two or more of these interactions.

If either mass matrix, $M_d$ or $M_u$, has two or more degenerate
elements, or if one or more of the mixing angles in the CKM matrix
vanishes, then with a physically-unobservable change of phases of the
quark fields, the CKM matrix can be made purely real and there is no
CP violation.  Thus if the combination 
\begin{eqnarray}
\label{jarlskog_det}
& d_{CP} = sin(\theta_{12}) sin (\theta_{23}) sin(\theta_{13}) \sin
\delta_{CP}  & \\
& \cdot (m_t^2 - m_c^2)(m_t^2 - m_u^2)(m_c^2 - m_u^2)(m_b^2 -
m_s^2)(m_b^2 - m_d^2)(m_s^2 - m_d^2) & \nonumber
\end{eqnarray}
vanishes, CP violation vanishes. The basis-invariant formulation of
this statement can be found in ref. \cite{jarlskog}.  Furthermore, it
is essential for baryogenesis that not only CP but also C be
violated.  Due to the chiral nature of the SU(2) gauge interaction,
left and right chiralities of quarks have different interactions, so
that ${\cal L}_{MSM}$ has both C-even and C-odd pieces.

The above remarks make it evident that in order for MSM processes to
produce a baryon asymmetry, all 3 quark generations, as well as
dynamics which distinguishes between chiralities and which involve
quark interactions with both neutral and charged
bosons,\footnote{I.e., not only the Higgs vev but also $W^{\pm}$ or
charged components of the Higgs field.} must play a significant role.
In the next section we will describe how both these elements are
incorporated in the mechanism we investigate.

The ``conventional wisdom'', that CP violation originating from the KM
phase is too small to be relevant to the observed baryonic asymmetry,
results from arguing\footnote{Here we give a popular version of the
more sophisticated treatment of ref. \cite{s:sm88}.} that the only
natural scale for the baryogenesis problem is the temperature of the
electroweak phase transition, $T
\sim 100$GeV. One might think that at this temperature the Yukawa
interaction can be treated as a perturbation, because quark masses are
small compared with the temperature. Then, since the baryon asymmetry
is a dimensionless number, the quantity (\ref{jarlskog_det}) should be
divided by something with the dimension of (mass)$^{12}$. The natural
mass parameter at high temperatures seems to be the temperature
itself, so that the asymmetry is argued to be at most
\eq
\frac{n_B}{s}\lsi \frac{d_{CP}}{N_{eff}T^{12}} \sim 10^{-20}.
\label{naive}
\en
This reasoning has been widely accepted, but, as has been the case
with many ``no-go'' theorems, proves not to be watertight when
examined carefully.\footnote{The present paper is not the first one to
look for and point out possible loopholes in this argument.  In
refs. \cite{s:sm88,bks} a mechansim of MSM baryogenesis was explored
based on the possible existance of a Chern-Simons condensate in the
high temperature phase and the decay of non-trivial fluctuations of
gauge and Higgs fields during the first order electroweak phase
transition.  There, a measure of CP-violation in the effective action
for the gauge fields in the expanding universe was found to be
\cite{bks}
\eq
\frac{\alpha_W}{\pi^4}[\frac{\alpha_W}{m_W^2}]^4\frac{s_2s_3
\sin\delta m_t^4
m_b^4 m_s^2}{\alpha_s^2(m_s^2 + s_2^2 m_b^2)}\sim 10^{-15}
\en
which is not analytic in Yukawa coupling constants.  More recently, in
\cite{s:msm} it was observed that for some processes occuring in the
hot plasma, Yukawa interactions cannot be treated as a perturbation
neither in the unbroken nor in the broken phase due to mixing effects,
so that the formal argument leading to the estimate (\ref{naive}) does
not hold.}

In order to reveal one important point of weakness in the above
argument, it is instructive to consider CP-violation in the $K^0$
system.  Here the CP-violating parameter is known to be quite large,
$\epsilon = 2~10^{-3}$. By analogy with the above discussion one could
say that since $K^0 \leftrightarrow \bar{K}^0$ oscillations are
described by the box diagram, the typical scale associated with this
process is the momentum in the loop, $p \sim M_W$, leading to the
conclusion $\epsilon \sim d_{CP}/M_W^{12} \sim 10^{-17}$. This is,
however, wrong by 14 orders of magnitude!

Of course, everybody knows why this ``derivation'' is incorrect.  The
mass of the W-boson is not the only scale of the problem here. There
are numerous other scales, such as the mass of the kaon itself.
Moreover, since $m_K$ is {\em smaller} than the masses of the $c$, $b$
and $t$ quarks and $m_K \sim m_s$, a perturbative expansion in the
quark masses does not work, so that detailed calculation is
necessary.  The box-diagram contribution was found\cite{egn:epsilon},
in the original KM basis, to be:
\begin{eqnarray}
&& \left| \frac{ Im \; M_{12}}{\Delta m} \right| \sim
\frac{s^2_1c_1s_2c_2s_3c_3 \sin \delta_{CP} (m^2_t - m^2_c)}{s^2_1 c^2_1
c^2_3 \left[ c^4_2 m^2_c + s^4_2 m^2_t + \frac{2s^2_2c^2_2m^2_t
m^2_c}{(m^2_t - m^2_c)} \ln \left( \frac{m^2_t}{m^2_c} \right)
\right]}\cdot \nonumber \\
 && \Bigl\{c^2_2 m^2_c \left[ \frac{m^2_t}{(m^2_t-m^2_c)^2}\ln \left(
\frac{m^2_t}{m^2_c} \right) - \frac{1}{m^2_t - m^2_c} \right]
\nonumber \\
&& + \; s^2_2 m^2_t \left[ \frac{m^2_c}{(m^2_t - m^2_c)^2} \ln \left(
\frac{m^2_c}{m^2_t} \right) \; + \; \frac{1}{m^2_t - m^2_c} \right]
\Bigr\}.  \label{epsilon}\\ \nonumber
\end{eqnarray}
Evidently, the dependence on mixing angles and quark masses is much
more complicated than in eq. \ref{jarlskog_det}.\footnote{Sometimes it
is said that any CP-violating quantity, in particular the BAU, must be
proportional to $d_{CP}$ since it has a basis-invariant
representation, the ``Jarlskog determinant''\cite{jarlskog}. This
reasoning is incorrect, however, because $d_{CP}$ is not the only
CP-violating quantity with a basis-independent representation.  Indeed
$\epsilon$ itself, given in a particular basis in eq. \ref{epsilon},
provides an example of a quantity besides $d_{CP}$ with a
basis-independent representation, as is obvious since it is a physical
observable.}

It is interesting to note that the expression for $\epsilon$ contains
{\em no} dependence on the charge -1/3 quark masses.  How then does CP
violation disappear when $m_d = m_s$? If the $d$ and $s$ quarks were
degenerate in mass, then kaons and pions would be degenerate, and the
expression for mixing between particles and antiparticles would
contain box diagrams connecting, e.g., $d \bar{d}$ with $d \bar{s}$
pairs, etc., in such a way that the sum would vanish (GIM
cancellation).  While it is surely true that CP violation must
disappear if any pair of like-charge quarks is degenerate in mass, we
see from this example that this does not mean that CP-violating
quantities are manifestly proportional to all pairs of mass-squared
differences.  The first term in a Taylor expansion in mass-squared
differences may be a very poor representation of the dependence on
masses, for physically relevant values of masses!

We can draw two lessons for cosmology from the kaon example.  First,
one should look for a process involving some small energy scale so
that perturbative expansion in the quark masses does not work. This
scale should be of the order of the strange quark mass or less, since
otherwise the contribution from the strange quark will tend to cancel
the $d$-quark contribution.  Second, the analysis must be concrete and
based on some specific mechanism, otherwise one cannot know which
scale is relevant.

There are many different energy scales at high temperatures. We shall
argue that the most important of them (from the point of view of CP
violation) may be the thermal momenta $p$ of the quarks.  While the
typical momentum in the heat bath is of the order of the temperature,
some fraction of the particles carry much smaller momenta and, for
them, CP-violation can be substantial.  In order to make a real
calculation of the effect, one must choose a mechanism for
baryogenesis.  In the next section we discuss some of them, and make a
rough estimate of the asymmetry which can be generated in a particular
one.

\mysection{Overview of MSM baryogenesis}
\label{sec:overview}

\hspace*{2em} A number of schemes for baryogenesis at the electroweak
scale have been
suggested\cite{krs85,s:m^14,s:sm87,s:sm88,mcl:prl,als:pl,als:np},
\cite{tz:prl,tz:np,mstv,ckn:L1,ckn:L2,ckn:Y1,ckn:Y2,dhss,kst}
(see also
reviews\cite{s:nobel,mrts,s:rev,dolgov:rev,ckn:rev,turok:rev}).  These
mechanisms rely heavily on the dynamics of the first order phase
transition \cite{linde:pl77,linde:np81,linde:np83,linde:rep} during
the spontaneous breaking of the SU(2)xU(1) gauge symmetry, which is
assumed to occur through the nucleation of bubbles of the new phase,
at a temperature of about 100 GeV . Inside the bubbles, the vacuum
expectation value (vev) of the Higgs field is non-zero and assumed to
be large enough that anomalous processes with B-violation are switched
off.  However in the high temperature phase outside the bubbles, the
electroweak symmetry is unbroken and the rate of B-violating sphaleron
reactions is high, so that a net baryonic number density cannot be
maintained.  Hence baryogenesis must be related to the presence of the
bubble wall.

Roughly speaking, the mechanisms for ew baryogenesis can be divided
into two categories. In the first
category\cite{s:sm88,als:pl,als:np,tz:prl,tz:np,mstv}, non-trivial
configurations of the gauge and/or Higgs fields (sphalerons or thermal
fluctuations) are supposed to have CP-violating interactions with the
moving domain wall, biasing the anomalous B non-conserving processes
in such a way that net fermionic number is produced when these field
configurations decay.  In the second class of mechanisms (``charge
transport baryogensis''\cite{ckn:L1,ckn:L2}), CP-non-invariant
interactions between fermions and the bubble wall lead to a separation
of some CP-odd charge by the bubble wall, which is then converted to
an asymmetry in the baryonic number by sphaleron processes in the
unbroken phase.

In ref. \cite{s:msm}, one of us [MS] observed that if the CP
violation in the interaction of thermal quarks with the bubble wall in
the Higgs field is strong enough, it could result in a direct
separation of baryonic number in the second type of mechanism
mentioned above, without the need for separation of a surrogate CP-odd
charge. Baryons\footnote{As opposed to antibaryons, on account of the
known sign of the present asymmetry.} inside the bubble survive till
the present time because the rate of B-violating reactions is
(required to be!) highly suppressed in the broken phase, while the
anti-fermions outside the bubble (partially) disappear through
equilibrium B-violating reactions.

In this paper we will elaborate this mechanism\cite{ckn:Y1,s:msm} in
more detail and compute the baryonic asymmetry in it in the framework
of the MSM. This does not mean that we insist that this mechanism is
the best one; others should be investigated as well.  However this is
a good first case, since the MSM CP-violation effects can be explored
in a very simple and physically transparent way.  If the wall is thin,
as suggested by recent work\cite{s:condensate}, this is likely to be
the dominant mechanism.  The most important new element of the present
work is the understanding of how the GIM cancellation, present when
quarks are taken to have typical thermal momenta, can be circumvented
in particular regions of momenta.  One such relevant region of
momentum is identified and its contribution is estimated.  Presumably
in the other mechanisms there may be a similar failure of the
GIM cancellation when the relevant regions of momenta are treated
sufficiently accurately.

Ref. \cite{s:msm} both noted the importance of including thermal
interactions with gauge and higgs particles in the plasma, and
proposed a specific mechanism for producing the asymmetry.  While our
work here is a natural extension of that work, it differs in several
important ways.  In ref. \cite{s:msm}, the asymmetry in reflection
probabilities of quarks and antiquarks was estimated, taking a
contribution near the top quark reflection threshold (the dominant
region of phase space) and conjecturing that $O(\alpha_s)$ loop
effects would produce phases which would cause the net asymmetry to be
non-vanishing.  In this paper, we identify a mechanism which evades
the GIM cancelation without invoking loop effects, and really
calculate the asymmetry in reflection probabilities.  We use a purely
quantum mechanical treatment of the scattering, noting that the ordinary
scattering phase shift provides the non-trivial CP-conserving phase
needed to interfere with the CP-violating phase to give a difference
in reflection probabilities between quarks and antiquarks.  Our treatment
of the fermionic excitations in the hot medium is more exact than that
of \cite{s:msm} because we include some loop effects in the broken phase
\cite{s:msm} which are necessary to get a quantitatively
accurate result.  In this paper we also relate the asymmetry in
reflection probabilities to the final baryonic asymmetry by studying
the baryon number diffusion in the vicinity of the bubble wall
(section \ref{sec:asym}).

\mysection{Rough estimate of the BAU from MSM CP violation}
\label{sec:rough}
\hspace*{2em}  We have $n_B/s \sim [ J_{CP}] \cdot [
f_{sph} ]$/[entropy], where $J_{CP}$ is the baryonic separation
current produced by quark interactions with the wall and $f_{sph}$ is
the sphaleron efficiency in removing the antibaryon excess in the
unbroken phase.  Before going to the technical details of the full
calculation in the latter portions of this paper, we give a
qualitative discussion of CP-violation in quark scattering from the
domain wall, ignoring higher order corrections in $\alpha_s$.  This
will permit us to estimate the left baryonic current, $J_{CP}$.

Some difference in reflection probabilities between quarks and
antiquarks is possible due to the interference between the
CP-violating phase in the coupling of the thermal quarks\footnote{By
working with the quasiparticles of the high temperature plasma,
interactions with gauge and higgs particles are included and
non-trivial CP violation is possible.  See section \ref{sec:thermal}.}
to the bubble wall, which changes sign in going from quarks to
antiquarks, and the ordinary (CP-conserving) scattering phase shift,
which is the same for quarks and antiquarks and is non-vanishing even
when $\delta_{CP} = 0$.  However how can we expect to ``evade'' the
GIM cancellation which lies at the heart of the conventional argument
that CP violation from the CKM matrix is far too small to account for
the observed baryonic asymmetry?  Evidently, an important variable for
the scattering problem is the momentum of the particle perpendicular
to the bubble wall, $p$.  The other essential scales are the mass,
$M$, of the particle and the inverse thickness of the wall, $a$.  The
component of the momentum parallel to the wall is conserved in the
scattering process and is unimportant in a qualitative
discussion\footnote{However parallel components play a non-trivial
role in the dynamics, contrary to the $T=0$ situation where no thermal
medium breaks Lorentz invariance, and could be quantitatively
significant, as discussed in section
\ref{sec:prediction} and appendix \ref{app:parallel}.}.

Let us discuss first the most typical situation, when $p \sim T,~ p
\gg M$, and $ p\gg a$. Then the interaction of fermions with the
domain wall is suppressed semiclassically by the factor $\exp(-\pi
p/a)$.  According to several estimates \cite{dlhll:pl,dlhll:pr} $a
\sim T/(40-10)$, so that for typical quark momenta in the plasma, $p
\sim T$, the light fermions do not scatter off the wall at all. The
only reflection coefficient which can be significantly different from
zero is that for the $t$-quark. It is clear, according to the
discussion in section \ref{sec:CPinSM}, that for this region of phase
space one cannot expect any non-negligible CP-violating effect.  If
the domain wall is very thin ($p/a \ll 1$), the reflection amplitude
is suppressed only by Yukawa coupling constants rather than by the
exponential factor.  Nevertheless, for $p \sim T$ the light quarks are
effectively degenerate due to the tiny difference between their masses
as compared to their typical momenta, so that their contributions to
the net separation of baryon number, when summed over all generations,
cancel nearly perfectly due to the GIM mechanism.  That is, even if
the phase in the CKM matrix means, say, that a $d$ is more likely to
reflect than a $\bar{d}$, this contribution to the baryonic asymmetry
will be nearly perfectly compensated by, say, the $\bar{s}$ being more
likely to reflect than the $s$, with the cancellation being perfect if
$m_s = m_d$.  In this region the ``conventional wisdom'' reasoning is
correct.

Let us consider now the region of phase space in which the momentum
perpendicular to the wall is low enough that the interactions of the
$s$ quark with the bubble wall are strong and its reflection
coefficient does not contain any powers of Yukawa coupling constants.
Roughly, we have
\begin{displaymath}
J_{CP} \sim [fraction ~of ~phase ~space] \cdot [ asymmetry ~of~
fluxes ]
\end{displaymath}
\begin{displaymath}
\cdot [CP ~violation] \cdot [dynamical~ details].
\end{displaymath}
The breaking of the GIM cancellation is most profound in the region of
the thermal spectrum in which the $s$ and $\bar{s}$ are totally
reflected, but the $d$ and $\bar{d}$ are partially reflected and
partially transmitted.  The fraction of phase space corresponding to
this situation is of order $\frac{m_s-m_d}{T}$, so that CP violation
vanishes when $m_s=m_d$, as required.

Another important factor arises because the contribution to the
baryonic current coming from quarks incident from the two sides of the
wall would exactly cancel if the Fermi-distributions on the two sides
of the wall were the same.  How the distributions near the wall differ
from the equilibrium distribution is not yet understood due to
uncertainties in the physics of bubble propagation and interaction
with the quarks.  However just from the motion of the themal medium
with respect to the wall, the flux in the wall rest frame of particles
of a given energy incident from the unbroken phase is generally
greater than the flux incident from the broken phase, producing an
asymmetry even for equilibrium distributions.  The asymmetry in the
fluxes is thus expected to be $ \gsi 2v ( p / T) n_F (1 - n_F)$ for small
wall velocity, $v$, but less than $\sim n_F$, as it would be if the
wall carried along all the quarks ahead of it, eliminating the flux
from the broken phase.

The natural measure of CP-violation is just\\ J $\equiv
sin(\theta_{12}) sin(\theta_{23}) sin(\theta_{13}) sin(\delta_{CP})$.
Global fits to determine CKM parameters place J in the range\cite{gn}
$(1.4-5.0)10^{-5}$.  Of course, there will be a further dependence on
quark masses and dynamical details, but this can give us a very rough
estimate to use as a guide for our expectations.  The analogous
estimate of $\epsilon$ is just J, so that for the kaon system the
dynamics increases the ratio of CP violating to CP conserving rates by
a factor of O(100).  An explicit analytic expression for the
additional quark mass and other dynamical dependence under certain
conditions is given in section \ref{sec:analytic}.

Putting together the factors above and dividing by the entropy
density, $s=2 \pi^2 N_{eff} T^3/45$, where $N_{eff}\sim 100$ is the
total number of particle degrees of freedom, one anticipates a baryon
asymmetry of order
\eq
|n_b/s| \sim (3-9)10^{-10}~f_{sph}f_{dyn}f_{flux},
\en
where $f_{dyn}$ accounts for additional dynamical effects which will
be included in a real calculation. Studies of the behavior of the
expanding bubble wall suggest that $v$ is in the range $0.1 - 0.9$, so
the flux asymmetry factor $f_{flux}$ can range from $10^{-1}p/T$ to
1.  For the one dimensional problem, total reflection of the
quasi-particles occurs for $p/T \sim 10^{-1}$.  We shall see in
section \ref{sec:asym} that $f_{sph}$ is estimated to lie in the range
$\sim 10^{-4} - 1$.  We see, then, that the CP violation present in
the CKM matrix may be sufficient to account for the observed baryonic
asymmetry of the universe. In any case, the conventional estimate
$\lsi O(10^{-21})$ must be discarded.

Predicting the sign of the asymmetry requires a quantitative
calculation.  As shall be seen from the results given in section
\ref{sec:results}, the crude estimate given in this section is roughly
correct in magnitude.  Moreover the sign predicted by the quantitative
calculation does agree with the observed positive sign, corresponding
to an excess in our universe of baryons rather than antibaryons.

Another region of phase space which {\it a priori} might seem
interesting is that in which the $c$ quark and antiquark are totally
reflected, but the $u$'s are not.  However since $c-u$ is not the most
degenerate like-charge pair, lifting their degeneracy dynamically is
less significant than for the $d-s$ pair.  As could be expected and is
born out in our quantitative results, the near degeneracy of the $d-s$
quarks asserts itself in a reduction of the degree of asymmetry in the
scattering of charge +2/3 quarks.  We also note that regions in which
only the heaviest quark is totally reflected, or none are totally
reflected, do not make important contributions to the separation of
baryon number, since in these regions the dynamics of the $d$ and $s$
quarks are essentially indistinguishable.\footnote{This can cease to
be true when higher order corrections, which mix momentum scales and
can generate phases which do not rely on total reflection, are
included.  Then it may happen that in the tradeoff between minimizing
the GIM cancelation and maximizing the volume of phase space, another
region can be more important.}

This qualitative discussion can only be considered as a guide for a
real calculation.  To be more quantitative, we must examine in greater
detail the mechanism of baryonic number separation by the domain wall,
which requires finding the correct excitations in the hot plasma, then
computing the reflection coefficients for their scattering off the
domain wall.  One thus determines the baryonic current which is
produced by a flux of equal numbers of quarks and antiquarks on the
domain wall, from either side.  In the next section, we imagine that
this current is known and we investigate the connection between this
baryonic current and the present-day $n_B/s$.  After devoting sections
\ref{sec:thermal} and \ref{sec:scattering} and a number of appendices to
developing the necessary technology to do the quantitative
calculation, we report the results of this calculation in section
\ref{sec:results}, confirming the heuristic estimates presented in
this section.

\mysection{Baryonic number separation and baryonic asymmetry}
\label{sec:asym}

\hspace*{2em}

The interactions of thermal quarks and antiquarks with the domain wall
are CP-noninvariant. Nevertheless, unitarity and CPT constraints
relate different transition amplitudes in such a way that the net
current of any C- or CP-odd number vanishes in thermal equilibrium
(see section \ref{sec:scattering} and ref. \cite{ckn:Y1}). Of course,
there is nothing surprising in this, since it would be too naive to
expect BAU generation in thermal equilibrium in any mechanism.

A first order ew phase transition, however, provides a deviation from
thermal equilibrium.  In our case the important manifestation is in
the movement of the domain wall.  Due to the interaction with the
medium, the domain wall is expected to move with a constant velocity,
which is estimated \cite{khl:wall,dlhll:pl,dlhll:pr,lmt:wall} to be $v
\sim 0.1-0.9$.  Thus in the wall rest frame there is a net flux of
particles and almost equal flux of antiparticles\footnote{In the
absence of CP-violation the fluxes would be precisely the same.}
flowing from the side of unbroken phase toward the side of broken
phase.  A crucial quantity for the mechanism of baryogenesis which we
study is the {\it difference} between the fluxes on the two sides of
the wall, viewed from the wall rest frame.  This difference of fluxes
depends on the extent to which the passage of the wall disturbs the
equilibrium distributions of the quarks.  Treatments of the
problem\cite{khl:wall,dlhll:pl,dlhll:pr,lmt:wall} based on a
perturbatively-calculated Higgs potential have envisaged a wall which
is sufficiently thick compared to a mean free path that the
equilibrium is approximately maintained.  That is, the velocity and
temperature of the medium is essentially the same on either side of
the wall.  We present below a calculation of $n_B(J_{CP})$ in this
case.  However recent work (c.f., \cite{s:condensate}) suggests that
non-perturbative effects may cause the transition in the Higgs vev to
occur much more abruptly than previously imagined, opening the
possibility that the quasi-equilibirum assumption is a poor
approximation.  We will return to the consequences of this possibility
below.

Let us go to the rest frame of the wall and assume that the medium
going through it with velocity $v$ has the same temperature inside and
ouside the bubble.  Since fluxes from the two sides are unequal, the
unitarity constraints do not apply and the asymmetry in reflection
coefficients can produce a non-zero baryonic current $J_{CP}$ (as well
as currents of other C- and CP- odd quantities such as left and right
baryonic number, $J_{CP}^L$ and $J_{CP}^R$).  If B-violating processes
are in equilibrium in the unbroken phase but not in the broken phase,
such currents imply the separation of baryonic number and other
numbers.

For small $v$, we will have an approximately static situation in the
wall rest frame, in which each of these currents, say $J_{CP}^L$, is
balanced by an opposite current through the wall due to an excess of
the appropriate density, in this case $L$ baryonic density.  Particle
densities will depend on the temperature and velocity in the usual
way, but will contain chemical potentials for the various CP- and C-
odd numbers which are non-vanishing in the broken phase. To determine
these chemical potentials, one must solve the kinetic equations in the
vicinity of the domain wall.  We shall write the kinetic equations in
a diffusive approximation, which is valid when the characteristic
length scale of the density variation is large compared with the mean
free path of the quarks. As will be seen $a~posteriori$, this
condition is satisfied if the velocity of the domain wall is not too
large ($v \lsi 0.3$), although it would be very nice to find the
relation between $J_{CP}$ and $n_B/s$ also for large velocities.

Quarks and leptons take part in many processes on both sides of the
wall with many different time scales.  In order to decide which
processes must be included in the equation describing diffusion, one
needs to define a relevant time scale and then include processes
taking place on shorter scales than that.  Let us follow the evolution
of a particle after it has been reflected from the domain wall towards
the unbroken phase.  Its typical distance from the bubble wall is
given by
\eq
\sqrt{D t} - vt,
\en
where the first term describes the random walk of the particle in the
rest frame of the plasma ($D$ is the diffusion coefficient for the
particle of interest) and the second term describes the motion of the
bubble wall.  This particle will be trapped by the bubble after a time
$t_D \sim D/v^2$, so one can neglect any process with a characteristic
time $\tau \gg t_D$. (Of course, B-violation must be included in any
case, since if it is absent, no baryonic asymmetry can be produced.)
The characteristic diffusion time $t_D$ is much smaller than the time
scale of the chirality breaking $L-R$ transitions coming from
inelastic scatterings with Higgs particles, since those are suppressed
by small Yukawa coupling constants.\footnote{Except for $t$-quarks,
however this can be neglected since the asymmetry resides in the
charge-1/3 quark sector.}  However the QCD sphaleron produces $L-R$
transitions at rate $\sim \frac{8}{3}(\frac{\alpha_s}{\alpha_W})^4
\sim 300$ times larger than the ew sphaleron rate\cite{mms:prd}.  Thus
even though the ew sphaleron only couples to $L$ chiral particles and
antiparticles, on account of the action of the QCD sphaleron, we can
simply discuss the total baryonic number, with small modifications in
the diffusion equations as compared to the case without the QCD
sphaleron.

Let $n_B (x,t)$ and $n_L(x,t)$ be the densities of baryonic and
leptonic numbers in the rest frame of the wall. We place the wall at
$x=0$ and take $x>0$ to be the broken phase where sphaleron processes
are switched off.  Then the diffusion equations are, for $x>0$
\eq
\frac{\partial}{\partial t}\left( \begin{array}{c}n_B\\ n_L
\end{array}\right)=
\left( \begin{array}{cc}
 D_B \frac{\partial^2}{\partial x^2}- v \frac{\partial}{\partial x} &
0\\ 0&D_L \frac{\partial^2 }{\partial x^2}- v \frac{\partial
}{\partial x} \end{array}\right)
\left( \begin{array}{c}n_B\\ n_L  \end{array}\right) ,
\label{diffun}
\en
but for $x<0$ we have
\eq
\frac{\partial}{\partial t}\left( \begin{array}{c}n_B\\ n_L
\end{array}\right)=
\left( \begin{array}{cc}
 D_B \frac{\partial^2}{\partial x^2}- v \frac{\partial}{\partial
x}-\frac{3}{2}\Gamma & -\Gamma\\ -\frac{3}{2}\Gamma&D_L \frac{\partial^2
}{\partial x^2}-
v \frac{\partial }{\partial x}-\Gamma \end{array}\right)
\left( \begin{array}{c}n_B\\ n_L  \end{array}\right) ,
\label{diffbr}
\en
where $\Gamma = 9 \Gamma_{sph}/T^3$ and $\Gamma_{sph}$ is the rate of
sphaleron transitions per unit time and volume.  (The relation between
the rate of B-violation and the sphaleron rate can be found in refs.
\cite{bs:higgs,ks}.)  $D_B$ and $D_L$ are diffusion constants for
quarks and leptons respectively\footnote{According to the argument of
ref. \cite{khl:Y}, any kinetic equation written in terms of {\em bare
particles} should also contain non-trivial corrections associated with
Debye screening of the hypercharge. These corrections, however, are
absent for equations (\ref{diffun},\ref{diffbr}) describing
quasi-particles.  Precisely on account of the Debye screening
phenomenon, quasiparticles do not carry hypercharge or any other
gauged quantum number, while they can have non-zero global numbers
like baryonic number and flavor.  A physical excitation in the high
temperature plasma with the global quantum numbers of a left quark
would, for instance, be actually composed of a left chiral quark,
gluons, electroweak gauge bosons and Higges, such that it is actually
an SU(3)xSU(2)xU(1) singlet.  This is discussed further in section
\ref{ss:uncertainties}.  Of course, the baryonic charges of
quasi-particles differ from the baryonic charges of the bare
particles. This fact is automatically taken into account in our
approach, since we define the various currents in terms of
quasi-particles (see section \ref{sec:thermal}). We thank S.
Khlebnikov for a discussion of this point.}.

These equations must be supplemented by a number of boundary
conditions.  At $x \rightarrow - \infty$ (the unbroken phase far from
the domain wall), $n_B$ and $n_L \rightarrow 0$ because we are
considering the case that the BAU does not exist prior to the ew phase
transition.  At $x \rightarrow + \infty$, $n_B$ and $n_L$ must be
finite.  The other boundary conditions are specified at $x=0$; we will
discuss them later.

We will assume that the velocity of the domain wall is low enough that
some ``steady state" solution to the kinetic equation can be
established.\footnote{Actually, this is only in order to enable us to
solve the problem relatively simply.  If the velocity is larger, the
effects are presumably larger, but since we find an interesting level
of asymmetry even in this very conservative approximation, we do not
attempt to extend the treatment to large velocities.  That is an
interesting and important subject to develop, however.} In other
words, in the rest frame of the domain wall the densities of particles
are time-independent.  For $x>0$ the only solutions consistent with
the boundary conditions are constants:
\eq
n_B(x) \equiv B_+,~~n_L(x) \equiv L_+.
\en
For $x<0$ they have the form
\eq
\left( \begin{array}{c}n_B\\ n_L  \end{array}\right)= \sum C_i \xi_i
\exp (k_i x),
\en
where $C_i$ are arbitrary constants, and $\xi_i$ and $k_i$ are
eigenvectors and eigenvalues to be found from the condition
\eq
\label{diffk}
\left( \begin{array}{cc}
 D_B k_i^2 - v k_i -\frac{3}{2}\Gamma & -\Gamma\\ -\frac{3}{2}\Gamma&D_L k_i^2-
v
k_i-\Gamma
\end{array}\right)\xi_i = 0 .
\en
One of the eigenvalues, say $k_4$, is zero independently of the
parameters.  It corresponds to a non-zero density of the conserved
number (B-L), so we put $C_4 =0$.  One can show that two other
eigenvalues, $k_1$ and $k_2$, have positive real part and the third
one, $k_3$, is real and negative. The latter corresponds to a growing
exponent as $x
\rightarrow - \infty$ and is therefore not consistent with the boundary
conditions. Hence, $C_3 = 0$. All three non-zero roots can be found
analytically.  They are solutions of the cubic equation
\eq
D_B D_L k^3 - v (D_L + D_B) k^2 - (D_B \Gamma + \frac{3}{2} D_L \Gamma - v^2) k
+ \frac{5}{2}v\Gamma =0.
\en
We shall write down here the relevant roots ($k_1$ and $k_2$) in two
limiting cases, when the dimensionless quantity $3 D_B \Gamma /v^2$ is
small or large.

(i) $3D_B \Gamma /v^2 \ll 1$. Then
\eq
k_1 = \frac{v}{D_B}(1 + \frac{3 \Gamma D_B}{2 v^2}),
\en
\eq
k_2 = \frac{v}{D_L}(1 + \frac{\Gamma D_L}{v^2}),
\en
\eq
\xi_1 = \left( \begin{array}{c}1\\ \frac{3\Gamma D_B^2}{2 v^2 D_L}
\end{array}\right),~~
\xi_2 = \left( \begin{array}{c}-\frac{\Gamma D_L}{v^2}\\1
\end{array}\right) .
\en

(ii) $3D_B \Gamma /v^2 \gg 1$. Then
\eq
k_1 = \frac{5v}{3 D_L},
\en
\eq
k_2 = \sqrt{\frac{3\Gamma}{2 D_B}} + \frac{v}{2D_B},
\en
\eq
\xi_1 = \left( \begin{array}{c}1\\ -\frac{3}{2}  \end{array}\right),~~
\xi_2 = \left( \begin{array}{c}1\\
\frac{D_B}{D_L}(1-\frac{v\sqrt{2}}{\sqrt{3\Gamma D_B}})  \end{array}\right) .
\en
In these expressions we have approximated $D_L \gg D_B$, since leptons
do not have strong interactions.

In order to define matching conditions at $x = 0$, one has to find the
total currents for baryonic and leptonic number at $x = 0$.  Let us
denote the densities of baryonic and leptonic numbers next to the
domain wall, in the unbroken phase, by $B_-$ and $L_-$.  Then, in the
absence of CP-violating effects, the baryonic and leptonic currents
flowing through the wall toward positive $x$ would be, respectively
\eq
j = \kappa(B_- - B_+) - \frac{1}{2} \left(D_B \frac{\partial (B_+ +
B_-)}{\partial x} - v (B_+ + B_-) \right)
\en
and
\eq
l = \kappa(L_- - L_+) - \frac{1}{2} \left( D_L \frac{\partial (L_+ +
L_-)}{\partial x} - v (L_+ + L_-) \right) .
\label{l}
\en
The factor $\kappa$ connects the density in a box to the current
flowing through one side of the box.  Essentially all quarks are
relativistic, including those which are responsible for our effect
which have small momentum perpendicular to the domain wall, because
even these typically have large parallel momenta.  Thus we have simply
\eq
\kappa = \frac{\int_0^{\frac{\pi}{2}}\cos\theta d \cos\theta}
{\int_0^{\pi} d \cos\theta} = \frac{1}{4}.
\label{kappa}
\en

In addition to the above currents arising from non-uniformities in
particle densities, there is an additional contribution to the
baryonic current due to the CP-violation present in the interaction of
quarks scattering from the bubble wall, denoted $J_{CP}$.  There is no
such contribution to the leptonic current, since CP is conserved in
the leptonic sector due to the masslessness of the neutrinos.
Altogether we have then for the baryonic and leptonic currents:
\eq
J_B = j + J_{CP},~~J_L = l.
\label{landj}
\en
Current continuity implies that the currents through the wall are
equal to those in the broken phase where there is no B or L
violation.  The currents in the broken phase are just due to the bulk
transport of the charge densities with velocity $v$, so that
\eq
J_B = v B_+,~~J_L = v L_+.
\label{bc1}
\en
Next equate the currents flowing into the unbroken phase to the total
rate of quantum number non-conservation due to sphaleron transitions,
\eq
-vB_+ = -v L_+ = \int_{-\infty}^0\Gamma(\frac{3}{2} n_B + n_L),
\label{bc2}
\en
which in turn satisfies
\eq
\int_{-\infty}^0\Gamma(\frac{3}{2} n_B + n_L) =D_B \frac{\partial B_-}{\partial
x} - v B_- = D_L \frac{\partial L_-}{\partial x} - v L_-
\label{bc3}
\en
because we require a static solution to (\ref{diffbr}).

{}From eqns (\ref{bc1}-\ref{bc3}) we learn $\frac{\partial L_-}{\partial
x} = -v (L_+ - L_-)/D_L$, which when combined with eqns (\ref{l},
\ref{landj}, and \ref{bc1}) implies $\frac{\partial L_-}{\partial
x} = 0$.  Thus $L_+ = L_-$ and $\frac{C_2}{C_1} =
-\frac{\xi_{21}k_1}{\xi_{22}k_2}$.  Similarly $B_+ - B_- =
J_{CP}/\kappa$; when combined with $B_+ = L_+$ this gives an equation
relating $C_1$ and $C_2$ to $J_{CP}/\kappa$.  Thus the baryonic number
density inside the bubble
\eq
B_+ = \frac{3}{5\kappa} J_{CP} f_{sph}(\rho),
\label{BAU}
\en
with $\kappa \sim 1/4$ (eqn. \ref{kappa}), $\rho = {3D_B \Gamma \over
v^2}$ and
\eq
f_{sph}(\rho) = \frac{5}{3}\frac{k_2-k_1}{k_2 (1-\xi_{11}/\xi_{21}) -
k_1 (1 -\xi_{12}/\xi_{22})}.
\en
$\xi$ is the matrix whose first and second columns are the
eigenvectors corresponding to eigenvalues $k_1,k_2$ respectively,
defined by eq.  (\ref{diffk}).  For $\rho \gg 1$, $f_{sph}(\rho) = 1$
and for $\rho \ll 1$, $f_{sph}(\rho) = \frac{5}{6}\rho$.  The physical
importance of $\rho$ is clear, since it represents the typical number
of sphaleron transitions to which a quark is exposed between the time
it enters the unbroken phase and the characteristic time at which the
bubble of low temperature phase envelops it.  Baryonic number density
vanishes when baryon-number violation is effectively turned off, i.e.,
$\rho \rightarrow 0$.  It also vanishes in thermal equilibrium, as required,
since $J_{CP} =0$ when $v=0$. $B_+$ is the net baryon number density
in the low temperature phase which, when divided by entropy, gives the
desired asymmetry, $n_B/s$.

We remind the reader that the relation of eq. \ref{BAU} is valid in a
quasi-equilibrium approximation, and provides a lower limit for a more
general situation.  For instance, if all the quarks were carried along
in front of the wall, the sphaleron rate would not matter at all, as
long as it is large compared to the expansion rate of the universe.

The next sections are devoted to determining $J_{CP}$.

\mysection{Fermionic excitations in the hot plasma and CP-violation}
\label{sec:thermal}

\hspace*{2em} At {\em zero} temperature and omitting loop effects,
the interaction of quarks with a (neutral) domain wall would be C- and
CP-conserving \footnote{The reason that loop corrections modify the
assertion that C is not violated is that they involve interactions of
the quarks with charged as well as neutral components of the Higgs
doublet. Furthermore, the ew gauge interaction distinguishes between
$L$ and $R$ in the covariant derivative, providing yet another source
of C violation.}.  Indeed, the Lagrangian ${\cal L}$ (see
eq.~\ref{ewlagr}) is CP- and C- invariant for any $x$-dependent Higgs
vev of the form
\eq
\phi = (0, v(x)),
\en
if all the gauge fields are zero. In other words, the separation of
any C-odd or CP-odd quantum number is not possible in this
approximation.

An even more general statement is true: as long as the Lagrangian for
quarks has the form
\eq
{\cal L} = i \bar{L} \not{\cal{D}} L + i \bar{R} \not{\cal{D}} R +
\bar{L} M (x) R + h.c.,
\en
where $M(x)$ is an arbitrary mass matrix and the operator ${\cal{D}}$
is the same for $L$ and $R$ terms, C is not violated. Since C relates
the left quark to the right antiquark, the reflection coefficients
will be the same for left quark as for right anti-quark and separation
of baryonic number (which is odd under C) is not possible.  However
chiral currents such as axial quark number, $\bar{\Psi} \gamma_{\mu}
\gamma_5 \Psi$, are C-even but CP-odd, so that they $can$ be produced
by this Lagrangian, as long as it violates CP.  Several extensions of
the standard model have been developed in refs.
\cite{ckn:L1,ckn:L2,ckn:Y1} making use of this fact to produce a BAU
in the electroweak phase transition.

The case of {\em non-zero} temperatures is quite different.  In the
high temperature plasma, quarks and antiquarks interact incessantly.
Each flavor and chirality of quark has a distinct interaction with the
Higgs particles in the heat bath, lifting the degeneracy between them.
Interactions with SU(2) gauge bosons further split $L$ and $R$
chiralities, and even their common interaction with gluons is
dynamically important since it affects their propagation.  As is well
known, it is essential to work in the correct basis of particle
states. In standard perturbation theory one simply takes the quadratic
part of the Lagrangian and finds particle states for it.  At zero
temperatures and densities this usually is successful for theories
with small coupling constants, since higher order corrections are
small.  Fermion masses, for instance, receive corrections of the form
$f^2 m_f$, where $f$ is a Yukawa coupling constant.  The high order
correction are small if $f \ll 1$.  However, at high enough
temperatures and densities, naive perturbation theory fails even for
theories with small couplings. The reason is that at high temperatures
an additional dimensionful parameter appears, namely the temperature
$T$ itself.  Corrections to masses can be large when the product of
Yukawa coupling and temperature is comparable to the zero temperature
mass.  In more physical language this means that the particles
appearing in the tree Lagrangian are not the actual particle
excitations of the problem under consideration. There are many
examples known from condensed matter and statistical physics in which
particle excitations have little in common with the fundamental
particles: phonons, sound waves, plasmon excitations, etc.

In order to find physical excitations one usually constructs the
effective Lagrangian for the theory incorporating high temperature
and/or high density effects, and then determines a better set of
fields for doing perturbation theory. This problem is not very well
defined mathematically, since it is not clear precisely what the word
`better' means. In practice, one usually calculates all mass operators
of the theory and constructs fields corresponding to the poles of
`exact' propagators (usually defined as an infinite sum of some subset
of graphs). One gets in this way the properties of one-particle
excitations of the medium and can consider the interactions between
them, which will be, hopefully, weak enough. Of course, if one would
be able to solve the problem exactly, the choice of variables would
not matter at all. However, since we are confined to using
perturbation theory for most problems, the starting approximation is
very important.

Although we can solve the quantum mechanical problem of quarks
scattering from the domain wall of Higgs vev without the use of
perturbation theory, we still must be careful in our choice of
particle states, in order to be able to ignore the (most significant)
interactions of the particles with the heat bath during the course of
their propagation through the domain wall.  That is, we must determine
what happens with the {\em fermionic excitations of the plasma} when
they go through the domain wall. First, we have to define them far
from the domain wall: outside the bubble, where the symmetry is
unbroken, and inside, where it is broken.

Quite an extensive literature exists on the fermionic excitations in a
hot plasma.  For the reader's convenience we will partially summarize
what is known about them from the literature, and then describe some
additional properties which, as far as we know, are not discussed in
the literature elsewhere.  We work to 1-loop accuracy in the
quasi-particle propagators.  This has the physical consequence
of neglecting inelastic scattering of the quasiparticles, which may
not be an adequate description of the problem, but is at least a first
step.  Improving this approximation is quite non-trivial for a number
of reasons.  For further discussion see section \ref{ss:uncertainties}.

\subsection{Unbroken phase}
\label{ss:unbroken}
\hspace*{2em} Fermionic excitations correspond to the poles
of quark propagators at high temperature.  For a review see, e. g.,
ref. ~\cite{Kalashnikov}. We work in Minkowski space and use the
following convention for the tree level Dirac operators:
\eq
\Sigma^0_{L,R} = \omega \pm \vec{\sigma}\cdot \vec{p}
\en
with $\omega > 0$ for particles and $\omega < 0$ for antiparticles,
$\sigma_i$ are the usual Pauli matricies. Then, in high temperature
approximation ($\omega$ and $|p| \ll T$) the one loop fermionic mass
operator in the gauge basis for the unbroken symmetry phase
is\cite{klimov,weldon}:
\eq
\Sigma^u_{L,R}(\omega,{\bf p}) = \Omega_{L,R}^2\left(\pm
\frac{\vec{\sigma} \cdot \vec{p}}{p^2}(1-F({{\omega}\over{p}})) -
\frac{1}{\omega}F({{\omega}\over{p}})\right),
\en
where
\eq
F(x) = \frac{x}{2}\left(\log(\frac{x+1}{x-1})\right).
\en
Note that Lorentz invariance is broken because the plasma rest frame
provides a preferred frame.  For the left quarks:
\eq
\Omega_{L}^2= \frac{2\pi\alpha_s T^2}{3} + \frac{3\pi\alpha_W
T^2}{8}\left(1 +
\frac{\sin^2\theta_W}{27}+\frac{1}{3}\frac{(M_u^2 + K M_d^2
K^{\dagger})}{M_W^2}\right).
\label{OmegaL}
\en
For the right quarks with charge $+2/3$:
\eq
\Omega_{R}^2= \frac{2\pi\alpha_s T^2}{3} +
\pi\alpha_W
T^2\frac{2\sin^2\theta_W}{9} +
\frac{\pi\alpha_W}{8}M_u^2\frac{T^2}{M_W^2},
\en
and for the right quarks with charge $-1/3$:
\eq
\Omega_{R}^2= \frac{2\pi\alpha_s T^2}{3} + \frac{\pi\alpha_W
T^2}{2}\frac{\sin^2\theta_W}{9} +
\frac{\pi\alpha_W}{8}{M}_d^2\frac{T^2}{M_W^2},
\en
where the first and the second terms come from the gluon and weak
gauge boson corrections, respectively, and the last term comes from
Higgs exchange.

An important point is that in spite of the fact that the vacuum
expectation value of the Higgs field is zero, particle excitations in
the unbroken phase are some specific mixture of the initial fields.
The physical fermionic fields (denoted by the bold letters) are three
component spinors in flavor space,
\eq
{\bf L} = O~\Psi_{Q_L},~~ {\bf U} = \Psi_{U_R},~~{\bf D} =
\Psi_{D_R},
\en
where the unitary matrix $O$ diagonalizes the matrix ${\Omega}_L^2$,
\eq
O{{\Omega}_L^2} O^{\dagger} \equiv {\omega_L^2} = diag.
\label{omega_L}
\en
Since $M_u \gg M_d$ the matrix $O$ is close to one.  In the right
sector the mixing is absent and we will use the notation
\eq
\Omega_{R} = \omega_{U,D}.
\label{omega_R}
\en
We use $\omega _0$ to represent any one of the $\omega_{L,R}$. The
dispersion relation for physical excitations has the form
\eq
\left[\left(\omega -\frac{\omega_o^2}{\omega}F({\omega\over p})\right)^2
-\left(p+\frac{\omega_o^2}{p}[1-F({\omega \over p})]\right)^2 \right]
=0.
\label{exactdisp}
\en
For each chrality, there are two solutions to this dispersion
relation.  For small momenta the spectra are
\eq
\omega^2(p)_{\pm} =\omega_{o}^2(1 \pm
\frac{2}{3}\frac{p}{\omega_{o}} +
\frac{7}{8}\frac{p^2}{\omega_{o}^2} + ...),
\label{lomom}
\en
while for high momenta, $p \gg \omega_{o}$, we have
\eq
\omega^2(p)_{+} = p^2 + 2\omega_{o}^2 -
\frac{\omega_{o}^4}{p^2}\log\frac{p^2}{\omega_{o}^2}, \; \; \;
\omega^2(p)_{-} = p^2(1+4\exp[-2\frac{p^2}{\omega_{o}^2}-1]).
\label{himom}
\en
The dependence of energy on $p$ is shown in Figs. \ref{dr_s_UB} and
\ref{dr_b_UB} for the strange and bottom quarks, in the approximation
of neglecting mixing.\footnote{There is no visible difference between
the strange and down quark dispersion relations on the scale of these
figures, but it is non-vanishing.} Unlike the situation at zero
temperature, the eigenstates are split due to their differing
interaction with Higgs particles in the plasma, even when the vev
vanishes.  Moreover at every energy there are two distinct collective
excitations having different momenta.  This phenomenon is analogous to
photon excitations in the plasma: in addition to the usual transverse
excitations, a longitudinal one also occurs.  We will call the mode
labeled $+(-)$ normal (abnormal) respectively.  Note the mass-gap
which these solutions exhibit.  It does not contradict the chiral
invariance of the underlying Lagrangian, but is connected with the
breaking of Lorentz invariance at non-zero
temperatures\cite{klimov,weldon}.  Higher order corrections to these
dispersion relations have been studied in a number of
papers\cite{smilga,brpis}.  It was shown in \cite{smilga} that the
abnormal branch is actually unstable at momenta $p > gT$.

The knowledge of the Dirac operator allows us to construct the
effective Lagrangian for the fermionic excitations in the plasma.  Let
us take for definiteness left chiral fermions and consider one
fermionic flavor\footnote{The treatment of right chiral fermions and
multiple flavors is completely analogous.}. Then, the effective
Lagrangian is:
\eq
{\cal L}_{eff} = L^{\dagger}(\Sigma^0_L + \Sigma^u_L)L.
\label{lagr}
\en

To become more familiar with the properties of the various
excitations, let us study them first for small momenta $p$. As we
shall see later, reflection amplitudes for particles are most
substantial in this kinematic range. At small momenta (see
eq.~\ref{lomom}) the energies of the fermions are close to the mass
gap, $\omega_0$. Expanding the Dirac operator for small $p$ and small
$\omega - \omega_0$ one obtains a linearised version of the
Lagrangian,
\eq
{\cal L}_{eff} = 2 i L^{\dagger}(\partial_0 -
\frac{1}{3}\vec{\partial}\vec{\sigma} + i \omega_0)L.
\label{linlagr}
\en
One can define in the usual way creation and annihilation operators
for the normal $(a_n^{\dagger},a_n)$ and abnormal
$(a_a^{\dagger},a_a)$ excitations, so that the part of the field $L$
which annihilates particles can be decomposed as\footnote{The part for
creation of antiparticles is obtained by expanding the Lagrangian for
small $\omega + \omega_0$.}
\eq
L(x) = \frac{1}{\sqrt{2}}\int
\frac{d^3k}{(2\pi)^3}\left(a_n^i(k)e^{-i\omega_+ t + ikx}u_i +
a_a^i(k)e^{-i\omega_- t + ikx}v_i\right),
\en
here the two-component spinors $u$ and $v$ obey the equations
\eq
(|k| + \vec{k}\vec{\sigma})u = 0,~~(|k| - \vec{k}\vec{\sigma})v = 0
\label{spinors}
\en
and
\eq
\omega_{\pm} = \omega_0 \pm \frac{1}{3}|k|,~~
\{a(k),a^{\dagger}(k')\}_+ = (2\pi)^3 \delta(k-k').
\label{dr_lin_eqn}
\en
{}From (\ref{spinors}) one can see that the two branches have different
relations between their chirality and their helicity.  For the normal
branch the chirality and helicity are equal while for the abnormal
branch the helicity is opposite to the chirality.

In the same way one can construct the effective Lagrangian for the
right chiral excitations. We do not write the corresponding equations
for this case; they can be derived from the equations for the left
particles using $\omega_0$ corresponding to right quarks.

The direction of the group velocity of the abnormal
excitation is opposite to the direction of $\vec{k}$,
since from eq.~\ref{lomom} the group
velocities of the two branches ($v = \frac {\partial\omega}{\partial
k}$) are
\eq
\vec{v}_+ =
\frac{\vec{k}}{3|k|},~~\vec{v}_- = -\frac{\vec{k}}{3|k|}.
\en
It is interesting and physically important that for both branches the
magnitude of the group velocity $\rightarrow$ 1/3 as the momentum
vanishes.  For large momentum, $k > \omega_0$, the group velocities of
both branches $\rightarrow$ 1, as can be found from eq.~\ref{himom}.

The left baryonic current is
\eq
J^0 = \int\frac{d^3k}{(2\pi)^3}(a_n^{\dagger}a_n + a_a^{\dagger}a_a)
\en
\eq
J^i=
\int\frac{d^3k}{(2\pi)^3}\frac{k^i}{3|k|}(a_n^{\dagger}a_n -
a_a^{\dagger}a_a).
\label{def_current}
\en
The minus sign in front of the abnormal mode contribution reflects the
fact that the abnormal excitation moves in the opposite direction to
$\vec{k}$.

We need to transform between the plasma rest frame and the wall rest
frame.\footnote{We
thank G. Baym for pointing out an error in our original discussion of
the following, and for calling our attention to ref. \cite{baym-chin},
where a detailed discussion of the Lorentz transformation properties
of quasiparticles can be found.} $(\omega(k),\vec{k})$ is a 4-vector,
so that the
dispersion relation in the moving frame is easily obtained from the
invariance of $(\omega(k),\vec{k})\cdot (1,v)\gamma$.  Denoting the
wall rest frame variables by $(\omega(k),\vec{k})$ and the
plasma-rest-frame energy by $\bar{\omega}$, we have
$\bar{\omega} = \gamma(\omega - \vec{k}\cdot\vec{v})$, where $\vec{v}$
is the velocity of the plasma with respect to the wall.  Since the phase
space volume is a Lorentz invariant, if we denote the Fermi
distribution in the plasma rest frame by $n_F(\bar{\omega})$, then
in the rest frame of the wall, where the plasma moves with velocity
$v$ normal to the wall, the number distribution is $n_F(\omega -
\vec{k}\cdot\vec{v})$.  Thus in the wall-rest-frame, particles with
momenta in the direction of the plasma have a higher density than if
the plasma were at rest with respect to the wall, as expected intuitively.

To give a simple example of the frame dependence of the dispersion
relation, let us work in linear approximation so the plasma-rest-frame
dispersion relation is given by eq. (\ref{dr_lin_eqn}), and let $v$ be
non-relativistic.  This gives
\eq
(\omega - v k_t) = \omega_0 \pm \frac{1}{3}\sqrt{k_{||}^2 + (k_t - v
\omega)^2}
\en
where $k_t$ is the component of the momentum parallel to $v$ (i.e.,
perpendicular to the domain wall) in the wall rest frame, and $k_{||}$
is the orthogonal component. From here one obtains the dependence
$\omega = \omega(k_{||},k_t,v)$.  Continuing with this simple example,
and taking $k_{||} = 0$, one finds the relationship between $\omega$
and $k$ in the wall rest frame, for the normal and abnormal
excitations respectively:
\eq
\omega_{n,a}= \omega_0(1 \mp \frac{1}{3}v) \pm \frac{k_t}{3}(1 \pm
\frac{2 v}{3}).
\en

{}From the definition of the current (\ref{def_current}) and these
considerations, we have for the flux factor in non-relativistic
approximation, but now for
a general dispersion relation and not just the linearized version:
\eq
\int\frac{d^3 k}{(2\pi)^3}n_F(\frac{\omega - v
k_t}{T})\frac{\partial\omega}{\partial k_t}.
\label{J^Lgroup}
\en

The formal expressions for the left fermionic current in terms of the
effective fields will be
\eq
J^0_L = L^{\dagger}(1 + \frac{\partial\Sigma^u_L}{\partial\omega})L,
\en
\eq
J^i_L = L^{\dagger}(\sigma_i + \frac{\partial\Sigma^u_L}{\partial
p_i})L
\label{J^L}
\en
and analogous relations for the right baryonic current.

\subsection{Broken phase}
\label{ss:broken}
\hspace*{2em} In order to determine the properties of the fermionic
excitations in the broken phase one has to calculate the Dirac
operator, which now will be a matrix in the space of right and left
fields. In the one loop high temperature approximation the Dirac
operator is
\eq
\left( \begin{array}{cc}
\Sigma^0_L + \Sigma^b_L & M \\
M^{\dagger}&\Sigma^0_R + \Sigma^b_{R} \end{array}\right).
\label{diracT}
\en
The mass term, $M$, is proportional to the scalar field vev, and
depends on whether one is considering the charge +2/3 ($U$) or
-1/3($D$) sector:
\eq
M_{U} = \frac{2g_w M_u}{M_W} \phi,~M_{D} = \frac{2g_w K M_d}{M_W}
\phi.
\en
In high temperature approximation, one-loop corrections to the
off-diagonal mass term describing left-right transitions, $M$,
vanish.

In the broken phase electroweak gauge bosons and quarks have non-zero
mass already in tree approximation so the one-loop contribution to the
Dirac operators, $\Sigma^b_{L,R}$, of any diagram has the generic form
\[
\int\frac{d^3k}{(2\pi)^3}\frac{n_F(\epsilon_F)}{\epsilon_F}\left(
\frac{\pm \vec{\sigma} \cdot \vec{p}}{p^2}[1-F({{\epsilon_F
\omega}\over{|p||k|}} )] -
\frac{1}{\omega}F({{\epsilon_F \omega}\over{|p||k|}})\right)+
\]
\eq
\int\frac{d^3k}{(2\pi)^3}\frac{n_B(\epsilon_B)}{\epsilon_B}\left(
\frac{\pm \vec{\sigma} \cdot \vec{p}}{p^2}[1-F({{\epsilon_B\omega}
\over{|p||k|}} )] -
\frac{1}{\omega}F({{\epsilon_B \omega}\over{|p||k|}})\right)
\en
where the $+$ sign is for left fermions and $-$ for right ones, $n_B$
and $n_F$ are Bose and Fermi distributions, and $\epsilon^2_{B,F} =
k^2 + m^2_{B,F}$ with $m_{B,F}$ denoting the appropriate bosonic or
fermionic mass.

In comparison with the unbroken case, the Dirac operator in the broken
phase has additional corrections of the type $m_B/T$ and $m_F^2/T^2
\log(T/m_F)$. Corrections of the first type do not influence quark
mixing in the broken phase and are numerically small in comparison
with the leading flavor independent terms.  They reduce the
$W^{\pm},~Z^0$ and Higgs contribution to $\omega_{L,R}^2$ compared to
the unbroken phase, causing the coefficient of the $\alpha_W
\partial/\partial t$ term in $\Sigma$ to be multiplied by the factor
$1 - \frac{4 M_B}{\pi T}$, where $M_B$ is the relevant boson's mass at
temperature $T$.  The coefficient of the $\alpha_W \partial/\partial
x$ term in $\Sigma^u$ also receives a correction, however it is less
important than the correction to the coefficient of $\omega$ since
$p/\omega$ is O(1/8).  Thus the net effect of these corrections can
be described by replacing eq. (\ref{OmegaL}) for $\Omega_{L}^2$ by:
\eq
{\Omega^b_{L}}^2 \approx \frac{2\pi\alpha_s T^2}{3} +
\frac{3\pi\alpha_W T^2}{8}\left((1- \frac{4 M_W}{\pi T}) +
\frac{ sin^2\theta_W}{27} +\frac{1}{3}\frac{(M_u^2 + K M_d^2
K^{\dagger})}{M_W^2}(1-\frac{4 M_H}{\pi T})\right).
\en
We have discarded the correction to the $sin^2\theta_W$ term because
the correction is suppressed by an additional factor of $sin^2 \theta_W$
since the photon does not get a mass.
The difference between $\omega_R$ and $\omega_R^b$ can be discarded
because the Higgs and ew gauge boson contributions there are totally
insignificant (except for the splitting between $b_R$ and $s_R$ where
the splitting will be reduced by the factor $(1-\frac{4 M_H}{\pi
T})$).  Except where specifically noted, corrections due to the
difference between $\omega^b_L$ and $\omega^u_L$ are not included in
the results reported in this paper.\footnote{These corrections were
not included in the first preprint version of this work.}

Fermionic mass corrections in the diagram with a {\em gluon} loop are
comparable in magnitude to the corrections coming from loops
containing Higgs exchange, which are responsible for the
vev-independent splittings between eigenstates.  (See
eq.~(\ref{OmegaL}).)  Thus they must be taken into account and a good
approximation to the Dirac operator in the broken phase is
\eq
\Sigma^b_{L,R} = \Sigma^u_{L,R} + \delta\Sigma_{L,R},
\en
with
\begin{eqnarray}
\delta\Sigma_{L,R} & = &   \frac{16 \pi \alpha_s}{3}
\int\frac{d^3k}{(2\pi)^3}\frac{n_F(\epsilon_k)}{\epsilon_k} \cdot
\nonumber \\
& & \left(\pm
\frac{\vec{\sigma}\cdot \vec{p}}{p^2}[1-F({{\epsilon_k\omega}
\over{|p||k|}})] -
\frac{1}{\omega}F({{\epsilon_k\omega}\over{|p||k|}})\right)-
\label{brcorr}
\end{eqnarray}
\[
\frac{16 \pi \alpha_s}{3}
\int\frac{d^3k}{(2\pi)^3}\frac{n_F(|k|)}{|k|}\left(\pm
\frac{\vec{\sigma} \cdot \vec{p}}{p^2}[1-F({{\omega}\over{|p|}})] -
\frac{1}{\omega}F({{\omega}\over{|p|}})\right) ,
\]
where for the left sector we have
\eq
\epsilon^2_k = k^2 + M^2_u + K M^2_d K^{\dagger}.
\en
For right quarks with charge $+2/3$, $\epsilon^2_k = k^2 + M^2_u$, and
for right quarks with charge $-1/3$, $\epsilon^2_k = k^2 + M^2_d$.

For sufficiently small momenta, $|p| \ll \omega$, and small fermionic
masses, $m_F \ll T$, this becomes
\eq
\delta\Sigma_{L,R}=  \frac{2 \alpha_s m_F^2}{3 \pi
\omega^2}\left( \pm (\log(\frac{\pi T}{m_F}) - \gamma -
\frac{1}{2})\vec{\sigma}\vec{p}+ (\log(\frac{\pi T}{m_F}) -
\gamma + \frac{1}{2})\omega \right),
\en
where $\gamma = 0.577$ is Euler's constant.  The expansion in small
fermionic masses is a poor approximation for the top quark, and the
integral (\ref{brcorr}) should be calculated precisely. Numerical
integration shows that the change due to including a non-zero value of
the top-quark mass in the diagram with a gluon loop is about 10\% of
the contribution of the diagram with the Higgs loop, and, therefore,
can be ignored. This is not the case for the light quarks where the
Higgs contribution is much smaller than in the heavy-quark sector.
For example, a non-zero charmed quark mass in the diagram with a gluon
loop induces a {\em negative} correction to $\omega_0$ in the broken
phase, which is numerically about factor $2.5$ times larger than the
corresponding diagram with Higgs exchange.

Finding the physical eigenstates is more complicated in the broken
phase than it was for the unbroken phase.  Even ignoring mixing
between different generations (for a discussion of mixing in the
broken phase see appendix \ref{app:mixing}), the dispersion relation
is a quite complicated function of $\omega$ and $p$:
\eq
det \left( \begin{array}{cc}
\Sigma^0_L + \Sigma^b_L & M \\
M^{\dagger}&\Sigma^0_R + \Sigma^b_{R} \end{array}\right)=0.
\en
As an example Fig. \ref{dr_b_B} shows the dependence of $\omega$ on
$p$ for a $b$-quark in the broken phase, neglecting mixing. For low
momentum or small vev, $\phi$, there is a one-to-one correspondence of
the normal and abnormal branches in the broken and unbroken phases. In
particular, the remarks regarding the group velocity of the
excitations apply to the broken sector also.

It is worthwhile studying the properties of the fermionic excitations
in the broken phase for small momenta $p$. In complete analogy with
the unbroken case one can write a linearized effective Lagrangian,
again taking for simplicity the one flavor case:
\eq
{\cal L}_{eff} = 2 i L^{\dagger}(\partial_0 -
\frac{1}{3}\vec{\partial}\vec{\sigma} + i \omega_L)L +
2 i R^{\dagger}(\partial_0 + \frac{1}{3}\vec{\partial}\vec{\sigma} + i
\omega_R)R + L^{\dagger}M R + h.c.
\label{linbroken}
\en
Now, left and right chiralities are mixed through the mass term. (We
continue to denote by $\omega_{L,R}$ the different $\omega_0$'s
corresponding to the $L$ and $R$ interactions, as in the previous
section.)  This Lagrangian describes four different particle
excitations with the dispersion relations
\eq
\omega = \frac{\omega_L + \omega_R}{2} \pm \sqrt{\frac{M^2}{4} +
\left(\frac{\omega_L - \omega_R}{2}\pm \frac{|k|}{3}\right)^2}.
\label{dispbrok}
\en
In the absence of the Higgs induced mass term, the correspondence of
the signs in eq.(\ref{dispbrok}) to the various particle excitations
is as follows for $|k| < \frac{3}{2}(\omega_L -
\omega_R)$: $(+,+)\leftrightarrow~L$, normal;
$(+,-)\leftrightarrow~ L$, abnormal; $(-,-)\leftrightarrow~ R$,
normal; $(-,+)\leftrightarrow~ R$, abnormal.  Note that in the
unbroken phase there is a coincidence in the left abnormal and right
normal branch at {momentum,energy} $= {\frac{3}{2}(\omega_L -
\omega_R),
\frac{1}{2}(\omega_L + \omega_R)}$.  At larger $|k|$ the
correspondence between signs in eq. ~\ref{dispbrok} and the
excitations is changed and $(-,-)\leftrightarrow~ L$, abnormal;
$(+,-)\leftrightarrow~ R$, normal, so that the levels cross. This
level crossing in the unbroken phase persists even if one takes into
account higher order perturbative corrections, since no transitions
between SU(2) doublet L and SU(2) singlet R are allowed due to the
unbroken gauge invariance.  In the broken phase the Higgs condensate
is non-zero, and these levels do not cross.\footnote{However it is
possible that off-diagonal terms in the Dirac operator are produced by
non-perturbative effects, even in the unbroken phase. If true, the
dispersion relation in the unbroken phase would more closely resemble
that of the broken phase and in particular the level crossing would be
removed.  This is discussed in appendix \ref{app:parallel}.}

With the use of the dispersion relations one can define the
contribution of the particles in the broken phase to the baryonic
current as well as to the total momentum of the media, following the
procedure of the previous section.

\subsection{Full effective Lagrangian}
\label{ss:currents_eff_fields}
\hspace*{2em} Equation (\ref{diracT}) allows one to write down an
effective Lagrangian for the fermions in the background of the
$x$-dependent scalar field. It has the form:
\eq
{\cal L}_{eff} = L^{\dagger}(\Sigma^0_L + \Sigma^b_L(x))L +
R^{\dagger}(\Sigma^0_R + \Sigma^b_R(x))L +{\cal L}_Y(x).
\label{lagrtot}
\en
where the $x$-dependence in the $\Sigma$'s comes from the
$x$-dependent masses of quarks in loop corrections to the propagator,
as discussed in the previous section.

This effective Lagrangian has an important property: it does not
conserve C and CP- symmetry, nor parity, provided the Higgs field is
$x$-dependent. As noted previously, this would not be the case if
thermal effects, namely the $W^{\pm},~Z^0$ and Higgs loop corrections
to the fermionic mass operator, were neglected. Moreover, due to the
fact that left and right mass operators are different, the interaction
of left handed particles with the domain wall is not the same as the
interaction of right handed antiparticles. Thus, separation of C- and
CP-odd quantum numbers is potentially possible, in contrast to the
zero temperature case. Of course, the CP-violating properties of this
Lagrangian are the same as that for the initial electroweak
Lagrangian, so that if (with three generations) there is a degeneracy
in the up or down sectors or some of the mixing angles are zero there
exists a transformation removing all complexities in the kinetic terms
and mass matrix simultaneously, and ${\cal L}_{eff}$ would be CP
conserving.  This however is not the case in nature.

With the help of this effective Lagrangian one can determine the
conserved baryonic current related to the problem.  The current with
components
\eq
J^{\mu}_B =J^{\mu}_L + J^{\mu}_R
\en
is conserved, where the left and right baryonic currents are defined
as in eq.(\ref{J^L}).

Having fixed the relevant properties of the quasiparticle excitations
in the hot plasma we can turn now to the question of their scattering
on the domain wall.

\mysection{Quark scattering from domain walls}
\label{sec:scattering}
\subsection{Preliminaries}
\hspace*{2em}
As we have argued in section \ref{sec:rough}, we expect the most
interesting part of phase space to be that in which the $s$-quark is
reflected from the domain wall.  Let us identify the momenta for which
this occurs.  We begin with a static domain wall, and discuss the
moving domain wall later on.

There are two conserved quantities in the interaction of the quark
with the domain wall, namely the energy, $\omega$, and the component
of the momentum parallel to the domain wall. For definiteness we will
first consider particles incident from the unbroken phase. Let us
consider the case when the total momentum is perpendicular to the
domain wall (the discussion of the more general case is contained in
appendix \ref{app:parallel}). Then (see Figs. \ref{dr_s_UB} and
\ref{dr_dsb_B}) at any energy $\omega > \omega_L =
0.502~T$,\footnote{Quoted numbers in this section are to clarify the
discussion and thus correspond to the figures, which do not include
mixing -- thus they do not correspond precisely to the results of the
real calculation where mixing is included.} there are 4 $s$-quark
excitations with different momenta, namely the left and right normal
and abnormal excitations.  At energies $\omega_L > \omega >
\omega^{min}_L$ (the minimum value of the energy of the left abnormal
branch) the normal left chiral excitation is absent and there are just
three types of $s$-quark excitations, $R_n, ~R_a,$ and $L_a$, however
for each value of the energy there are actually two distinct $L_a$
excitations having different momenta.  In the unbroken phase,
$\omega^{min}_L = 0.463~T$.  In the range $\omega_R = 0.459~T < \omega
< \omega^{min}_L$ there are just two excitations, $R_n$ and $R_a$,
while in the range $\omega^{min}_R < \omega < \omega_R$ there is only
a single type of excitation, the right abnormal one, with however two
distinct momentum states for each energy.  As will be seen below
(section \ref{ss:currents}), in order to compute the total baryonic
current in the 1-d problem it is sufficient to determine the
reflection coefficients for quarks (and, of course, antiquarks)
incident from the unbroken phase, which become $L$ upon reflection,
since all other contributions can be obtained from these.  Consistent
with angular momentum conservation, we have the following
possibilities:
\\
(i) $\omega > \omega_L$ : $R_a \rightarrow L_a,~R_n \rightarrow
L_a,~R_a \rightarrow L_n,~R_n \rightarrow L_n$,\\ (ii)$\omega^{min}_L
< \omega < \omega_L$ : $R_n \rightarrow L_a, ~R_a \rightarrow L_a$.\\
(iii) For $\omega <\omega^{min}_L$ no processes involving left chiral
particles are possible.\\

The interaction of the $s$-quark with the domain wall is strongest
when it is totally reflected. To find this region of $\omega$,
consider the dispersion curves in the broken phase. Due to the small
value of the $s$-quark mass in comparison with temperature, they look
almost the same as the dispersion curves in the unbroken phase, shown
in Fig. \ref{dr_s_UB}, except for a shift in $\omega^b_L$ relative to
$\omega^u_L$. The only qualitative difference is the absence of the
intersection of the right normal branch with the left abnormal one
near $\omega = \frac{1}{2}(\omega_L + \omega_R), ~~p = \frac{3}{2}
(\omega_L - \omega_R) \sim 6$ GeV.  This region is shown ``close up''
in Fig. \ref{dr_dsb_B}. Hence, for
\eq
1/2(\omega_L + \omega_R - m_s) < \omega < 1/2(\omega_L + \omega_R +
m_s)
\label{interval}
\en
instead of four solutions to the $s$-quark dispersion relations (as in
the unbroken phase) we have in the broken phase only two, which we can
designate left and right abnormal branches, with momenta about $40$
GeV.  Since chirality is not conserved in the presence of the
Higgs-induced mass, the labeling of these states is just a matter of
convention, which we fix by analogy with the zero-mass case.

Let us consider first what happens in the broken phase if we send,
say, a right normal $s$ or $\bar{s}$ from the unbroken phase towards
the domain wall, with energy just above the interval (\ref{interval}),
case (ii).  Helicity is conserved in transmission, so two transmitted
waves are possible: $R_n$, with practically the same momentum as the
incident $R_n$, and another, $L_a$, with a much higher momentum, $\sim
40$ GeV.  The transmission probability for production of the high
momentum mode ($L_a$) is semiclassicaly exponentially suppressed by
the factor $\exp(\pi p/a)$ \footnote{$a$ is the inverse wall
thickness.}, so that one can neglect this process.  Therefore, only
the right normal excitation will be transmitted to the broken phase
when a right normal excitation is incident.  For the same reason the
reflection probability for producing a left abnormal state from an
incident $R_a$ is exponentialy suppressed, since in this case the
reflected particle has large momentum in comparison with the momentum
of the incident particle. In other words, only the reflection process
$R_n \rightarrow L_a$ and the transmission process $R_n \rightarrow
R_n$ have non-negligible amplitudes.  If the energy is well above
(\ref{interval}) the reflection coefficient will for all cases be
small, since then the $s$-quark mass can be neglected.

Thus we can deduce that in the interval of energies (\ref{interval}),
the reflection coefficient for an incident $R_n$ $s$-quark is
essentially unity.  This is because in this energy range there is no
physical $R_n$ excitation of the $s$-quark in the broken phase, so
that the transmission process $R_n \rightarrow R_n$ is not allowed,
while at the same time the transmission coefficient for producing the
high momentum $L_a$ excitation, and the reflection coefficient for
$R_a \rightarrow L_a$, are still exponentially suppressed.  This is
the part of phase space where one can expect to have a substantial
contribution to the left chiral current, due to an asymmetry in the
reflection coefficients for $s,\bar{s}$ which can be significantly
different from the asymmetry in reflection coefficients for
$d,\bar{d}$ as long as the $d$-quark is not also totally reflected.
Thus to be precise, the energy range of interest is
\begin{eqnarray}
1/2(\omega^b_L + \omega^b_R - m_s) < \omega < 1/2(\omega^b_L + \omega^b_R -
m_d); \nonumber \\ 1/2(\omega^b_L + \omega^b_R + m_d) < \omega <
1/2(\omega^b_L + \omega^b_R + m_s).
\end{eqnarray}

Another region in which the reflection coefficients can be large, and
thus produce a net quark-antiquark asymmetry, occurs at a slightly
lower energy where there is a coincidence of solutions to the
dispersion relation for $s_L$ or $d_L$ and $b_R$ excitations, as can
be seen from Fig. \ref{dr_dsb_B}.  Due to off-diagonal pieces in the
matrix $ M $ (see eq. (\ref{diracT})), especially $m_b sin
\theta_{23}$ which mixes $b$ and $s$, these levels actually repel in
the broken phase, producing another region of total reflection in
which an (abnormal) $s_L$ incident from the unbroken phase is
reflected to a (normal) $b_R$ or vice versa.  This phenomenon produces
the lower energy region of non-vanishing asymmetry which will be
evident in the figures of section \ref{sec:results}.

The momentum of the $s$ and $d$ quarks in both these energy ranges is
small compared with $\omega_L$ and $\omega_R$.  This allows one to
expand fermionic self-energies with respect to momentum and keep only
the first term in $p$.  Thus the scattering can be described in terms
of first order differential equations, as will be discussed next.

\subsection{Basic equations}
\hspace*{2em} In section \ref{sec:thermal} we derived the effective
Dirac equation describing the interaction of quark excitations with
the scalar field.  Let the scalar field of the domain wall be
\eq
\phi = \phi_0 f (vt - r),
\en
where $\phi_0$ is the vacuum expectation value of the Higgs field
inside the bubble at $T=T_c$ , $v$ is the velocity of the bubble wall,
and $r$ is the bubble radius.  Letting $t=0$ at the moment of bubble
formation at $r=0$, $f = 0$ for negative argument and $f = 1$ for
positive argument larger than the wall thickness.  For sufficiently
large bubbles, the domain wall can be considered as planar and
perpendicular to, say, the $x_3$ axis, and in the rest frame of the
wall we have the Higgs vev:
\eq
\phi (x) = \phi_o F(x_3), \; \;  \mbox{with} \; \; F (- \infty) = 0, \;
F (+ \infty) = 1.
\label{profile}
\en

We will consider the scattering problem in the rest frame of the wall.
In order to solve the problem of reflection from the domain wall one
has to specify boundary conditions. Particles incident from the
unbroken phase and particles transmitted to the broken phase have a
wave function proportional to
\eq
\exp(-i\omega t + i \vec{k}\vec{x}),
\en
with $\omega >0$ and $\frac{\partial \omega}{\partial k_3} > 0$, while
reflected particles have a wave function proportional to
\eq
\exp(-i\omega t + i \vec{k}\vec{x}),
\en
with $\omega >0$ and $\frac{\partial \omega}{\partial k_3} < 0$. The
conditions on $\frac{\partial \omega}{\partial k_3}$ guarantee that
the direction of the group velocity has the correct sign.

The equations for antiparticles can be derived from those for
particles by CP conjugation. They differ from the equations for
particles only in one place: everywhere the CKM matrix $K$ appears, it
should be replaced by its complex conjugate, $K^*$. The boundary
conditions for antiparticles are the same as for particles.

The Dirac operator (\ref{diracT}) is highly non-local in space and
time and it is extremely difficult to solve the Dirac equation as it
stands. For a domain wall at rest the non-locality in time does not
matter very much, since energy is conserved so that every time
derivative can be eliminated by $\frac{\partial}{\partial t}
\rightarrow -i\omega$. The same applies to derivatives with respect to
the coordinates parallel to the domain wall, since the momentum
parallel to the wall is conserved: $\frac{\partial}{\partial x_2}
\rightarrow ik_2,~
\frac{\partial}{\partial x_1} \rightarrow ik_1$. However, the
dependence of the equation on $x_3$ is still non-trivial.  The
following physical consideration simplifies the problem substantially.
It is clear that scattering amplitudes should depend strongly only on
$k_3$, being the component of the momentum of the particle normal to
the bubble surface.  As we argued in the discussion of CP violation,
we are particularly concerned to treat reliably the regime of energy
in which the $s$ quark is reflected but the $d$ is not.  We shall see
that in this regime the perpendicular momenta are small compared to
$\omega$, the energy, so that we can expand the fermionic kinetic term
operator with respect to $k_3$ and take into account only the first
term. ( Improvement of this approximation is discussed in Appendix
\ref{app:quadratic}.) This results in a first order differential
equation which can be solved.  This kind of approximation is not good
at all for the up quark sector due to the large value of the top mass.
However, as we have argued in the previous section and shall confirm
below, the magnitude of CP-violating effects in the up-quark sector is
not important.

In this paper we solve the scattering problem only for the case when
the momenta of the fermions are normal to the bubble wall.  This
simplifies the problem significantly while giving results which are at
least qualitatively applicable.  A discussion of the formalism for the
general case is given in Appendix \ref{app:parallel}.  We will return
to the quality of this approximation in the discussion of
uncertainties in Section
\ref{sec:prediction}.  Henceforth we consider the 1 dimensional
scattering problem and denote the coordinate normal to the wall,
previously called $x_3$, simply by $x$.

The system of equations describing the reflection of fermions from the
domain wall is then: {\scriptsize \eq
\left( \begin{array}{cc}
\omega(1+\alpha_L+\beta_L)+ i\frac{\partial}{\partial x}(1+\alpha_L)
&{\cal M}_{U,D}\\ {\cal
M}_{U,D}^{\dagger}&\omega(1+\alpha_{R}+\beta_{R}) -
i\frac{\partial}{\partial x}(1+\alpha_{R}) \end{array} \right) \cdot
\en
\[
\left( \begin{array}{c}L\\ R \end{array}\right) =0
\] }
where $L$ and $R$ correspond to up and down components of
two-dimensional Weyl spinors which have 3 flavor components.  In the
plasma rest frame, and neglecting for the moment corrections due to
mass insertions in one-loop diagrams, the 3x3 diagonal matrices
$\alpha$ and $\beta$ are defined as
\eq
\alpha_{L,U,D}= \frac{1}{2}\beta_{L,U,D} =
-\frac{1}{3}\frac{\omega_{L,U,D}^2}{\omega^2}
\label{define_alpha}
\en
and
\eq
{\cal M}_{U,D}=O~M_{U,D},
\en
with $\omega_{L,U,D}$,$ M_{U,D}$ and $O$ as defined in section
\ref{sec:thermal}.  As in that section, we use the subscript $R$ to
generically denote the relevant $U$ or $D$ subscript for the right
chiral states.  Expressions for finite plasma velocity are given in
appendix \ref{app:velocity}.  Aside from a small (but important)
correction due to $W^{\pm}$ and Higgs masses, and quark
mass-insertions in a gluon-loop, discussed in section \ref{ss:broken},
the above expressions for $\alpha$ and $\beta$ are valid
in the broken as well as unbroken phases.  For the quantitative
calculations reported below we use the exact expressions including the
change in $\alpha$ and $\beta$ in going from unbroken to broken
phases.

These equations correspond to left chirality particles incident from
the unbroken phase (right particles are reflected in this case) and
right chirality particles incident from the broken phase (with left
particles reflected).  If the sign of the $i\frac{\partial}{\partial
x}$ term is reversed one gets the complementary cases. The treatment
of both equations is similar, so we will deal in this section with one
equation only, in the down quark sector to be concrete.

We introduce new variables
\eq
\label{PsifromLR}
\Psi = R^{-1}\left( \begin{array}{c} L \\ R \end{array} \right)
\en
where $R$ is a diagonal matrix
\eq
R=\left( \begin{array}{cc}R_{LL} & 0\\ 0&R_{RR} \end{array}\right),
\en
\[
R_{LL}= (1+\alpha_L)^{-1},~~ R_{RR} = -(1+\alpha_R)^{-1}.
\]
In terms of these new variables the equations can be rewritten in the
convenient form
\eq
\frac{\partial}{\partial x}\Psi = i D R \Psi,
\label{basic}
\en
where the hermitian matrix $D$ is defined by
\eq
D=\left( \begin{array}{cc}
\omega(1+\alpha_L+\beta_L) & {\cal M}
\\ {\cal M}^{\dagger}&\omega(1+\alpha_R+\beta_R)  \end{array}
\right).
\en
The expression for the baryonic current (see eq. \ref{J^L}) in terms
of these new variables is simply
\eq
J^B = \Psi^{\dagger}R\Psi .
\en

Consider first the asymptotics of the solutions at $x \rightarrow
-\infty$ and $x \rightarrow +\infty$. We can write
\eq
\Psi (x \rightarrow \pm\infty)  \rightarrow
e(\pm\infty)\exp(ip(\pm\infty)x)\Psi_{\pm},
\en
where $p(\pm\infty)$ are the eigenvalues of the matrix $D(\pm\infty)R$
and $ e(\pm\infty)$ is the matrix whose columns are the eigenvectors
of $D(\pm\infty)R$:
\eq
DR~e=e~p.
\label{DRe=pe}
\en

One can introduce also a scattering matrix (which is determined by
solving the Dirac equation) mapping the incident onto the outgoing
wave function:
\eq
\Psi_+ = V \Psi_-
\en

First we note that at $x \rightarrow -\infty$ all eigenvalues of $DR$
are real and are given by
\eq
p_i(-\infty) \equiv p^u_L = \omega
\frac{1+\alpha_L+\beta_L}{1+\alpha_L},~~i\leq 3
\label{pLun}
\en
\eq
p_i(-\infty) \equiv p_R^u = -\omega
\frac{1+\alpha_R+\beta_R}{1+\alpha_R},~~i > 3.
\label{pRun}
\en
The first three of these correspond to left chiral incident particles
and the last three to reflected right chiral outgoing
particles.\footnote{Since $\alpha_{L,R}$ and $\beta_{L,R}$ are
negative and proportional to $\frac{1}{\omega^2}$, for small enough
$\omega$ the sign of the momenta will change, but not the group
velocity.  Thus our asymptotic behavior is correct for all $\omega$.}

We cannot write the corresponding analytic expressions for the
particle momenta in the broken phase in closed form due to the mixing
of the quark generations.  We denote them as
\eq
p_i(+\infty) =p^b_L,~~i\leq 3 \; \; p_i(+\infty) = p^b_R,~~i > 3.
\en
These momenta are not necessarily real due to the fact that the matrix
$DR$ is not hermitean. Complex eigenvalues appear in complex conjugate
pairs (see Appendix \ref{app:dirac}). Physically, a pair of complex
conjugate eigenvalues corresponds to complete reflection of some
particular quark flavor eigenstate.

Let us order the particle momenta in the broken phase in such a way
that the first three correspond to propagating modes with positive
group velocity or to non-propagating modes with positive imaginary
part (this will produce an exponentially decaying wave function) and
the last three have negative group velocity or negative imaginary
part.

\subsection{Reflection and Transmission Coefficients}
\label{ss:refln_coefs}
\hspace*{2em} It is not difficult to relate the  scattering
matrix, $V$, to the reflection and transmission amplitudes. The wave
function corresponding to the reflection (and transmission) of a left
particle incident from the unbroken phase of, say, the first flavor,
has the form at $x
\rightarrow -\infty$
\eq
\Psi_{-} = \left( \begin{array}{c}
1\\0\\0\\r_{11}^u\\r_{21}^u\\r_{31}^u
\end{array} \right)
\en
and at $x \rightarrow +\infty$
\eq
\Psi_{+} = \left( \begin{array}{c}
t_{11}^u\\t_{21}^u\\t_{31}^u\\0\\0\\0
\end{array} \right)
\en
where $r^u$ and $t^u$ are the reflection and transmission coefficients
to be determined.  The overall phase is irrelevant to us, since we use
in the end only the magnitudes of $r_{ij}$ and $t_{ij}$.

The wave functions for right handed particles of the first flavor,
incident from the broken phase, have the asymptotic behavior, again
dropping an irrelevant phase:
\eq
\Psi_{-} = \left( \begin{array}{c}
0\\0\\0\\t_{11}^b\\t_{21}^b\\t_{31}^b
\end{array} \right), ~~x \rightarrow -\infty
\en
\eq
\Psi_{+} = \left( \begin{array}{c}
r_{11}^b\\r_{21}^b\\r_{31}^b\\1\\0\\0
\end{array} \right),~~x \rightarrow +\infty
\en
where $r^b$ and $t^b$ are the reflection and transmission coefficients
for the particles incident from the broken phase.  Note that the
labeling of which flavor is ``first'', etc. is arbitrary and not
related between broken and unbroken phases, except in the limit of no
mixing.

If we denote
\eq
V =
\left( \begin{array}{cc}
V_{LL} & V_{LR}\\ V_{RL} & V_{RR} \end{array} \right)
\en
then the reflection and transmission coefficients are:
\eq
r^u = -V_{RR}^{-1}V_{RL},~~t^u = V_{LL} - V_{LR}V_{RR}^{-1}V_{RL},
\label{ru}
\en
\eq
r^b = V_{LR}V_{RR}^{-1},~~t^b = V_{RR}^{-1}.
\label{rb}
\en

The determination of the scattering matrix $V$ is discussed in
Appendices \ref{app:dirac} and \ref{app:observables}.  Let us denote
the various reflection coefficients of interest as follows:
$(r^u_{LR})_{ij}$ is the reflection coefficient for left particle of
flavor $j$ incident from the unbroken phase, which becomes upon
reflection a right particle of flavor $i$. $r^b_{RL}$ is the
reflection coefficient matrix for right particles incident from the
broken phase, etc.  Thus for instance the $r^u$ given in the equation
above would be more precisely denoted as $r^u_{LR}$.

\subsection{CPT properties of amplitudes}
\label{ss:cpt_amps}
\hspace*{2em} CPT-invariance puts a number of constraints on the
reflection and transmission amplitudes. To find them, let us write all
the equations we have for our problem.\\ 1. For left particles
incident from the unbroken phase we have:
\eq
\frac{\partial}{\partial x}\Psi_1 = i D R \Psi_1
\label{psi1}
\en
2. Left (chirality!) anti-particles obey:
\eq
\frac{\partial}{\partial x}\Psi_2 = i D^* R \Psi_2
\label{psi2}
\en
3. Right particles obey:
\eq
\frac{\partial}{\partial x}\Psi_3 =- i D R \Psi_3
\label{psi3}
\en
4. Right (chirality!) antiparticles obey:
\eq
\frac{\partial}{\partial x}\Psi_4 = - i D^* R \Psi_4.
\label{psi4}
\en

One can see that equations (\ref{psi1}) and (\ref{psi4}) as well as
(\ref{psi2}) and (\ref{psi3}) are related by complex conjugation.
Therefore, the reflection and transmission coefficients for, say, left
incident particles are related to those for right incident
antiparticles. Let us find this relation.

As we noted in the previous subsection, and develop in detail in
appendix \ref{app:dirac}, the solution to equation (\ref{psi1}) can be
written in the form
\eq
\Psi_1 = eEV\Psi_0
\en
where $\Psi_0$ is some constant vector and $E$ is a diagonal matrix
whose entries are $\exp [i \int p dx]$ .  Evidently, then,
\eq
\Psi_4 = e^*E^*V^*\Psi_0^*.
\en
However in this expression at $x \rightarrow - \infty$, the first
three exponentials in $E$ correspond to waves going from right to left
while the last three correspond to the incident right anti-particles.
To make the calculation of the reflection coefficients transparent, we
wish to return to our convention in which the first three describe
(now antiparticle) waves going from left to right.  Therefore we
introduce the matrix $T$, with $T^2 = 1$, of 0's and 1's which
reshuffle the positions of the eigenstates relabeling the eigenvalues
in the broken phase in the following way:
\eq
\Psi_4 = e^*T^2E^*T^2V^*T^2\Psi_0^*, \; \;  TpT = p^*.
\en
Then, when $x \rightarrow + \infty$ the last three exponentials in
$TE^*T$ will correspond to the waves transmitted to the broken phase,
as desired. If some of the particles undergo complete reflection from
the domain wall then $ T \neq 1.$ If we denote
\eq
\left( \begin{array}{cc}
\bar{V}_{LL} & \bar{V}_{LR}\\
\bar{V}_{RL} & \bar{V}_{RR}  \end{array} \right)
\equiv TV^*,
\en
we see that the reflection coefficients for right chirality
antiparticles incident from the unbroken phase are determined by the
same $V$ which determined the reflection coefficients of $L$ particles
incident from the unbroken phase:
\eq
\bar{r}^u_{RL} = -\bar{V}_{LL}^{-1}\bar{V}_{LR},~~\bar{t}^u = \bar{V}_{RR}-
\bar{V}_{RL}\bar{V}_{LL}^{-1} \bar{V}_{LR},
\en
\eq
\bar{r}^b = \bar{V}_{RL}\bar{V}_{LL}^{-1},~~\bar{t}^b =
\bar{V}_{LL}^{-1}.
\en
This means, in particular, that there is no need to separately solve
all four initial equations; one can solve first two of them and CPT
will fix all other matrix elements.

\subsection{Unitarity constraints}
\hspace*{2em}From any of the equations one finds the general
statement that the `probability' current (coinciding in our case with
baryonic current) is conserved, namely
\eq
\frac{d}{dx}\Psi^{\dagger}R \Psi = 0.
\en
{}From here one can find a number of useful relations between reflection
and transmission amplitudes. In particular, the following relation
holds true at $x \rightarrow +\infty$:
\eq
V^{\dagger}E^*e^{\dagger}ReEV =V^{\dagger}R^bV = R,
\en
where the matrix $R^b$ is defined by
\eq
R^b =e^{\dagger}Re.
\en
It is a diagonal matrix when there is no total reflection and contains
non-diagonal pieces when some of the eigenvalues are complex. From
here one finds
\eq
{(t^u)}^{\dagger}R^b_{LL} t^u - {(r^u)}^{\dagger}R_{RR}r^u = R_{LL}
\en
\eq
{(r^b)}^{\dagger}R^b_{LL}r^b - {(t^b)}^{\dagger}R_{RR} t^b = -
R^b_{RR}
\en
which reflects the fact that one can calculate the baryonic current
either from transmission or from reflection coefficients.

Combining CPT symmetry and unitarity one finds the following relations
between reflection and transmission coefficients:
\eq
|(r^u_{LR})^{ij}|^2 R_{RR}^{ii}/R_{LL}^{jj} =|(\bar{r}^u_{RL})^{ji}|^2
R_{LL}^{ii}/ R_{RR}^{jj},~~
\en
\eq
|(r^b_{LR})^{ij}|^2 (R^b_{RR})^{ii}/(R^b_{LL})^{jj}
=|(\bar{r}^b_{RL})^{ji}|^2 (R^b_{LL})^{ii}/ (R^b_{RR})^{jj}.~~
\label{cpt_relns_rflcoefs}
\en
These relations guarantee the vanishing of any C and CP-odd currents
through the domain wall in thermal equilibrium, as will be seen when
they are used in the explicit expressions for currents given in the
next section.

\subsection{Baryonic current in terms of reflection coefficients}
\label{ss:currents}
\hspace*{2em}One can show (see Appendix \ref{app:flux_facs}) that
the thermal averages of the currents of interest are, for the one
dimensional problem:

1. If we send towards the domain wall an equal number of left quarks
and antiquarks from the unbroken phase, with the distribution of
quasiparticle momenta given by the fermi distribution in the unbroken
phase, their contribution to the net baryonic current is:
\begin{eqnarray}
\label{JuLR}
\langle J^u_{LR} \rangle & = & \int
\frac{d\omega}{2\pi}Tr\bigg(n_F({\omega,v})_{uL}
(R_{LL})^{-1}[(r^u_{LR})^{\dagger}R_{RR}r^u_{LR} \nonumber \\ & - &
(\bar{r}^u_{LR})^{\dagger} R_{RR}\bar{r}^u_{LR}]\bigg),
\end{eqnarray}
where the $\bar{r}$'s are the reflection coefficients computed with $K
\rightarrow K^*$.
\\2. The contribution of right quarks and antiquarks incident from the
unbroken phase is:
\begin{eqnarray}
\label{JuRL}
\langle J^u_{RL} \rangle & = & \int
\frac{d\omega}{2\pi}Tr\bigg(n_F({\omega,v})_{uR}
(R_{RR})^{-1}[(r^u_{RL})^{\dagger}R_{LL}r^u_{RL} \nonumber \\ & - &
(\bar{r}^u_{RL})^{\dagger} R_{LL}\bar{r}^u_{RL}]\bigg).
\end{eqnarray}
3. `Left' particles incident from the broken phase, now using the
equilibrium momentum distributions appropriate to the broken phase,
contribute:
\begin{eqnarray}
\langle J^b_{LR} \rangle & = & -\int
\frac{d\omega}{2\pi}Tr\bigg(n_F({\omega,v})_{bL}
(R^b_{LL})^{-1}[(r^b_{LR})^{\dagger}R^b_{RR}r^b_{LR} \nonumber \\ & -
& (\bar{r}^b_{LR})^{\dagger} R^b_{RR}\bar{r}^b_{LR}]\bigg).
\label{JbLR}
\end{eqnarray}
4. `Right' particles incident from the broken phase contribute:
\begin{eqnarray}
\langle J^b_{RL} \rangle & = & -\int
\frac{d\omega}{2\pi}Tr\bigg(n_F({\omega,v})_{bR}
(R^b_{RR})^{-1}[(r^b_{RL})^{\dagger}R^b_{LL}r^b_{RL} \nonumber \\ & -
& (\bar{r}^b_{RL})^{\dagger} R^b_{LL}\bar{r}^b_{RL}]\bigg).
\label{JbRL}
\end{eqnarray}
Here $n_F(\omega,v)_{uL,uR,bL,bR}$ are the distributions of the
particles in the rest frame of the wall defined in Appendix
\ref{app:flux_facs}.  Note that only
propagating modes in the broken phase contribute to the asymmetry, so
that the trace in eqs. \ref{JbLR} and \ref{JbRL} is taken only over
eigenstates having real $p_L^b$ and $p_R^b$.

The total CP-non-invariant component of the baryonic current through
the surface is given by
\eq
\langle J^B \rangle = \langle J^u_{LR} + J^u_{RL} + J^b_{LR}+
J^b_{RL} \rangle.
\en
It is the same in the broken and unbroken phases, since baryonic
number is conserved in the interaction with the scalar field. Left and
right fermionic currents are not conserved, and cannot be defined in
the broken phase. In the unbroken phase they are equal to
\eq
\langle J^L \rangle = \langle J^u_{RL} + J^b_{LR} \rangle
\en
and
\eq
\langle J^R \rangle = \langle J^u_{LR} + J^b_{RL} \rangle.
\en
QCD sphaleron transitions violate chirality, so that on time scales
long compared to $\Gamma_{QCDsph}^{-1}\sim 100/T$ these currents are
not separately conserved, even in the unbroken phase.

In thermal equilibrium with $v = 0$, all distributions of the
particles have the standard form $n_F(\omega)= (\exp(\omega/T) +
1)^{-1}$. In this limit, using
eqs. \ref{cpt_relns_rflcoefs}\footnote{Since the range of $\omega$ for
which $\Delta(\omega)$ is non-zero is very narrow and $n_F$ does not
change significantly in this range, we take it out of the integral and
evaluate it at the central value of $\omega$: $\bar{\omega}$.}
\begin{eqnarray}
\langle J^u_{RL} \rangle & = &  -\langle J^u_{LR} \rangle =
-\langle J^b_{LR} \rangle \nonumber \\ & = & \langle J^b_{RL} \rangle
=
\int\frac{d\omega}{2\pi}\Delta(\omega)n_F(\omega)\simeq
\frac{n_F(\bar{\omega})}{2\pi} \Delta_{int}
\label{JuRL_from_Delta}
\end{eqnarray}
where
\eq
\Delta(\omega)=
Tr((R_{RR})^{-1}[(r^u_{RL})^{\dagger}R_{LL}r^u_{RL}-
(\bar{r}^u_{RL})^{
\dagger}
R_{LL}\bar{r}^u_{RL}]).
\label{Deltaom}
\en
thus all C and CP-odd currents vanish due to the CPT and unitarity
relations between different amplitudes given in eqns.
\ref{cpt_relns_rflcoefs}, as expected.

The main quantity of interest for us is $\Delta(\omega)$, so in the
next section we turn to the problem of finding the reflection
coefficients.  If we are satisfied with the linear term at small $v$,
we can use the differential equation for $v=0$. $J_{CP}$ will be
non-zero due to the difference between fermi distributions for the
particles incident from broken and unbroken phases, when $v \ne 0$.
In fact, however, keeping the $v$ dependence in the differential
equation (see appendix \ref{app:velocity}) is quantitatively
significant for $v$'s in the expected range, $v > 0.1$, as is reported
in section \ref{sec:results} where we parameterize the velocity
dependence of $\Delta_{int}$ which enters $J^L_{CP}$ as $\Delta_{int}(v)
\equiv \Delta_0(1 + \zeta v)$.  In appendix \ref{app:velocity} it is shown
that for $J^R_{CP}$, $\Delta_{int}(-v)$ enters.

At small but non-zero plasma velocity $v$ in the wall rest frame, we
saw in section \ref{ss:unbroken} that
\eq
n_F(\omega,v)= \frac{1}{exp[\frac{\omega}{T}(1-\frac{ k v}{\omega})] +
1}.
\en
For the various contributions to the current enumerated in eqns
(\ref{JuLR}-\ref{JbRL}), we encounter different
$k$'s.  For instance for $J^u_{RL}$, $k\equiv k^u_R$ is the momentum of the
$R$ quasiparticle incident from the unbroken phase.  The main contribution in
this region comes from $s_R$ or $b_R$ being reflected to $s_L$ or
$d_L$.  The flavor-dependence of the $R-$quasiparticle momenta can be
neglected as it is
a higher order effect, so $k$ is just the momentum of the right normal
branch at $\bar{\omega} = 1/2(\omega_L^b + \omega_R)$\footnote{Recall
that there is no need to distinguish between $\omega^u_R$ and
$\omega^b_R$.}, which from eqn (\ref{dr_lin_eqn}) is
$k^u_R \equiv \frac{3}{2} (\omega^b_L - \omega_R)$.
For $J^b_{RL}$ we have right normal branch excitations incident from the
broken phase.  Thus $k$ for this case has the opposite sign as for $J^u_{RL}$.
Moreover since the right branch does not shift in going from broken to
unbroken phase, the magnitude of $k$ is the same as in the previous
case, so $k=-k^u_R$.  For $J^u_{LR}$ and $J^b_{LR}$ the incident
excitation is left abnormal.  As we saw in section \ref{ss:unbroken},
the abnormal branches have $k$ opposite in sign to the group velocity,
which must be positive for particles incident from the unbroken phase
and negative for particles incident from the broken phase.  Thus for
$J^b_{LR}$ we have $k= + k^b_L$.  Since total reflection occurs at the
value of $\bar{\omega}$
for which levels cross in the broken phase, $k_L^b = k^b_R$.
Similarly, for $J^u_{LR}$ we have $k=-k^u_L$, where $k^u_L \equiv
3 ( \omega_L^u - \bar{\omega})$.  Neglecting the shift in going from
the unbroken to broken phase for the average of $\omega_L$ and $\omega_R$
compared to the shift in that difference, we have from section
\ref{ss:broken} that $(\omega^b_L - \omega_R) = (1 - \frac{4
M_W}{\pi T})(\omega^u_L - \omega_R)$, so that $k^u_L = 3/2(\omega^u_L
- \omega_R)(1+\frac{4 M_W}{\pi T})$.  Combining the above with
eq. (\ref{JuRL_from_Delta}) gives:
\eq
J_{CP} = \frac{\Delta_{int}}{2 \pi} \left( n_F(\bar{\omega} - k^u_R v_u) -
n_F(\bar{\omega} + k^u_L v_u) + n_F(\bar{\omega} + k^u_R v_b) -
n_F(\bar{\omega} - k^u_R v_b) \right),
\label{JfromDelta}
\en
where the terms come in order from $J^u_{RL}$, $J^u_{LR}$, $J^b_{RL}$,
and $J^b_{LR}$ and $v_u,~v_b \geq 0$ are the velocities of the plasma
in the wall rest frame in the unbroken and broken phases.  In general
the temperature differs slightly in going through the wall and that
can be taken into account if quantitatively relevant.  If the actual
distribution functions are not those of thermal equilibrium, say
because the wall carries along particles in front of it,
then the true distributions may be substituted for the equilibrium
distributions we have assumed.

For the quasi-equilibrium assumption
of equal temperatures and equal velocities, we can make the small
velocity expansion of eq. (\ref{JfromDelta}).  In this case,
contributions from $J^u_{RL}$ and $J^b_{RL}$ cancel to leading order
in the flavor dependence of the fluxes, leaving:
\eq
J_{CP} = \frac{\Delta_{int}}{2
\pi}n_F(\bar{\omega})(1-n_F(\bar{\omega}))(k^u_L - k^u_R)v.
\label{JfromDelta_smallv}
\en
Inserting the expressions obtained above for $k^u_R$ and $k^u_L$, and
using $(\omega_L^u - \omega_R) \approx \frac{3\pi\alpha_W T^2}{16
\bar{\omega}}$, gives
\eq
J_{CP} = \frac{\Delta_{int}}{2
\pi} n_F(\bar{\omega})(1-n_F(\bar{\omega}))\frac{9 \alpha_W T}{4
\bar{\omega}}v.
\label{JfromDelta_smallv_final}
\en
We can also consider, as an extreme alternative, the case when the
distribution functions vanish in the broken phase, corresponding to
all quarks (in the relevant energy range) being swept along by the
wall.  Now the $v$ dependence of $\Delta_{int}$ (see appendix
\ref{app:velocity}) is of the same order as the $v$ dependence from
the fluxes.  Taking $\Delta_{int}(v) = \Delta_0(1 + \zeta v)$, we
find:
\eq
J_{CP} = \frac{1}{2 \pi}\left(n_F(\bar{\omega} - k^u_R v_u)\Delta_{int}(v) -
n_F(\bar{\omega} + k^u_L v_u) \Delta_{int}(-v) \right)
\label{JfromDelta_vb=0}
\en
\[
\approx \frac{\Delta_0}{2 \pi} n_F(\bar{\omega})\left(2 \zeta
+ (1-n_F(\bar{\omega}))\frac{3(\omega_L^u -
\omega_R)}{\bar{\omega}} \right)v,
\]
again having expanded for small $v$.

\mysection{Some analytical results}
\label{sec:analytic}

\hspace*{2em} Some properties of the fermions' reflection from the
domain wall can be studied analytically. Namely, the case without
fermionic mixing allows an analytical solution for some specific
profiles of the domain wall, while the real case with mixing can be
solved perturbatively in a thin wall approximation.

\subsection{No mixing.}
\hspace*{2em} If mixing between different quark flavors is absent and
the kinetic term is independent of the vev\footnote{I.e.,
$\alpha_{L,R}$and $\beta_{L,R}$ are $x$-independent, as is the case when
mass corrections to loops in the broken phase are dropped.},
then the quark scattering problem simplifies significantly.  We have
just two differential equations for the scattering problem, which can
be transformed to the equation for a hypergeometrical function for
some profiles of the domain wall. For instance, for
\eq
F^2(x) = \frac{1}{1+\exp(-ax)}
\label{profile_tanh}
\en
the problem can be converted to one which is solved in, e.g.,
\cite{landau}.  We do not go to the
details of the derivation and only present the result.  The
differential equations describing the scattering of the fermion on
the domain wll can be written as
\eq
(p_L^u + i \frac{\partial}{\partial x})L - M_R R = 0,
\en
\[
M_L L + (p_R^u + i \frac{\partial}{\partial x})R = 0.
\]
where $M_{L,R} = M/(1 + \alpha_{L,R})$, and $p^u_{L,R}$ are as
defined in eqns (\ref{pLun}) and (\ref{pRun}.  $M$ in this expression
is the higgs-induced mass in the broken phase, $\equiv M_{T=0} \left(
\frac{vev(T)}{vev(T=0)} \right)$.  Introduce
\eq
\bar{\omega} \equiv \frac{\omega}{2}\left(\frac{1+\alpha_L +
\beta_L}{1+\alpha_L}+\frac{1+\alpha_R + \beta_R}{1+\alpha_R}\right) \equiv
\frac{p^u_L - p^u_R}{2},
\en
and
\eq
B \equiv \bar{\omega}^2 - \frac{M^2}{(1+\alpha_R)(1+\alpha_L)}.
\en
There are two solutions to this differential equation. They are:
\eq
L_1(x) = \exp[i(\frac{(p^u_R + p^u_L)}{2} + \sqrt{B})x] F \left(\frac{1}{2} -
\frac{i}{a}( \bar{\omega} + \sqrt{B}),\frac{i}{a}( \bar{\omega} - \sqrt{B}), 1
-
\frac{2i}{a}\sqrt{B}; - e^{-a x}\right)
\en
and
\eq
L_2(x) = \exp[i(\frac{(p^u_R + p^u_L)}{2} - \sqrt{B})x] F \left(\frac{1}{2} -
\frac{i}{a}( \bar{\omega} - \sqrt{B}),\frac{i}{a}( \bar{\omega} + \sqrt{B}), 1
+
\frac{2i}{a}\sqrt{B}; - e^{-a x}\right).
\en
The right-handed component of the wave function corresponding to
these solutions is given by:
\eq
R_j(x) = \frac{1}{M_R}(p_L^u + i \frac{\partial}{\partial x})L_j(x).
\en
Here $F(\alpha,\beta,\gamma;z)$ is the usual hypergeometric function,
whose asymptotic behavior fixes the reflection coefficients.  The
result can be cast into the form:
\eq
|r_{L\rightarrow R}|^2 =
\left( \frac{1+\alpha_R}{1+\alpha_L} \right) \gamma(\omega)
\en
\eq
|r_{R\rightarrow L}|^2 =
\left( \frac{1+\alpha_L}{1+\alpha_R} \right) \gamma(\omega)
\en
with
\eq
\gamma(\omega) = \frac{sh2\pi(\bar{\omega} -
\sqrt{B})/a}{sh2\pi(\bar{\omega} + \sqrt{B})/a}
\en
for $B>0$, and $\gamma = 1$ for $B < 0$.

If the energy of the fermion is much higher than the barrier height,
then:
\eq
|r_{L\rightarrow R}|^2 =
\frac{2\pi M^2}{(1+\alpha_L)^2 a\bar{\omega}}\exp(-\frac{2\pi
\bar{\omega}}{a}).
\en
As expected, reflection from the domain wall is suppressed by the
semiclassical exponent in this case.
Note that this dependence of the amplitude on $M$ can be interpreted
as reflecting the perturbative coupling of the fermion to the Higgs
potential.  On the other hand, in the energy region where total
reflection occurs ($B < 0$) the amplitudes are of order 1. In this
energy regime perturbation theory in Yukawa coupling constants does
not work.

Complete reflection of some flavor corresponds to the case when $B <
0$.  With $m_t = 150$ GeV, $m_c = 1.6$ GeV, $m_u = 5$ MeV, $m_b = 5$
GeV, $m_s = 0.15$ GeV and $m_d = 10$ MeV, $T = 100$ GeV, and the
temperature-dependent Higgs vev such that $m_W(T_c)=50$ GeV, i.e.,
vev$= 150$ GeV, the regions of total reflection (ignoring flavor
mixing) are:
\begin{enumerate}
\item for $t$-quark:    $~~~\omega < 117.764$ GeV,
\item for $c$-quark:    $~~~47.892 < \omega < 48.886$ GeV,

\item for $u$-quark:    $~~~47.384 < \omega < 48.387$ GeV,

\item for $b$-quark:    $~~~49.491 < \omega < 52.531$ GeV,

\item for $s$-quark:    $~~~48.146 < \omega < 48.237$ GeV,

\item for $d$-quark:    $~~~48.188 < \omega < 48.194$ GeV.
\end{enumerate}

It is interesting to note that, contrary to expectations based on the
zero-temperature dispersion relation, the regions of complete
reflection are in general limited by some minimal and maximum values
of energies.  The reason is the peculiar dispersion relation of the
thermal excitations in the hot plasma: as $p$ is increased from $p=0$,
the energy of the abnormal branch initially decreases with increasing
momentum, as shown in figs. \ref{dr_s_UB} and \ref{dr_b_UB}.  We also
note that the region of $d$-quark reflection lies inside the region of
$s$-reflection, and $u$-quark reflection lies inside the region of
$c$-reflection.  The interesting region of $\omega$ is shown in
fig.~\ref{dr_dsb_B}.  As we shall see in the section on numerical
results, the region of $\omega$ in which the momentum of $b_R$ is
nearly degenerate with that of $s_L$ or $d_L$ also produces total
reflection, with a $b_R$ incident from the unbroken phase being
reflected as an $s_L$.

The consideration of this subsection confirms that the interesting
part of phase space is the region where reflection is substantial.
Consideration of the thin wall approximation, with CP-effects taken
into account, allows one to be even more specific.

\subsection{Thin wall approximation}
\hspace*{2em} Many properties of the fermionic interaction with the
domain wall can be seen already in thin wall approximation. In this
sub-section we will suppose that the function $F$ in equation
(\ref{profile}) has the step form
\eq
F(x) = \theta(x).
\en
Now the solution of the Dirac equation for all $x < 0$ is
\eq
\Psi = \exp(i p^u x)\Psi_0,
\en
and for $x>0$
\eq
\Psi = e \exp(i p^b x)V \Psi_0.
\en
We will suppose that the matrix of eigenvectors and eigenvalues are
known in both the unbroken and broken phases and that the first three
eigenvalues in the broken phase correspond to the transmitted waves,
as has been our convention.  The aim is to determine the scattering
matrix, $V$. The wave function is continuous at $x=0$, giving the very
simple relation
\eq
V = e^{-1} .
\label{V=e-1}
\en
The reflection coefficients, which are determined by $V$ (see
eqs. \ref{ru} and \ref{rb}), can therefore be expressed via the $x
\rightarrow +\infty$ eigenvalue matrices.  E.g., for incident $L$
quark by using $Ve = 1$ we find
\eq
r^u = e_{RL}(e_{LL})^{-1},~~r^b = - (e_{LL})^{-1} e_{LR}.
\label{rfrome}
\en

Now let us consider the propagation of the antiparticles. The
corresponding equation for them is
\eq
\frac{\partial}{\partial x}\Psi = i D^* R \Psi,
\label{basic1}
\en
with their eigenvalue problem in the broken phase giving
\eq
D^*R\bar{e} = \bar{e}\bar{p}^b.
\label{pbar}
\en
As demonstrated in appendix \ref{app:dirac}, one can prove that the
set of $p^b$ and $\bar{p}^b$ are the same.  This is fundamentally a
CPT theorem result.  We choose a basis in such a way that $p^b =
\bar{p}^b$.

With these simple relations we can investigate in which ranges of
energy CP-violating effects can be important.  Suppose first that all
particle modes can propagate in the broken phase. This means that all
eigenvalues are real in the broken phase.  Making then a complex
conjugation of eq.(\ref{pbar}) and comparing to eq. \ref{DRe=pe} one
finds that
\eq
\bar{e} = e^*
\en
and, therefore, $\bar{V} = V^*$. This means that reflection
coefficients for particles are precisely the same as those for
antiparticles. Therefore, in thin wall approximation, one cannot
expect any CP-violation effects in this region of the phase space.

The other extreme case is also quite simple. Suppose that all
eigenvalues are complex so that all fermions are completely reflected
from the domain wall. Clearly no separation of baryonic number can
occur here!

Now, let one fermionic flavor be reflected while all others are
transmitted.  This means, that, e.g., $p^b_1$ and $p^b_4$ are complex and
$p^b_1 = (p^b_4)^*,~Im p^b_1 > 0$ (by our convention).  Then, complex
conjugation of equation \ref{DRe=pe} for particles {\em does not} give
(\ref{pbar}). Instead, we get
\eq
\bar{e} = e^*T,
\en
where
\eq
T^2 = 1,~~TpT = p^*, ~~T \ne 1.
\en
So, we have
\eq
\bar{V} = T V^*,
\en
and reflection coefficients for particles and antiparticles are
different (see (\ref{rfrome})). The same conclusion is true also for
the case when two quark flavors are reflected.

It would be nice to have an analytical expression for the asymmetry
covering the whole range of quark masses and mixing angles.
Unfortunately, it does not generally exist even for the case of the thin wall.
Since we have (eq. \ref{V=e-1}) $V = e^{-1}$, simply finding the eigenvectors
in the broken phase is sufficient to solve the problem.  However doing this
requires finding analytically the roots
of a sixth order polynomial, which is not generally possible as is well
known.  However we can proceed perturbatively in the mixing and obtain
some useful insight.

Rather than deal with the 6 by 6 matrix $e$, we derive an equation for the
reflection coefficient $r^u$, a 3 by 3 matrix. From the eigenvalue equation
(eq. \ref{DRe=pe}) in the broken phase, $DRe = e~p^b$, one can
show\footnote{These results are for the case that the kinetic term is
vev-independent, i.e., $\omega^b_L = \omega^u_L$.  More generally, the
$p^u_{L,R}$ which appear should be replaced by $\tilde{p^b}_{L,R}$,
the broken phase eigenmomenta for ${\cal M} = 0$.}
\eq
p_L^u e_{LL} + {\cal M}R_{RR}e_{RL} = e_{LL} p_L^b
\label{fir}
\en
\eq
{\cal M}^{\dagger}R_{LL} e_{LL} + p_R^u e_{RL} = e_{RL}p_L^b.
\label{sec}
\en
Now, multiply eq.(\ref{fir}) by the matrix $e_{LL}^{-1}$ from the
right and eq. (\ref{sec}) by $e_{RL}^{-1}$ from the left. Then insert
$p_L^b$ defined by the second equation into the first equation. The
result of these transformations is:
\eq
p_L^u + {\cal M}R_{RR}e_{RL} e_{LL}^{-1} = e_{LL}e_{RL}^{-1}{\cal
M}^{\dagger} R_{LL} + e_{LL}e_{RL}^{-1} p_R^u e_{RL} e_{LL}^{-1}.
\en
Then, using the fact (eq. \ref{rfrome}) that the matrix $r$ of reflection
coefficients for $R$ particles incident from the unbroken phase can be
written $r = e_{RL} e_{LL}^{-1}$, we arrive at:
\eq
r p_L^u - p_R^u r + r {\cal M} R_{RR} r = {\cal M}^{\dagger} R_{LL}.
\label{thinwall}
\en

We will solve this equation perturbatively.  Let us start first from the case
without mixing.  The eigenvalue problem for one flavor, using the
linear Dirac equation since $p$ is small, can be written (see section
\ref{sec:scattering}) as
\eq {\scriptsize
det\left( \begin{array}{cc}
\omega(1+\alpha_L+\beta_L)- p (1+\alpha_L) & - M\\
M&\omega(1+\alpha_R+\beta_R) +p (1+\alpha_R)
\end{array} \right) = 0. }
\en
The solution, using notation defined in the previous subsection, is:
\eq
p_{\pm} = \bar{\omega} \pm \sqrt{B}.
\en
The reflection coefficient for $R \rightarrow L$ with no mixing
is
\eq
r_0 = \frac{1+\alpha_R}{M}(\bar{\omega} \pm \sqrt{B}),
\label{r0}
\en
with the root chosen so that $r \rightarrow 0$ when $M \rightarrow 0$.

The physical case with mixing can be solved by perturbation theory in
mixing angles.  Equation (\ref{thinwall}) provides an ideal recursion
procedure for that.  For regions of $\omega$ with flavor-diagonal total
reflection one writes $r = r_0 + \delta r,~ {\cal M} = M +
\delta M$, where $\delta M$ is the non-diagonal part of the mass
matrix ${\cal M}$, and defines $\delta r$ order by order in mixing
angles. The expansion parameters are roughly
\eq
\frac{(\delta M)_{ij}}{d_{kl}}
\en
whee
\eq
d_{kl} = (p^u_L)_k - (p^u_R)_l + (r_0 M R_{RR})_{kk} + (r_0 M R_{RR})_{ll}.
\en
The procedure is straightforward, and the calculation can be
done with the use of Mathematica or Maple.  For the region of $\omega$ with
off-diagonal $b_R - s_L$ reflection, one first rotates the basis and the
equations become more complicated, but present
no fundamental problem.  $\bar{r}$ is obtained in the same
way as $r$ with $\delta_{CP} \rightarrow -\delta_{CP}$.  Since in regions of
total reflection $r_0$ is complex (see eq. \ref{r0}), $r \neq \bar{r}$.
A non-zero asymmetry first
appears in the third order of perturbation theory, as expected, since
mixing between all three generations is essential.

The full expressions are too lengthy to be quoted here, however if we make
further approximations, valid for $p/\omega << 1$, we can obtain a more compact
result as follows.
We present here the result of the computation of the asymmetry in the
region where only the $s$-quark is reflected. We shall work in the
approximation in which the velocity of the domain wall is equal to
zero and mass corrections to the left-left and right-right transitions
in the broken phase are neglected.  Modifications of the result for
more realistic cases is discussed later.  Recall that the quantity
$\Delta$ in which we are interested is defined by eq.(\ref{Deltaom}).
It is convenient to redefine reflection coefficients in the following
way:
\eq
r = \tilde{r} K^{\dagger}O^{\dagger} R_{LL}.
\en
In terms of $\tilde{r}$, eq. (\ref{thinwall}) for the reflection
coefficients is:
\eq
\tilde{r} K^{\dagger}O^{\dagger} p_L^u O K - p_R^u \tilde{r}  +
\tilde{r} K^{\dagger}O^{\dagger}R_{LL} O K M R_{RR}\tilde{r} = M^{\dagger},
\label{thinwall1}
\en
where $M$ is the diagonal down quark mass matrix at $T_c$ (recall that
${\cal M}= O  K M$ for the charge -1/3 sector). The
expression for $\Delta$ is a bit more complicated in this basis and
contains the matrices $K$ and $O$.  There are two non-diagonal entries in
equation (\ref{thinwall1}). Namely, the second term on the left contains the
momenta of charge -1/3-quarks in the unbroken phase rotated by the matrices $O$
and $K$, and the last term contains the rotated matrix $R_{LL}$.  Now let
us take advantage of the fact that reflection of the $s$-quark occurs
at small incident particle momenta, so that an expansion with respect
to $p/\omega_0$ can be used. In zeroth order in this expansion,
$R_{LL} = -R_{RR} = 3/2$, and
\eq
\Delta = \frac{9}{4}Tr[\tilde{r}^{\dagger}\tilde{r} - \bar{\tilde{r}}^{\dagger}
\bar{\tilde{r}} ].
\en
One perturbs in the non-diagonal pieces of
$K^{\dagger}O^{\dagger} p_L^u O K$, which are to a good approximation
\eq
- \frac{3\pi\alpha_W T^2}{16} \frac{K^{\dagger}M_u^2 K}{\omega M_W^2}.
\label{nondiag}
\en

In the limit $m_b,m_s \ll \omega,~ m_d = 0,~r_{33}m_b \gg m_s$, and
$p^u_L-p^u_R$ for the $b$-quark
$ \ll m_b$ (which are actually quite good approximations as long as $m_t \sim
150$ GeV) the
expression simplifies to:
\eq
\Delta(\omega) = -2 \left(\frac{\pi \alpha_W T^2}{8 \omega M_W^2}\right)^3
\frac{m_t^4 m_c^2 s_{12}s_{23}s_{13}sin \delta_{CP}}{ m_b^2 m_s}
Im(r_s),
\label{asympt}
\en
where all masses are taken at high temperature and $r_s$ is the reflection
coefficient for the strange quark given by eq. \ref{r0}.

The dependence of this result on quark masses has a simple physical
interpretation.  Let us work in the basis of $\tilde{r}$, where the
bubble wall presents a diagonal barrier, i.e., the basis of zero
temperature physical quarks.  The main contribution to the amplitude
for $s \rightarrow s$, i.e., the contribution which is present even
when mixing angles are zero, we call $r_s$.  When there is total
reflection the reflection coefficient has unit magnitude but a
non-trivial phase, so $r_s = e^{i \delta_r}$.  However the
incident $s$ quark actually is a mixture of the broken phase
eigenstates which we could call $d_b,~s_b,~b_b$.  Thus there are
additional contributions to the amplitude for $s \rightarrow s$ coming
from ``paths'' in which another quark is present as an intermediate
state.  CP violation is first encountered when one considers paths
such that, for instance, the $b_b$ component of the $s$ reflects, and
is projected back onto the $s$ via a $d$.  Thus we have an amplitude
which is of the form
\eq
e^{i \delta_r} + A_{(s~in~b)} \cdot r_b \cdot A_{(b~in~d)} \cdot
A_{(d~in~s)}.
\label{3step}
\en
The path in which an intermediate $d$ is reflected makes a
negligible contribution as long as the $d$ is not totally reflected,
since then $r_d \sim m_d$ and is thus very small compared to $r_b$.
Note that as long as the $b$ is not totally reflected, $r_b$ is purely
real.  From standard perturbation theory, the amplitudes $A_{(i~in~j)}$
are just the relevant mixings divided by the level separations.  Thus
from eq. \ref{nondiag} we have, e.g., $A_{(d~in~s)} \sim
\frac{-3\pi\alpha_W T^2}{16\omega M_W^2} m_c^2 s_{12}$, divided by the
level separation, $m_s$.  From $A_{(s~in~b)} \cdot A_{(b~in~d)}$ we have
$(\frac{-3\pi\alpha_W T^2}{16\omega M_W^2})^2 m_t^4 s_{23} s_{13} e^{i
\delta_{CP}}$, divided by the 2-3 and 1-3 level splittings.  Some care
must be taken to determine them, using eqs (\ref{nondiag}), but one finds
$\frac{r_b}{d_{23}d_{31}} \sim 1/m_b^2$, which is anyway the naive guess when
the momentum of the bottom quark is small.  Therefore the relevant
portion of the amplitude for $s$ quark reflection is
\eq
\sim e^{i \delta_r} + c \left(\frac{3\pi\alpha_W T^2}{16\omega
M_W^2}\right)^3
\cdot \frac{m_c^2 m_t^4 s_{23}s_{13} s_{12} }{m_s m_b^2} e^{i \delta_{CP}},
\label{amp_refl}
\en
where $c$ is a constant of order 1.
Taking the absolute value squared of this, minus the corresponding
quantity with $\delta_{CP} \rightarrow - \delta_{CP}$ for the
antiparticles, then produces the form seen in the actual analytic
result above.  Of course to get the overall coefficient and do the sum
over flavors requires the real calculation outlined earlier in this
section.

This heuristic derivation is useful for realizing that the quark
masses which appear in the denominator of (\ref{asympt}) are just the
usual level-splitting denominators in perturbation theory.  It shows
us that as long as the levels are split by $m_s \neq m_d$, so that
there is a region in which the $s$ but not the $d$ quark is totally
reflected, the asymmetry is actually enhanced by a near-degeneracy in
the levels, since that increases the inter-generational mixing which
is essential to the asymmetry.  Clearly, when the masses of the $b$
and $s$ quarks go to zero, perturbation theory breaks down and
eq. (\ref{asympt}) does not hold.  Furthermore, we see from yet
another point of view how the total reflection of one or two quark
eigenstates is crucial to obtain a non-vanishing asymmetry: had there
been no total reflection, the only phase in (\ref{amp_refl}) would
have been the CP violating phase $\delta_{CP}$, and there would be no
asymmetry in rates -- as is familiar from $B$ physics, where the
necessary interference is between a CP violating phase and a strong
interaction final state phase shift.  In the case at hand, the only
source of a phase shift to interfere with the CP violating phase is
the phase shift which develops when the reflection is total, as is
evident from eq. (\ref{r0}).  The resultant asymmetry is proportional
to the product $sin \delta_{CP} sin \delta_r$, and the shape of the
upper peak we will see in the numerical results reflects the shape of
the function $sin \delta_r \equiv Im(r_s)$ as the reflection phase
shift moves from $0$ to $\pi$.

Taking into account the mass corrections to the left-left part of the
Dirac operator, discussed in subsection \ref{ss:broken}, requires one to
replace $m_c^2$ in eq. (\ref{asympt}) with
\eq
m_c^2 \rightarrow - m_c^2[\frac{16 \alpha_s M_W^2}{3 \pi^2\alpha_W
T^2} (\log\frac{\pi T}{m_c} - \gamma + \frac{1}{2}) -1].
\en
This is because the dispersion relation, when modified for this
correction, can be rewritten in its original form with that
substitution.  The correction due to bosonic masses in the broken
phase can be incorporated by replacing $\alpha_W$ in eq.
(\ref{asympt}) with $\alpha_W (1 - \frac{4 M_W}{\pi T})$.

For non-zero but small velocities the result increases
by a factor $[(1-v/3)(1+v)]^3$ for $J^u_{RL}$ and $J^b_{LR}$ while for
$J^u_{LR}$ and $J^b_{RL}$ the same factor appears with $v \rightarrow -v$,
due to the velocity dependence of the
coefficients $\alpha_{L,R}$ and $\beta_{L,R}$ (see appendix
\ref{app:velocity}).  For physical quark masses, the analytic result
presented here reproduces the exact numerical result in the region of
the upper peak for which eq. (\ref{asympt}) is applicable, to about
factor-of-2 accuracy.  Keeping terms of order $p/\omega$ is necessary
to do better than this.

The total asymmetry also includes a contribution at a slightly lower
energy from $b \rightarrow s$ reflection which is comparable in
magnitude to this contribution and of the opposite sign.  Thus as
parameters change, the sign of the total predicted asymmetry can
change as shall be seen in the next section, and for a quantitative
description we must do the full calculation.

We see that the asymmetry in the up-quark sector is suppressed by a
factor of roughly $ \frac{m_b^6 m_s^2}{m_t^6 m_c^2} \sim 10^{-11}$ and is
therefore numerically unimportant, as argued heuristically in section
\ref{sec:CPinSM}.

\mysection{Numerical Results for $\Delta$}
\label{sec:results}

\hspace*{2em}  The analytic results of the previous section are
helpful for our comprehension, but we wish to go beyond a perturbative
expansion in the mixing, in particular not to miss level-crossing
phenomena, and also to investigate the dependence of the result on
wall thickness.  This must be done numerically.

In this section we present the results of our full computation of the
baryonic asymmetry current.  We solve numerically the 1-dimensional
quantum mechanical scattering problem for the thermal Dirac equation,
(\ref{diracT}), using the methods described in Appendices
\ref{app:dirac} and \ref{app:num_int}.  We neglect the shift in the
broken phase kinetic term due to mass corrections for thermal
$W^{\pm}$ and Higgs.  The differential equation is
solved twice, once using $K$ and once using $K^*$.  As described in
sections \ref{ss:refln_coefs} and \ref{ss:cpt_amps}, this enables us
to determine all the reflection and transmission coefficients which we
need, for particles and antiparticles of both chiralities, incident
from either the broken or unbroken phase.  We can then determine the
various contributions to the baryonic current, using the formulae
of section \ref{ss:currents}.  In this section, we present our results
for this current.

We take as our ``canonical'' values of parameters: $m_t = 150
\rm{GeV}$, $m_c = 1.6 \rm{GeV}$, $m_u = 0.005 \rm{GeV}$, $m_b = 5
\rm{GeV}$, $m_s = 0.15 \rm{GeV}$, $m_d = 0.01 \rm{GeV}$, $ s_{12} =
.22, s_{23} = .05, s_{13} = .007$, $sin(\delta_{CP}) = 1$, $M_W(T_c) =
50$ GeV, $T_c = 100$ GeV,\footnote{While $T_c$ and $M_W(T_c)$ are not
precisely known, from (\ref{asympt}) one sees that it is the ratio
$M_W(T_c)/T_c$ which is important.  This ratio is constrained to be
near $1/2$ from the requirement that the sphaleron rate in the low
temperature phase is sufficiently suppressed.} and
inverse-wall-thickness (see eq. \ref{profile_tanh}) $a = T/10$.  Our
result is proportional to $\sin\delta_{CP}$.  As we shall illustrate
with a series of figures, the total asymmetry current depends
sensitively on some quantities, and little on others.  Figures
\ref{std}-\ref{th2301} shows $\Delta(\omega)$ as a function of
$\omega$ in the range of $\omega$ for which the total asymmetry is
non-negligible, for several choices of masses and mixing angles.

For the canonical values of the masses and mixings, the region of
total reflection of the $s$, found in the section on analytical
results, is $\omega = 48.15-48.24$ GeV.  This can be seen to coincide
with the region of the upper pair of peaks in Fig. \ref{std}.  The ``notch''
between them, from $\omega = 48.188-48.194$ GeV, is the region in
which both the $s$ and $d$ is totally reflected.  As can be seen from
figure \ref{dr_dsb_B} and was discussed in section \ref{sec:scattering}, the
broad peak of
opposite sign at lower $\omega$ corresponds to a region of $\omega$ in
which the momentum of the $b_{R^n}$ becomes degenerate with the
momentum of the $s_{L^a}$ or $d_{L^a}$ somewhere in its traversal of
the bubble wall, inducing resonant level crossing and total reflection
to an $s$ or $d$ quark.

The interpretations we have given above of the peaks are born out by
their dependence on the quark masses and CKM parameters, as seen by
comparing the various figures to one another.  The width and positions
of the upper peaks are very insensitive to changes in the mixing
angles, or to changes in any masses other than $m_s$ or $m_d$.  By
contrast, the position, width, and shape of the lower peak is very
sensitive to the mixing angles and $m_t$.  The height of the upper
peaks is given to within $\sim 30 \%$ by the thin wall analytic
expression (\ref{asympt}), and varies qualitatively correctly as the
relevant masses and mixings are varied.

In the regions between the peaks, the asymmetry is less than
$O(10^{-12})$, consistent with zero given our numerical precision.
This confirms our expectation that the asymmetry is negligible except
in a region in which either the $s$ or the $d$, but not both, is
totally reflected.

The character of the lower peak is sensitive to the top mass and the
intergenerational mixing, particularly $\theta_{23}$.\footnote{ One
can crudely estimate its width and location in analogy to the argument
used for the upper peak, as follows.  Since it occurs on account of
the level crossing of the $s_L$ and $b_R$ dispersion curves, one can
approximately use eq. (\ref{interval}) replacing the mass of the
$s$-quark with the mixing-induced contribution of the $b$-quark,
so that the width of the peak is $\sim m_b sin(\theta_{23})$.}
Moreover it has the opposite sign and the same order of magnitude as
the upper peaks.  Thus as $m_t$ and $\theta_{23}$ are varied, the sign
of the total asymmetry, integrated over $\omega$, can change.  This
can be seen by comparing Figs. \ref{std}, \ref{mt90}, \ref{mt210} and
\ref{th2301}.  The integrated asymmetry, $\Delta$ is shown in Fig.
\ref{mtdep}, over the range $90 < m_t < 250$ GeV.  For $m_t=90$ GeV, the
sign of the asymmetry current is negative, then changes to positive
(for positive $sin(\delta_{CP})$) for larger $m_t$.  Its magnitude
increases by a factor of 4 in going from $m_t = 150 \rightarrow 210$
GeV, and decreases by a factor of 2 (50) in going from $m_t = 150
\rightarrow 130(110)$ GeV.  Figure \ref{th23dep} shows the sensitivity
of the integrated asymmetry to $\theta_{23}$, which is far from
``perturbative'' in the physically interesting region, due to growth
of the lower peak.  On the other hand, for values of $sin(\theta_{13})
\lsi .01$, the baryonic current is linear in $sin(\theta_{13})$, so we
give no figures for that.  We have checked that when charge +2/3
sector masses are made degenerate, e.g., $m_u \rightarrow m_c$, the
asymmetry also vanishes.  This occurs through a reduction in the size
of the asymmetry for all $\omega$.\footnote{If $m_u$ is varied from
0.005 to 1.4 GeV the asymmtery decreases by a factor of 6, although
neither a linear nor quadratic dependence on mass differences provides
a perfect fit over this entire mass region, as is not very
surprising.} When $m_d \rightarrow m_s$ the notch between the upper
peaks grows until there is nothing left, with the magnitude of the
asymmetry inside the peaks staying approximately constant and the
magnitude of the negative peak vanishing like $m_s^2 - m_d^2$.  We
give also, in Fig. \ref{mbdep}, the dependence of the total asymmetry
on $m_b$, mainly for its conceptual interest, since the value is well
enough known to not be a major source of uncertainty.  It is
interesting however that had the bottom mass been 2 GeV lower, CKM CP
violation would have been insufficient to account for the BAU in this
mechanism, at least for 3 generations.

An important result, shown in Fig. \ref{adep}, is the relative
insensitivity of the asymmetry current to $a$, the inverse thickness
of the wall, for wall thickness up to $\sim (20 ~\rm GeV)^{-1}$, and
then strong increase for large wall thickness.  As previously noted,
the result does not vanish for zero wall thickness.  Furthermore, as
could have been anticipated, the positions of the peaks are
independent of $a$.  A thick wall enhances the asymmetry, although
the quantum mechanical approach becomes invalid when the wall
thickness is greater than the quasiparticle scattering length.

Modifying the expressions for $\alpha_{L,R}$ and $\beta_{L,R}$ to
correspond to their finite velocity expressions, given in appendix
\ref{app:velocity}, we can rerun the program to have an idea of the dynamical
effect of finite wall velocity.  The left baryonic asymmetry increases
as $v$ is increased, as shown in fig. \ref{vdep}.  Using the linear
approximation to the dispersion relation, the asymmetry reaches a
maximum at $v=0.25$, where $\Delta_{int}$ is a factor of 4 greater than
at $v=0$.  We cannot trust the results of the linear approximation for
larger $v$, because the value of $p/\omega$ corresponding to the
region of total reflection increases rapidly for $v \gsi 0.25$, with
$p/\omega \gsi 0.4$ for $v \gsi 0.3$.  However blindly using the small
$p/\omega$ and high-$T$ approximation formulae, even outside the
region where they can be trusted, suggests that $\Delta_{int}$ may not
continue to increase, and may even decrease, for $v>0.3$.  Thus
although the figure suggests that $\Delta_{int}$ will continue rising
for larger $v$, that may be misleading and further work is necessary
to know what actually happens for larger $v$.  For $v \lsi 0.25$ we
can use the linear approximation $\Delta_{int}(v) = \Delta_0(1 + \zeta v)$ with
$\Delta_0 = 2.3~10^{-6}$ GeV and $\zeta = 12$. We remind the reader
that for non-zero $v$, $J^R_{CP}$ is determined by the usual equations
but with $\Delta$ determined using $-v$ (see appendix \ref{app:velocity}).

As noted in appendix \ref{app:parallel}, a remnant of GIM cancelation
between the asymmetry for $s(\bar{s})_R \rightarrow d(\bar{d})_L$ and
that for $s(\bar{s})_R \rightarrow s(\bar{s})_L$ suppresses the total
asymmetry by a factor $O(10^{2-3})$ in comparison to the asymmetries
in individual flavor channels.  Figure \ref{std_i} demonstrates this
point, for ``canonical'' masses, mixing and $v=0$, by showing the
individual $s_R \rightarrow d_L$ vs. $\bar{s}_R \rightarrow \bar{d}_L$
asymmetry, and the sum of it plus the $s_R \rightarrow s_L$
asymmetry.  If there is some mechanism which we have not incorporated
in this calculation which lifts this degeneracy, e.g., by a
flavor-dependent modification of flux factors, then the result could
be 2-3 orders of magnitude larger than we find.

The quantitative results presented in this section raise some
interesting questions.  Firstly, how does the sensitivity to $m_t$ and
$m_b$ arise, given that the heavy quarks might be expected to decouple
from the scattering of the light quarks?  Since at least three
generations are required for there to be CP violation, it is clear
that they cannot decouple altogether, but one would like to understand
the mechanism through with they exert their influence.  Secondly, how
do the properties of the charge +2/3 quarks manage to affect the CP
violation in the scattering of the charge -1/3 quarks?  For instance
we could have removed the KM phase from the charge -1/3 sector and
put it in the charge +2/3 sector, where one might have thought that it
could not affect the scattering of the charge -1/3 quarks at all, and
would have transferred the effect to the +2/3 reflection coefficients,
which on account of the larger splitting between $c$ and $u$ might
have produced a bigger effect.  Of course this does not happen, since
the physics must be independent of the convention as to which sector
contains the KM phase.  Moreover an appropriate measurement at the
time of the ew phase transition could have determined that the
baryonic asymmetry was in the charge -1/3 sector.  Similarly, it must
be\footnote{and, we checked, is} the case that if two quarks of the
charge +2/3 sector were degenerate in mass, the asymmetry in the
reflection coefficients in the charge -1/3 sector would disappear.
But how do these consistency requirements on the scattering get
enforced?  Appendices \ref{app:mixing},\ref{app:observables}, and
\ref{app:quadratic}, are devoted to developing the technology
necessary to answer these questions.  Using those techniques, one can
see that heavy quarks do not participate significantly in the
scattering, but do participate in the fixing of the eigenstates, and
play their crucial role this way.  The independence of the convention
as to which charge sector contains the KM phase arises because it is
the {\em change} in the relative phases of the eigenstates when the
bubble wall is traversed which matters, and this is independent of
convention.  It is non-vanishing in the charge +2/3 sector, just much
smaller than for the charge -1/3 sector.  Likewise, the masses of the
charge +2/3 quarks play a role in fixing the charge -1/3 eigenstates,
both in the broken and unbroken phases.

\mysection{Prediction for $n_B/s$ and Discussion of Uncertainties}
\label{sec:prediction}

\subsection{Prediction}
\label{ss:prediction}
\hspace*{2em}

In eq. \ref{BAU}, we found the relation
between the left-baryonic current and the ultimate baryon number
density in the low-temperature phase:
$n_B = \frac{12}{5} J_{CP}f_{sph}(\rho)$, in
quasi-static approximation.  We have given in the previous section our
results for $\Delta_{int}$, which determines this current, in the
1-dimensional problem.  From it, we might attempt to estimate the
current corresponding to the full 3-d problem. As explained in
appendix \ref{app:parallel}, this cannot be done reliably.
Instead we divide the 1-dimensional current by the 1-dimensional
entropy corresponding to the known particle content at the
temperature of the ew phase transition:
\eq
s_{1-d}=\frac{73 \pi T}{3} = 76.44 T,
\en
and insert a factor $f_{3d}$ as a reminder of this source of uncertainty.

The results for $\Delta_{int}$ shown in the figures of the previous
section must be multiplied by $sin \delta_{CP}$.  We can reexpress
$sin \delta_{CP}$ in terms of\\
J $\equiv sin(\theta_{12})sin(\theta_{13})
sin(\theta_{23})sin(\delta_{CP})$,
in order to reduce the overall
uncertainty due to CKM angles and phase, since this combination is
relatively better constrained than the individual angles and phase.
The ``one-sigma'' range on J, $(1.4-5.0)10^{-5}$, obtained from a
global fit\cite{gn} to data, allows us to replace $sin \delta_{CP}$ in
$\Delta_{int}$ by
\eq
\left(\frac{{\rm J}}{.22 \times .05 \times .007}\right) = (0.2 - 0.7).
\en

We have seen that the final asymmetry is sensitive to the asymmetry in
the fluxes from broken and unbroken phases, and have considered two
extreme models of this.  In the first, the only deviation from perfect
equilibrium comes from the non-zero velocity of the plasma with
respect to the bubble wall, taking it and the temperature to be the
same on both sides.  In this case, using
eq. (\ref{JfromDelta_smallv_final}) to find $J_{CP}$ from
$\Delta_{int}$ and taking $m_t=150$ GeV, and otherwise ``canonical''
values as defined in the previous section, we obtain to leading order
in $v$:
\begin{equation}
n_B/s \approx \frac{3}{73 \pi} J_{CP} \frac{12}{5} f_{sph} f_{3d} =
(2-8)~10^{-5}
\frac{\Delta_0}{\bar{\omega}} ~v ~f_{sph}~f_{3d} = (1-4) 10^{-12} v
{}~f_{sph}~f_{3d}.
\label{result_qequib}
\end{equation}
For the other extreme of no flux from the broken phase, we use eqn
(\ref{JfromDelta_vb=0}) to determine $J_{CP}$ from $\Delta_{int}$ and find
\begin{equation}
n_B/s \approx  (3 - 9) 10^{-10} v ~f_{sph}(\rho)~f_{3d}.
\label{result_vb=0}
\end{equation}

The range in the parentheses reflects the uncertainty in J, the
product of sines of CKM angles.  Letting $m_t$ be $135 ~(180)$ GeV
multiplies the above results by about $ 2/3 ~(2)$.  We shall discuss
other sources of uncertainty below.

First, however, let us note that comparison with the observed
asymmetry, $n_B/s \sim (4-6)10^{-11}$, is quite encouraging.  Not only
is the magnitude in the right ballpark, the sign is correct as
well.\footnote{To be precise, only the sign of the product $B sin
\delta_{CP}$ is actually measured at present, where $B$ connects quark
matrix elements to hadronic matrix elements for the K meson.  If it is
positive as is generally believed to be the case, then J is positive
and our prediction has the same sign as observation, namely a
baryonic, not antibaryonic excess.  Fortunately, the sign of $sin
\delta_{CP}$ can be separately determined experimentally, given the
correct set of measurements\cite{nq}.} This is highly non-trivial,
given the intricate dependence of $\Delta(\omega)$ on quark masses and
mixing angles as discussed in section \ref{sec:results}.  However
until a number of uncertainties in our calculation and in the theory
of the electroweak phase transition are removed, and quasiparticle
behavior in that environment is better understood, this result can
only be taken as indicative that minimal standard model physics can be
responsible for the observed baryon excess of the universe.  We now
turn to a discussion of the factors $f_{sph}$ and $f_{3d}$, and a
critique of the calculation which we have done to obtain $\Delta_{int}$.

\subsection{Sphaleron Efficiency Factor, $f_{sph}(\rho)$}
\label{ss:suppression}
\hspace*{2em}
The sphaleron efficiency factor $f_{sph}(\rho)$ depends
a lot on whether the wall perturbs the fermionic distribution
functions in its vicinity.  Recall that if the
quasi-equilibrium approximation employed in section \ref{sec:asym} is
valid, $f_{sph}(\rho)=1$ when $\rho \equiv \frac{3 D_B \Gamma}{v^2}
\gg 1$, while for $\rho \ll 1, ~f_{sph}(\rho)= \frac{5}{6}\rho$.  In
order to estimate $\rho$ we need rather detailed knowledge of the high
temperature environment: the sphaleron rate, velocity of the domain
wall, and, to estimate $D_B$, the mean free path of the
quasiparticles.

Assuming the wall does not disturb the quark distribution functions,
which is correct for walls thicker than a typical mean free path, the
rate of sphaleron transitions in the unbroken phase directly
enters the expression for the final baryonic asymmetry through the
function $f_{sph}(\rho)$, with $\rho = {3D_B \Gamma \over v^2}$ and $\Gamma
= 9 \Gamma_{sp}/T^3$.\footnote{See section \ref{sec:asym}.} Only two
estimates of the sphaleron rate in the unbroken phase have been
made\cite{armc:zero,aaps:pl,aaps:np} so far. In the first, an attempt
was made to analytically compute the rate of sphaleron transitions,
taking some sphaleron-like configuration (namely, the standard
instanton in $A_0 = 0$ gauge at $t=0$) with a fixed size and half
integer topological number, and then integrating over the size of this
configuration.  This yielded the estimate\cite{armc:zero}:
\eq
\Gamma_{sph} \sim 20(\alpha_W T)^4 \sim 2~10^{-5}T^4.
\en
It is not clear how reliable the analytical calculation is, since the
corrections to it are not under control.\footnote{Estimates of the
prefactor for the sphaleron rate in the broken phase differ by many
orders of magnitude\cite{sphaleron:dpy,sphaleron:cm,sphaleron:bj}.}

In ref.\cite{aaps:pl,aaps:np}, lattice simulations of the sphaleron
transitions were performed.  A lower limit was found to
be:\footnote{The values of $\kappa$ in table 4 of ref. \cite{aaps:np}
must be multiplied by the factor 4.4, due to an arithmetic mistake in
eq. 50, which should be replaced by $N(t) = 0.01 \kappa
\frac{N^3}{\beta_G^4}$.}
\eq
\Gamma_{sph} \gsi 0.45 (\alpha_W T)^4 .
\en
The actual sphaleron rate may be much larger than this lower limit, and to
find it with better accuracy much larger lattices and larger values of
the lattice coupling $\beta_G$ must be used.  In addition to the usual
lattice artifacts (finite spacing and finite size effects) which make
it difficult to accurately estimate the sphaleron rate on available
computers, there is also a subtle problem connected with the
renormalization of the Debye screening mass in 3 dimensions.  This
problem, however, is not supposed to affect the main conclusion of
ref. \cite{aaps:np}.\footnote{J. Ambjorn and M. E. Shaposhnikov, in
preparation.}  We shall take $\Gamma_{sph} = 10^{-5 \pm 1}T^4$ for our
estimates below, although given the many uncertainties and
inadequacies of the existing estimates, the actual value may well wind
up outside this range.

The parameter $\rho$ also depends on the velocity of the bubble wall
which has about one order of magnitude uncertainty
\cite{khl:wall,dlhll:pl,dlhll:pr,lmt:wall} $v \sim 0.1-0.9$, and on
the quark diffusion constant, $D_B$, which has
never been calculated.  Roughly speaking, $D_B$ is the quark mean free
path which has been estimated to be $ \lambda \sim 4/T$
\cite{dlhll:pl,dlhll:pr} based on strong interaction scattering
cross-sections. Another similar estimate of the mean free path comes
from the calculation of the damping rates of the quasiparticles
\cite{smilga,brpis}, $\lambda \sim (0.15 g_s^2 T)^{-1} = 5/T$.  Yet
another estimate can be had by taking the viscosity of a 5-flavor
quark-gluon plasma as determined in \cite{bmpr:viscosity} and dividing
by the typical energy density.  This gives $\lambda \sim 4/T$.  In
appendix \ref{app:D_B} we discuss the specific case of relatively low
momentum quasi-particles and include the effect of Debye screening.  We
conclude that $D_B \sim (3-5)/T$ and $\lambda \sim (4-25)/T$.

Combining the above results gives
\eq
\rho \sim  10^{-3 \pm 1}/v^2.
\en
As can be seen, one can get either no suppression of the baryonic
asymmetry, or 4 orders of magnitude of suppression.  In the latter
case, MSM baryogenesis seems practically impossible, unless there is a
fourth family of quarks. Clearly, a more precise determination of the
sphaleron rate in the unbroken phase is of crucial importance.

We also recall to the reader that our derivation (section
\ref{sec:asym}) of the relation $n_B = \frac{12}{5} J_{CP}f_{sph}(\rho)$
assumed almost-equilibrium conditions, while in fact if the domain wall
is much thinner than a typical mean free path this may be a poor
approximation.  In the extreme case that quarks in front of the wall are at
rest with respect to it, with a void immediately behind the wall, then
one has simply $n_B = J_{CP}/\kappa$, from the definition of $\kappa$
($\sim 1/4$, see section \ref{sec:asym}).  In this case the
spahleron rate only needs to be greater than the expansion rate of the
universe to do the necessary job.  While we do not advocate this
extreme example as a good description, it does serve to emphasize the
necessity of understanding how the bubble wall affects the quark
distribution functions in its neighborhood, in order to determine the
magnitude of the final asymmetry.  Clearly, a treatment which goes
beyond the quasi-static approximation is needed.

To summarize this section, we have seen that it is reasonable to
suppose that there is no significant suppression from
$f_{sph}(\rho)$.  However there could instead be a huge suppression.
We can take $10^{-4} < f_{sph} \leq 1$.  From appendix \ref{app:D_B}
we have also seen that the effective collision length of the relevant
quasiparticles may be as large $\sim 25/T$, although a value as small
as $4/T$ cannot be ruled out.

\subsection{$3d$ versus $1d$ Prediction, $ f_{3d} $ }
\label{ss:3d}
\hspace*{2em}
We have developed the technical tools for carrying out the QM
scattering problem for arbitrary $p_{||}$ (see Appendix
\ref{app:parallel}), but this calculation is substantially more work,
and requires solving other problems first.  The contribution coming
from the region of $p_{||} \gsi p_{\perp}$ is much more sensitive to
the issue of the asymmetry in the fluxes than is the small $p_{||}$
contribution, because the velocity perpendicular to the wall is much
smaller in this case.  In addition, we should improve our
approximation methods, especially the high-$T$ expansion, in order to
obtain a valid Dirac operator for large momentum particles and to deal
with large velocities of the plasma with respect to the wall rest
frame.

Even if the Dirac operator we have obtained here were valid
under the more extreme circumstances encountered in the full $3d$
problem, it would be difficult to reliably deduce even the order of
magnitude of the $3d$ result, from the $1d$ results obtained so far.
This is because the level crossing structure is quite different in the
two cases, and one needs $R-R$ and $L-L$ reflection amplitudes as well
as the $L-R$ and $R-L$ amplitudes required in the $1d$ case.  These
complications appear as likely to give an enhanced result as a
decreased result since, for instance, the residual GIM cancelation
mentioned in section \ref{sec:results} seems to be lifted for at least
some of the kinematic regions relevant to large
$p_{||}$.\footnote{When $p_{||}=0$, with the approximations of the
present work, the degeneracy between the momenta in the unbroken phase
of $s_L$ and $d_L$ is almost perfect: $(p_s -p_d)/(p_s +p_d)\sim
10^{-3}$, due to the dominance of the thermal inertia when $p$ is
small.  This degeneracy in the flux factors determines the extent of
GIM compensation between the individual asymmetries in the final
result, causing the $p_{||} = 0$ asymmetry summed over flavors to be
$O(10^{-3})$ times a typical individual asymmetry.  A number of
effects might lift this, either associated with the full treatment of
large $p_{||}$, or coming from non-perturbative effects on the
propagation which we have not taken into account.  In our numerical
study of the one dimensional system (see section \ref{sec:results}),
the asymmetries in the individual processes are found to be quite
large, of the order of $10^{-1}$ for $v=0.25$.  Thus the net
contribution could be 2-3 orders of magnitude larger than when the
cancelation mentioned occurs.  To minimize the proliferation of
symbols, this possible source of enhancement will be included in
$f_{3d}$, even though it may also appear in a more complete treatment
of the $1d$ problem.}  Therefore for the time being we must retain the
uncertainty factor $f_{3d}$, which we believe is in the range $\sim
(10^{-3} - 10^{+3})$.\footnote{In the first preprint version of this
paper we had not yet included the effects of QCD-sphaleron-induced
$L-R$ transitions, nor the effects of mass corrections to bosonic
propagators in the broken phase.  When these are included, the naive
estimates of $f_{3d}$ which we attempted then break down, and must
simply be discarded.  See appendix
\ref{app:parallel} for details.}

\subsection{Other Uncertainties in the Calculation of $ J_{CP} $ }
\label{ss:uncertainties}
\hspace*{2em}
One source of uncertainty is the calculation of the thermal Dirac
operator in section \ref{sec:thermal}.  We used and even extended the
state-of-the art calculations of the quark propagator, which are done
in 1-loop, high-temperature expansion.  However one can think of
various effects which are clearly physically important and which are
not included in this approximation.  Simply going to higher-loop
approximation and to higher order in mass-insertions could make a
quantitative difference in the predictions, given their strong
sensitivity to the small splittings between eigenstates.  Moreover,
lattice studies suggest that non-perturbative effects may be very
important.  We can expect that these effects could also modify the
propagation of the fermionic quasiparticles and their inclusion could
modify the result for $\Delta_{int}$, although we have not identified any
effect of this sort which would modify the result by more than
numerical factors of order 1.

Related to working to lowest non-trivial order in coupling constants
in obtaining the quasiparticle propagator, is our purely
quantum-mechanical approach to fermion scattering from the domain
wall.  This neglects non-diffractive collisions of the fermions on the
particles in the plasma which occur while they are scattering from the
domain wall in the Higgs vev.  While this effect is technically of a
higher order (2-loop), we must ask whether it is physically
unimportant or not.  One can neglect the influence of these processes,
provided the effective collision length $\lambda_{inel}$ of the quarks in the
plasma is much larger than the bubble wall thickness $a^{-1}$ and
smaller than the imaginary part of the momenta of totally reflected
particles, in the broken phase.

The first requirement, that the quasiparticles can scatter
through the wall without having a collision, depends on the wall
thickness.  Due to uncertainties in the effective potential, the wall
thickness is
poorly known.  For instance using a perturbative calculation of the
effective potential yields a wall thickness of
$(10-40)/T$\cite{lmt:wall,dlhll:pl,dlhll:pr}, while recent work
including nonperturbative effects in the unbroken phase indicates that
the wall may be much thinner than this, $\sim
1/T$\cite{s:condensate}.  Evidently, if the wall is as thin as the
latter estimate, our calculation is valid, but if its thickness is
$40/T$, our calculation can only be considered an indication of the
possible order of magnitude of the asymmetry.  For a thick domain
wall, a better approximation to the problem would be consideration of
the plasma in the background of a (slowly) varying uniform scalar
field.

The second requirement for the validity of our quantum mechanical
approach, $Im(p_t) \lambda$ large compared to 1, arises because
we specify boundary conditions at infinity in which the coefficient of
the growing exponential is set to zero while the falling exponential
is kept.  This is not physically sensible if there is a collision at a
distance $\lambda$ from the wall, with $Im(p_t) \lambda$ not large
compared to one.  From the discussion in section \ref{sec:analytic},
one can see that in the region of $s$-quark total reflection,
$Im(p_t)\sim m_s$ so that for this region at least, the condition is
not met.  This is an aspect of the present calculation which must be
improved before the result can be considered quantitatively reliable.
However if it is the flavor-decoherence length, $\lambda_{fd} \sim
\frac{\alpha_s M_W^2}{\alpha_W m_c^2} \lambda_{inel}$, which proves to
be the relevant quantity, there may be no problem with our method
since $m_s \lambda_{fd} \sim 200$.  Unfortunately, improving the
calculation to systematically include these effects seems quite
non-trivial, since one must also include at this order emission and
reabsorption of thermal quanta during the scattering itself.  We have
not found a convincing method for estimating the consequences of these
effects.  Since they are formally of a higher order in $\alpha_s$,
including them may not qualitatively change the conclusions.

\subsection{Effect of a Fourth Generation}
\label{ss:fourth_gen}
\hspace*{2em} Presently feasible experimental measurements are generally
insensitive to a possible fourth generation of quarks, unless there is
a large difference between the $t'$ and $b'$ masses.\footnote{Of
course a fourth generation neutrino must be heavy enough to not have
influenced too much the $Z^0$ width.} This is not the case for the
baryon asymmetry produced in the MSM.  If there is another generation,
then the relevant GIM cancelation will be between the
second-most-degenerate pair of generations, namely the second and
third generations.  Then the $s-b$ degeneracy will limit the magnitude
of the CP violation.  Taking the analytic thin-wall result of eq.
(\ref{asympt}) as a guide, suggests an enhancement over the
3-generation prediction by a factor
\eq
\sim(\frac{m_{t'}}{m_t})^4(\frac{m_{t}}{m_c})^2(\frac{m_{b}}
{m_{b'}})^2\frac{s_{23}s_{24}s_{34}}{s_{12}s_{23}s_{13}},
\en
where we used the fact that $\Delta(\omega)$ is non-zero over a range
$\sim m_s(m_b)$ in the two cases, respectively, and replaced all CP
violating sines by 1.  For $\frac{m_{t'}^2}{
m_{b'} m_t} = 10$, and taking the ratios of the sines of the mixing
angles to be $\sim 1$, this produces an enhancement by a factor $\sim
10^3$.  It would appear to be unnatural for the contribution of a
fourth generation to be very much less than this, at least given our
present inability to account for quark masses and mixings, so that we
can consider three-generation results to be a lower bound on the
barynic asymmetry produced in the MSM.  If refinements in the theory
of MSM baryogenesis significantly lower the prediction in comparison
to observation, it could signal the existence of a
hitherto-unsuspected fourth generation.

\mysection{Conclusion}
\label{sec:conclusion}
\hspace*{2em}
We have shown that the baryonic asymmetry of the universe may be a
natural consequence of the CP-violation present in the minimal
standard model.  Within our present treatment of this problem we have
found that:\\ $\bullet$ The MSM prediction is very sensitive to quark
masses and mixings, for instance changing sign for $m_t < 110$ GeV and
increasing by a factor of eight when $m_t$ increases from $130
\rightarrow 210$ GeV.\\
$\bullet$ For $m_t = 150$ GeV and a plausible choice of wall velocity
($v=0.25$), we found
\begin{eqnarray*}
n_B/s \sim (0.1 - 10)~ 10^{-11} v ~f_{sph}~f_{3d}.
\end{eqnarray*}
The range reflects the experimental uncertainty in J, the product of
sines of CKM angles, and in the asymmetry in the flux factors.
Varying $m_t$ from 135 to 180 GeV would introduce a factor
$\frac{2}{3}$ to 2.  The greatest uncertainties in this prediction
come from dynamical aspects of the electroweak phase transition which
are still unclear.  To remove the uncertainty from the flux factor
requires knowing the quark distributions in the vicinity of the wall,
specifically, understanding how the passage of the wall affects these
distributions.  The sphaleron conversion efficiency, $f_{sph}$, is 1
if the rate of sphaleron transitions in the unbroken phase is large
compared to the typical time the quarks remain in the unbroken phase
before being overtaken by the expanding bubble of low temperature
phase.  While $f_{sph}$ may be 1, it could also be as small as
$10^{-4}$, in which case the known CP violation of the MSM is
inadequate to explain the observed BAU.  $f_{3d}$ is very hard to
estimate.  We believe it lies between $10^{-3}-10^{+3}$; it also
depends sensitively on the flux asymmetry.

While the present result can be considered only a preliminary
indication of the true prediction, the possible consistency in
magnitude with the observed $n_B/s \sim (4-6)10^{-11}$ is
encouraging.  The sign of the prediction is very sensitive to quark
masses and mixing angles, but is correct when these quantities lie in
their experimentally allowed ranges.  Many of the uncertainties which
affect the magnitude of the final asymmetry, such as the sphaleron
efficiency and the effect of the bubble wall on the quark
distributions in its neighborhood, and our simplified
quantum-mechanical treatment neglecting inelastic effects, do not
affect the sign of the prediction.  Thus it is heartening that the
prediction of the sign is correct, because this is likely to not
change as the calculation is improved.

In any scenario of electroweak baryogenesis the Higgs potential must
be such that it produces a strongly first order phase transition, and
such that sphaleron transitions after the phase transition are
suppressed.  When non-perturbative effects are better understood in
the MSM, this should imply a firm upper bound on the Higgs
mass\cite{kajantie,s:condensate}.  In
addition to the LEP experimental lower bound, there is a theoretical
lower bound resulting from requiring the $T=0$ vacuum to be stable,
which for a MSM Higgs is actually more stringent:
\eq
m_H > 75 + 1.64(m_t - 140),
\en
in GeV, for $130 < m_t < 150$ GeV\cite{sher:vac_stability}.  Combining
the theoretical bounds will either exclude, or precisely predict, the
mass of the MSM Higgs, if the minimal standard model with no
extensions whatever can be responsible for the baryonic asymmetry of
the universe.

If the upper bound from requring the sphaleron rate in the low
temperature phase to be small enough is violated, it does not mean
that electroweak baryogenesis must be rejected or that the mechanism
we have developed in this paper cannot be responsible for the BAU.  It
could instead indicate, for instance, that the Higgs sector is more
complicated than in the MSM, so that the upper bound on the Higgs mass
is relaxed. The real test, eventually, of whether the phenomenon we
have discussed is responsible for the observed baryonic asymmetry of
the universe, will be in its quantitative comparison with the measured
sign and magnitude of the BAU.  Once the dynamical aspects of the ew
phase transition are well enough understood, this can be done.

Our work underlines the importance of a reliable and precise observational
determination of $n_B/\gamma$, which fixes $n_B/s \approx
\frac{1}{7} n_B/n_{\gamma}$.  The most
recent comprehensive analysis\cite{nucleosyn} quotes the range $ 2.8~
10^{-10} < n_B/n_{\gamma} < 4.0~ 10^{-10}$.  However many aspects of
the determinations of the primordial abundances are complex and
controversial, and the true uncertainty may be larger than reflected
in these error bars.  For instance if the primordial $^4He$ abundance,
$Y_P$, proved to be $0.228 \pm 0.005$ as claimed in ref. \cite{pagel},
i.e., below the Big Bang Nucleosynthesis (BBN) ``lower limit'' of
$0.236$, some change in determinations of $D$ and $D + ^3He$ or in the
simple, homogeneous BBN theory would be necessary for
self-consistency, since we know now that there are three light
neutrinos.  Such changes could cause the prediction for $n_B/s$ to
move outside the $(4-6)10^{-11}$ range.  For instance using only $Y_P
= 0.228 \pm 0.005$ and three neutrinos would lead to\cite{nucleosyn}
$n_B/s = 2~10^{-11}$.

If the minimal standard model is responsible for baryogenesis, we will
be able to use a well-determined value for $n_B/s$ to quantitatively
test our understanding of the dynamics of the electroweak phase
transition.  This could eventually be as powerful a test of our
dynamical understanding, as nucleosynthesis has been for later stages
of cosmology.  Conversely, anticipating the day when the physics of
the electroweak phase transition can be considered understood, one can
even imagine being able to constrain the particle content, masses and
mixing of the MSM on account of the sensitivity of the BAU to these
quantities.  As noted in section \ref{sec:prediction}, a fourth
generation would characteristically increase the asymmetry by a large
factor in comparison to the 3-generation prediction.  Just as the
theory of nucleosynthesis, combined with measurement of the relative
abundances of primordial nuclei, led to the orrect conclusion that
there are three light neutrinos, we may one day be able to rule out
the existence of a fourth generation, or infer properties it must
have, by comparing the observed baryonic asymmetry to the predicted
one.

\mysection{Acknowledgements}
\hspace*{2em}
We have benefitted from stimulating, provocative, and useful
discussions with many colleagues, especially J. Ambj\o rn, G. Baym,
A. Cohen, E. Farhi, J.-M. Frere, B. Gavela, S.  Khlebnikov, R. Kolb,
P.  Langacker, A. Linde, L. McLerran, Y. Nir, J. Orloff, O. Pene,
J. Reppy, A.  Ringwald, J. Sellwood, S. Shenker, I. Tkachev,
M. Turner, and L.  Wolfenstein.  The contributions of A. Terrano were
particularly important.

\appendix

\mysection{The solution of the Dirac equation}
\label{app:dirac}
\hspace*{2em} In spite of the fact that the problem of
finding the reflection coefficients can be given a transparent
formulation, as discussed in section \ref{sec:scattering}, it is not
so easy to solve.  In this appendix we will describe a method suitable
for high accuracy numerical solution.

First we note that that the problem we want to solve is a problem with
boundary conditions rather than a Cauchy problem. We should be able to
separate incoming and outgoing waves, exponentially rising and
decaying functions. There are many different scales, and there is no
way to separate them lookig at the numerical solution for the wave
function.

The linear character of the differential equations is very helpful.
The following trick converts the boundary condition problem to the
Cauchy problem. Let us look for solutions of eq.(\ref{basic}) in the
form
\eq
\Psi(z) = e(z) E(z) V(z)\Psi_0,
\en
where $\Psi_0$ is a constant vector, $e$ is a matrix constructed from
the eigenvectors of the matrix $D(z)R$:
\eq
D(z)Re(z) = e(z)p(z),
\label{ev}
\en
$p(z)$ is a diagonal matrix of eigenvalues of the matrix $D(z)R$,
\eq
E=\exp(i\int_{z_0}^z dz p(z)),
\en
and $z_0$ is some arbitrary point. Here $z$ is a complex variable.

We shall suppose that $F(z)$ is an analytic function of the complex
variable $z$ in some region of the complex $z$-plane including the
real axis.  Most numerical studies we have done are based on the
following choice of the domain wall profile
\eq
F^2 = \frac{1}{1+\exp(-ax)}
\en
There is no serious motivation of this particular choice of the domain
wall structure. It resembles, however, some basic features of the
expected behavior of the scalar field near the domain wall. Namely,
when $x \rightarrow \infty$
\eq
F \rightarrow 1 - \frac{1}{2} \exp(-ax),
\en
so that the parameter $a$ can be identified with the effective Higgs
mass in the broken phase, while for $x \rightarrow -\infty$
\eq
F \rightarrow \exp(-a|x|/2),
\en
incorporating the expectation that the effective Higgs mass in the
unbroken phase is generally smaller than in the broken phase. The
other advantage of this choice is that, for it, we can find the
reflection and transmission coefficients analytically for the case
without mixing (see section \ref{sec:analytic}) and compare them with
the numerical solution, checking in this way the correctness of the
numerical calculations.

In order to get an equation for $V$ let us consider in more detail the
properties of eigenvectors and eigenvalues of the matrix $D(x)R$ on
the real axis.  One can show that for any real $x$ the set of
eigenvalues {$p$} obeys the following properties.  We have either\\
\hspace*{2em} 6 real eigenvalues\\
or\\
\hspace*{2em} 4 real eigenvalues and 1 complex conjugate pair \\
or\\
\hspace*{2em} 2 real eigenvalues and two conjugate complex pairs   \\
or\\
\hspace*{2em} 3 conjugate complex pairs.

The proof: The equation for determining the eigenvalues of the matrix
$DR$ can be written in the form $det (DR-p) = 0$, where $p$ here is
any one of the eigenvalues, not a matrix.  It can also be written in
the form $det(RDR - Rp) = 0$, since $det~R \neq 0$.  The equation for
the complex conjugate of the eigenvalue has the same form, due to the
hermiticity of the matrices $RDR$ and $R$: $det(RDR - Rp^*)=0$.
Therefore the set $\{p\}$ coincides with the set $\{p^*\}$, proving
the statement.

This fact has a transparent physical meaning: if all eigenvalues are
real, all particle states can propagate in the background of the
scalar field $\phi(x)$, while if two eigenvalues are complex then
those states cannot propagate, etc.  From this result one can derive
the following orthogonality conditions for the eigenvectors. Let us
denote by $e_i$ the eigenvectors of the matrix $DR$, and define the
matrix $e = (e_1, ..., e_6)$.  It is easy to see that if $p_i \neq
p_j^*,~ i\neq j$, the vectors $e_i$ and $e_j$ are orthogonal in the
following sense:
\eq
e_j^{\dagger}Re_i =0.
\en
Therefore, if $e_i^{\dagger}Re_i \neq 0$, then $p_i$ is real and given
by
\eq
p_i = \frac{e_i^{\dagger}RDRe_i}{e_i^{\dagger}Re_i}.
\en
If, for some $i$ and $j$, $p_i = p_j^*$, then $e_i^{\dagger}Re_i =0$
and $e_j^{\dagger}Re_j =0$. However, generally speaking,
$e_i^{\dagger}Re_j
\neq 0$.

It is convenient to introduce also another eigenvalue matrix $f$
obeying the equation on the complex axis
\eq
fRD = \tilde{p}f.
\label{cptev}
\en
One can easily see that eigenvalues defined by
eq.(\ref{ev},\ref{cptev}) are the same.  To prove it, we notice that
that the eigenvalues of the problem eq.(\ref{ev}) can be found from
the equation
\eq
\label{evalDR}
det(DR - pI) = 0,
\en
and the eigenvalues of the other one from the equation
\eq
\label{evalRD}
det(RD - \tilde{p}I) = 0.
\en

Taking into account the fact that $detR \neq 0$, one sees that the
roots of (\ref{evalDR}) and (\ref{evalRD}) are the same. Therefore,
one can choose a basis in which the matrices $p$ and $\tilde{p}$ are
identical. The important property of the matrices $f$ and $e$ which
motivates the introduction of $f$ is that they are orthogonal
everywhere in the complex plane in the sense that
\eq
fRe = diagonal~ matrix.
\en
This follows from the the relation
\eq
pfRe = fRep.
\en
A convenient normalization is
\eq
fRe = R.
\en

On the real axis, where the operator $D$ is hermitean, $f$ and $e$ can
be related as
\eq
f = Te^{\dagger}
\en
where
\eq
T^2 = 1,~~ Tp^*T = p.
\en

Then, the equation for $V$ has the following form:
\eq
\frac{\partial V}{\partial z} = -(RE)^{-1}fR\frac{\partial
e}{\partial z}E V.
\en
One can also find the equations for the evolution of eigenvectors and
eigenvalues:
\eq
\frac{dp_i}{dz} = \frac{1}{R_{ii}}\left(fR\frac{dD}{dZ}Re\right)_{ii}
\en
\eq
\frac{de}{dz} = eR^{-1} A, ~~\frac{df}{dz} = - AR^{-1} f,
\en
where
\eq
A_{ii} = 0,~~A_{ij} = -\frac{1}{p_i -
p_j}\left(fR\frac{dD}{dZ}Re\right)_{ij}.
\en
One can see that the matrix $A$ is singular when $p_i - p_j = 0$;
therefore the complex contour should be chosen in such a way that $A$
is regular everywhere along it. The advantage of this formalism is
that we have explicitly separated the waves corresponding to the
different flavors and helicities. The other helpful feature is that
these equations are local in the sense that the matrix $V$ changes
only in the vicinity of the domain wall, provided all $Im~p_i
(+\infty) < a/2$.  In this case the exponential tail of the domain
wall is stronger than the exponentially rising functions appearing in
the complete reflection case. If some $Im~p_i(+\infty) > a/2$, then
the equation for $V$ does contain an exponentially rising term and it
is difficult to achieve high accuracy. The formalism for treating this
case is described in appendix \ref{app:observables}. It was necessary
for the numerical analysis of the charge +2/3 sector in the region of
momenta in which the $c$ quark is totally reflected but the $u$ is
not.

It is easy to find the eigenvectors and eigenvalues of the operator
$DR$ far from the domain wall in the unbroken phase. Eigenvalues are
given by (\ref{pLun},\ref{pRun}) and the matrices $e$ and $f$ are just
unit matrices.

It is convenient to choose the initial condition for the matrix $V$
\eq
V(z_0) = 1.
\en
Then, choosing $z_0$ to lie on the real axis and far from the domain
wall ($z_0 \rightarrow -\infty$), if one can determine the asymptotic
value of the matrix $V$ for $z \rightarrow +\infty$ on the real axis,
all reflection and transmission coefficients are determined as shown
in section \ref{sec:scattering}.

Even in the case without total reflection it is useful to integrate
the equation for $V$ along a complex contour, due the the physical
phenomenon of level crossing.  In spite of the fact that $\Psi(z)$ is
an analytic function in the same region of the complex $z$-plane as
the potential is analytic, including the real axis, the matrix $V$ has
different analytical properties. For example, if at some $z$, some of
the eigenvalues of the matrix $DR$ are degenerate, the eigenvectors
corresponding to those eigenvalues are not uniquely defined and, in
general, are singular at this point.  Furthermore, eigenvalues of the
matrix $DR$ have branch cut singularities on the real axis for the
case of the complete reflection of some fermionic flavor. Therefore,
the equation for $V$ has singularities of various types at the points
where $p$ and $e$ have singularities. Of course, these singularities
would be canceled in the expression for $V$, but this fact does not
help when solving for $V$.

So, our strategy is to solve the equation for $V$ on some contour in
the complex plane of the variable $z$, lying in the region of
analyticity of the function $F(z)$. The initial and final points of
the contour lie on the real axis, far from the domain wall. For
example, for the function $F$ in eq. (\ref{profile_tanh}), a suitable
contour is
\eq
z ~=~ x~ -~ \frac{i}{b~ cosh(c \cdot x)}
\en
where $b>a/\pi$ to avoid singularities of the function $F$, and $c$ is
an arbitrary number. The function $\Psi$ at the end point does not
depend on the choice of contour, on account of the analyticity.
However, the function $V$ aswell as the normalization of the
eigenvectors can depend on the contour. Nevertheless, observables,
such as the baryonic current, are, of course, contour independent (see
below).

Suppose now that we have chosen some contour and have calculated at
the end point the matrix $V$ as well as the eigenvector matrix and
eigenvalues, taking for definiteness the case of particles incident
from the unbroken phase.  In order to find the reflection amplitudes,
we first must decide which eigenvalues correspond to acceptable
boundary conditions and which must be excluded.  For complex
eigenvalues it is simple: the exponentially dying wave function ($Im ~
p > 0$) can exist in the broken phase ($+\infty$), while the
coefficient in front of the exponentially rising wave function ($Im ~p
< 0$) must be equal to zero.  The situation with propagating waves
(real eigenvalues) is more complicated.  In order to decide which one
is allowed in the broken phase, one should calculate the group
velocities corresponding to the various eigenvalues. Those with
positive group velocities at $x \rightarrow + \infty$ are acceptable
transmitted waves, while those with negative group velocities
correspond to waves traveling in from $+\infty$ and must have zero
coefficient in the solution. In the unbroken phase the sign of the
group velocity corresponding to the eigenvalue $p_i$ is the same as
the sign of the matrix element
\eq
(e^{\dagger}R e)_{ii}.
\en
One can check that this is also true in the broken phase, at least
when the mass of the fermion is not much larger than its momentum,
which is the region of interest.

Having decided which waves are allowed in the broken phase, let us
relabel the eigenvalues $p$ in such a way that the first 3 eigenvalues
at $+ \infty$ correspond to transmitted waves. This relabeling causes
columns of the matrix $V$, and rows of the matrix $e$, to be
interchanged:
\eq
\Psi = eEV\Psi_0 = ePPEPPV\Psi_0
\en
where the matrix $P$ with the property $P^2 = 1$ 'reshuffles' the
eigenvalues in the desired way.  If we denote
\eq
PV =
\left( \begin{array}{cc}
V_{LL} & V_{LR}\\ V_{RL} & V_{RR} \end{array} \right),
\en
then the reflection coefficients are determined by eq.
(\ref{ru},\ref{rb}).

\mysection{Numerical integration}
\label{app:num_int}
\hspace*{2em} Even taking an optimistic view that the asymmetry could
be as large as $sin(\theta_{12}) sin(\theta_{13})sin(\theta_{23})
sin(\delta_{CP})\sim 10^{-(4-5)}$ in the most favorable regions of
energy, it is clear that very precise calculations are required.
Furthermore we wish to investigate the dependence of the result on
many parameters including quark masses and mixing angles and the wall
thickness, and we must determine the asymmetry as a function of energy
with a fine grid spacing in order that the integrated asymmetry be
accurately determined.  Thus an integration method is required which
is at the same time efficient and accurate.

We have used C++ as a programing language, in order that complex
numbers and matrices could be treated as natural units while retaining
the benefits of C.  We adopted the Burlirsch-Stoer integration
algorithm described in Numerical Recipes\cite{num_recipes}, although
we wrote our own programs in order to use C++ functionality and the
customized matrix manipulation procedures we required.  The Numerical
Recipes routines we mimicked were odeint, bsint, mmid, and rzextr.
The Burlirsch-Stoer method is well adapted to our situation: most of
the non-trivial variation occurs in a range which is small compared to
the full integration range, so that an adaptive stepsize is required,
while rational function extrapolation enhances the precision in a
minimal number of steps.  We checked, by comparing our numerical
results with the exact analytic solution for the case with no mixing
(see section \ref{sec:analytic}), that the actual precision of the
numerical integration was what it was supposed to be, even when the
precision was required to be 1 part in $10^{14}$.

We require reflection coefficients to be known, typically, with an
accuracy of one part in $10^8$, in order to be able to take
differences between the quark and antiquark sectors, sum up over all
the flavors, and still have a result which is accurate to one part in
a thousand or better.  Depending on the energy and other parameters,
we could achieve a final precision on the total asymmetry at each
energy of one part per mil by running our integrations at a precision
of $10^{-8}-10^{-10}$.  We verified that our results are independent
of the precision of the integration, at this level, under these
conditions.

An extremely important check of our results was to verify that they
are independent of the complex contour chosen.  We integrated along
the complex contour
\begin{equation}
z ~=~ x~ -~ \frac{i}{b~ cosh(c \cdot x)}
\end{equation}
and varied $b$ between 1 and 100 and $c$ between 0 and 2.  We also
checked that the result does not change as the initial and final
points of the integration, ($x_{min},~x_{max}$), are
varied.\footnote{It is essential to begin and end the integration on
the real axis, but the contributions of the segments between
$x_{min},~x_{max}$ and $z(x_{min}),~z(x_{max})$ can be represented
analytically.} For routine use we took $x_{min}=-60/a$ and
$x_{max}=+30/a$, where $a$ is the inverse wall thickness in GeV.  When
the mass and mixing parameters are such that there is level crossing,
one cannot allow the contour to be too close to the real axis, or the
kernel of the differential equation becomes singular.  However even in
these cases we verified that we could vary $b$ and $c$ each by factors
of 5 without changing the final asymmetry by more than one part in a
thousand, the typical precision of our numerical calculation for the
overall result, as discussed above.

The continuity with $\omega$ of the asymmetry, shown in the figures of
section \ref{sec:results}, testifies to the quality of the numerical
integration, since roundoff and many other types of errors would be
uncorrelated in the runs at each different $\omega$, and would
therefore show up as jitter in the $\omega$-dependence.

\mysection{Velocity dependence of the reflection coefficients}
\label{app:velocity}
\hspace*{2em}

The equation describing the reflection of left fermions (incident from
the unbroken phase) from the moving domain wall, in the rest frame of
the plasma, for small momenta of fermions is:
\eq
\left( \begin{array}{cc}
\omega(1+\alpha_L+\beta_L)+ i\frac{\partial}{\partial x}(1+\alpha_L)
&{\cal M}(\gamma(x + vt))\\ {\cal M}^{\dagger}(\gamma(x +
vt))&\omega(1+\alpha_{R}+\beta_{R}) - i\frac{\partial}{\partial
x}(1+\alpha_{R}) \end{array} \right) \cdot
\label{vel_dep}
\en
\[
\left( \begin{array}{c}L\\ R \end{array}\right) =0,
\]
where $L$ and $R$ correspond to the upper and lower components of
two-dimensional Weyl spinors which have 3 flavor components.
$\gamma=1/\sqrt{1-v^2}$, $t$ is time and $v$ is the velocity of the
domain wall.  Positive $v$ corresponds to the wall propagating into
the unbroken phase, as is the case physically. The 3x3 diagonal
matrices $\alpha$ and $\beta$ are defined to be
\eq
\alpha_{L,R}= \frac{1}{2}\beta_{L,R} =
-\frac{1}{3}\frac{\omega_{L,R}^2}{\omega^2}.
\label{alpha,beta}
\en
Due to the explicit time dependence of the Higgs field, the energy
$\omega$ in this equation is a time derivative: $\omega \rightarrow i
\partial/\partial t$, rather than a c-number. It is more convenient to
solve this equation in the rest frame of the domain wall, where the
energy of the fermions interacting with the classical scalar field is
conserved. In order to go to the rest frame of the wall one can make
the standard Lorentz transformation of coordinates $x \rightarrow
\gamma(x + vt), ~t \rightarrow \gamma(t + vx)$ together with the
transformation of spinor fields:
\eq
\left( \begin{array}{c}L\\ R \end{array}\right)\rightarrow
\left( \begin{array}{cc}
\Lambda_L&0\\
0&\Lambda_R \end{array} \right) \cdot
\left( \begin{array}{c}L\\ R \end{array}\right),
\en
with $\Lambda_L = (\frac{1+v}{1-v})^{\frac{1}{4}},~\Lambda_R =
(\frac{1-v}{1+v})^{\frac{1}{4}}$. Now, keeping only the linear term in
space derivatives (which is correct for small enough momenta of
incident fermions) one obtains the equation
\eq
\left( \begin{array}{cc}
\omega(1+\tilde{\alpha_L}+\tilde{\beta_L})+ i\frac{\partial}{\partial
x}(1+\tilde{\alpha_L}) &{\cal M}(x)\\ {\cal
M}^{\dagger}(x)&\omega(1+\tilde{\alpha_{R}}+\tilde{\beta_{R}}) -
i\frac{\partial}{\partial x}(1+\tilde{\alpha_{R}}) \end{array}
\right) \cdot
\label{vel_indep}
\en
\[
\left( \begin{array}{c}L\\ R \end{array}\right) =0
\]
with
\eq
\tilde{\alpha_L}= \alpha_L(1-3v-2v^2)(1-v),~~
\tilde{\beta_L} = 2\alpha_L(1+v)^2(1-v)
\en
and
\eq
\tilde{\alpha_R}= \alpha_R(1+3v-2v^2)(1+v),~~
\tilde{\beta_R} = 2\alpha_R(1-v)^2(1+v).
\en

The consideration of right particles incident from the unbroken phase
goes along the same lines. Now, one can solve these equations by the
methods described in the paper and in the appendices.  The only
important difference is that the reflection coefficients entering
$J^u_{RL}$ and $J^b_{LR}$ are obtained from the same equations with $v
\rightarrow -v$.  It is necessary to solve for $V$ twice, with $D(v)$
and $D(-v)$ since when $v \ne 0$ parity relates the equation for right
particles to the one for left particles with $v \rightarrow -v$,
modifying eqns. (\ref{psi1}) and (\ref{psi3}).

For the case of incident $R$ particles (whose reflection contributes
to the left baryonic current) and $v \gsi 0.4$, some momenta become so
large that the techniques of appendix \ref{app:observables} are
required.  One must also take care not to leave the regime of
applicability of the approximations which have been made in obtaining
eqns. \ref{vel_dep}, \ref{alpha,beta}, and \ref{vel_indep}.  For
instance, a small $p$ with respect to the wall, corresponding to
$s$-quark reflection, may come from a plasma-rest-frame momentum which
is large, for large enough $v$.  In practice we can safely work to $v
\sim 0.25$ with these approximations, for the case of interest.

\mysection{Flux Factors in 1-Dimension}
\label{app:flux_facs}
\hspace*{2em} The calculation of the flux factors is a bit
non-trivial in this scattering problem, since we deal with
quasiparticles rather than particles.  This means that we must take
care regarding such things as wave function normalization, etc.  Let
us first fix these factors for left chiral particles incident from the
unbroken phase. We recall that the current of interest is
\eq
j = \Psi^{\dagger}R\Psi.
\en
This current is conserved ($\partial_x j =0$) so that we can find it
wherever it is most convenient, e.g., at $x \rightarrow - \infty$. If
we send a quark of the first flavor toward the domain wall, then the
initial wave function is given by
\eq
\Psi_{-} = \eta_1\left( \begin{array}{c}
\\0\\0\\r_{11}^u\\r_{21}^u\\r_{31}^u
\end{array} \right)
\en
where $\eta_1$ is the normalization factor to be determined later.
Then the current is
\eq
j = |\eta_1|^2 \left({(R_{LL})}_{11} + (r^{\dagger}R_{RR}r)_{11}\right).
\label{eta}
\en

The normalization factor $\eta$ should be chosen in such a way that we
have just one particle in the initial state. To find it, let us
consider the initial left flux in the unbroken phase.  According to
eqs.(\ref{J^L},\ref{J^Lgroup}) it is given, in the one-dimensional
case we are considering, by
\eq
\int \frac{dk_1}{2\pi}\frac{1}{{(R_{LL})}_{11}}n_F.
\label{jl1dim}
\en
Comparing (\ref{jl1dim}) and (\ref{eta}) one finds $\eta_1 =
1/{(R_{LL})}_{11}$. Now, integrating the result (\ref{jl1dim}) with
respect to the momentum of the initial fermion and changing the
integration variable from $dk_1$ to $d\omega$ we arrive at eq.
(\ref{JuLR}).

The calculation of the flux factors in the broken phase is precisely
the same, with the obvious substitution $R \rightarrow R^b$, and we do
not present it here.  This amounts to assuming that the distribution
functions of the quarks in the broken phase are just the equilibrium
thermal distribution functions of the broken phase.  For a
sufficiently slow bubble wall this is a good approximation.

We have found the flux factor for the one dimensional problem. In the
real 3-dimensional problem, one must integrate over the components of
the momentum parallel to the surface. As is argued in Appendix
\ref{app:parallel}, the reflection coefficients may strongly depend on
the parallel components of the momentum, and the solution of the exact
equations is required. We do not attempt to solve here the problem of
parallel motion. Instead, we will just use the 1-dimensional entropy
of the plasma when we estimate the final baryonic asymmetry.

\mysection{Parallel momenta}
\label{app:parallel}
\hspace*{2em} In the main body of the paper we deal with the case in
which the momenta of the fermions are perpendicular to the domain
wall. In this appendix we construct the formalism for the more general
case and discuss the influence of parallel momenta on the asymmetry.
Evidently, when $p_{||} \ne 0$ angular momentum can be conserved in
the scattering while at the same time $L-L$ and $R-R$ reflection can
occur, in addition to $L-R$ and $R-L$ reflection.  Thus the matrix
equations will be 12x12 instead of 6x6.  In the familiar situation
without a plasma, one can Lorentz boost to a $p_{||}=0$ frame where
the problem can be reduced to one involving just $L-R$ and $R-L$
reflection.  However Lorentz invariance is lost in the plasma, so that
a trivial boost in the direction parallel to the domain wall does not
produce the problem we have already solved and the additional $R-R$
and $L-L$ amplitudes are physically important.

The study of the new problem can be divided into steps. The first step
is the construction of a formalism allowing one to compute the
reflection coefficients in this more complicated case. The second step
is the derivation of the general expressions for the asymmetry
current, assuming that the reflection coefficients are known as a
function of the energy and $p_{||}$.  The third step is the kinematic
analysis to determine in which part of phase space the asymmetry can
be substantial.  We do not proceed in this paper to the final step of
actually computing the reflection coefficients for finite $p_{||}$
quantitatively.

{\bf Reflection coefficients.} Here we will construct a transformation
which allows the problem to be studied by the 1-dimensional methods
given above.  We shall work in the approximation in which the
component of the momentum perpendicular to the surface is small (the
region relevant to CP violation), while the component of the momentum
parallel to the wall is arbitrary.  At first order in derivatives with
respect to $z$, the coordinate normal to the wall, the Dirac equation
is { \scriptsize
\[
K\Psi =
\]
\eq
\left( \begin{array}{cc}
\omega(1+\alpha_L+\beta_L)+ (i\sigma_3\frac{\partial}{\partial
z}-\vec{\sigma}\vec{p_{||}})(1+\alpha_L) &{\cal M}\\ {\cal
M}^{\dagger}&\omega(1+\alpha_{R}+\beta_{R}) -
(i\sigma_3\frac{\partial}{\partial
z}-\vec{\sigma}\vec{p_{||}})(1+\alpha_{R}) \end{array} \right)\cdot
\en
\[
 \left(
\begin{array}{c}L\\ R  \end{array}\right) =0
\] }
where $\vec{p_{||}}$ denotes the momentum parallel to the wall, i.e.,
transverse to $z$.  Now although $p_t$ is small, the total momentum
need not be small so $\alpha$ and $\beta$ cannot be simplified as in
eq. (\ref{define_alpha}).  We must use:
\eq
\alpha_{L,R} =
-\frac{\omega_{L,R}^2}{p^2}(1-F(\frac{\omega}{p})),
\en
\eq
\beta_{L,R} = -\frac{\omega_{L,R}^2}{p^2}F({\omega \over p}).
\en
Note that the spinors $L,R$ here have 2 components for each of the
three flavors.  In the $p_{||}=0$ case the equations decouple and we
could reduce to a description in which $L,R$ have just one component
for each flavor.  Here that will not be possible.

We want to find some transformation to the new variables in which this
equation has a diagonal form in the unbroken phase, so we can apply
the methods which have been already developed. In other words, we want
to remove somehow the matrix $\vec{\sigma} \cdot \vec{p_{||}}$ from these
equations. We introduce an analog of the spinor Lorentz
transformation, which is different for the left and right sectors:
\eq
\Lambda K \Lambda \Psi\prime = 0,~~\Psi\prime = \Lambda^{-1}\Psi,
\en
\eq
\Lambda =
\left( \begin{array}{cc}
\Lambda_L & 0\\
0 & \Lambda_R \end{array} \right)
\en
where $\Lambda_L$ and $\Lambda_R$ are given by
\eq
\Lambda_L = \exp(\frac{1}{2}\vec{\sigma} \cdot \vec{n}\Theta_L), \; \;
\Lambda_R = \exp(\frac{1}{2}\vec{\sigma} \cdot \vec{n}\Theta_R),
\en
Here $\vec{n} = \vec{p_{||}}/|p_{||}|$ is the direction of the
momentum parallel to the wall, and $\Theta_L$ and $\Theta_R$ are
matrices in flavor space to be determined.  One finds
\eq
v_{L,R} = th \Theta_{L,R} = \mp
\frac{p(1+\alpha_{L,R})}{\omega(1+\alpha_{L,R}+\beta_{L,R})}
\en

The equation in terms of the new variables is {\scriptsize \eq
K\prime\Psi\prime = \left( \begin{array}{cc}
\omega(1+\alpha_L+\beta_L)/\gamma_L +
i\sigma_3\frac{\partial}{\partial x}(1+\alpha_L) &\Lambda_L{\cal
M}\Lambda_R\\
\Lambda_R{\cal
M}^{\dagger}\Lambda_L&\omega(1+\alpha_{R}+\beta_{R})/\gamma_R -
i\sigma_3\frac{\partial}{\partial x}(1+\alpha_{R}) \end{array}
\right)
\label{ppar}
\en
\[
\left( \begin{array}{c}L\\ R \end{array}\right) =0
\]}
where
\eq
\gamma_{L,R} = (1-v_{L,R}^2)^{-1/2} = ch \Theta_{L,R}
\en
In a sense, $v_{L,R}$ is the velocity of the Lorentz boost, and
$\gamma_{L,R}$ is the analog of the usual $\gamma$ factor.

As we see, in this more complicated case the equations for the upper
and lower components of the 2-dimensional (for each flavor) spinors
$L,R$ do not decouple and one must solve the complete system of 12
differential equations.  (Note that for the vacuum case where $\alpha$
and $\beta$ are zero the equations do decouple, as expected from
Lorentz invariance.)  To keep the analogy with the one-dimensional
case it is convenient to ``reshufle" the rows and coulums of the
matrix $K\prime$ in such a way that its first 6 rows correspond to
particles moving from left to right and the other 6 to particles
moving in the opposite direction. After this reshufling, in full
analogy with the one dimensional case, one can introduce matrices $D$
and $R$,
\eq
R = \left( \begin{array}{cc} R_{++} & 0\\ 0 & R_{--}
\end{array} \right)
\en
so that the equation for non-zero $p_{||}$ has the form (\ref{basic}).
We do not solve this equation in the present paper.

{\bf Flux factors.} Suppose that the reflection coefficients are
determined from equation (\ref{ppar}) and are now functions of energy
$\omega$ and parallel momentum $p_{||}$. Let us take for definiteness
an initial particle in the unbroken phase of type $i$ ($i$ is the
index according to eq.(\ref{ppar})) and compute the baryonic flux
coming from its interaction with the domain wall. We denote the matrix
of reflection coefficients by $r(p_{||},\omega)_{ji}$ or $r$ for
short.  Now, using the analysis of the one dimensional case contained
in Appendix \ref{app:flux_facs}, one obtains for the asymmetry
current, summed over all final states:
\eq
\langle J \rangle =\int \frac{d\omega k_{||} dk_{||}}{(2\pi)^2}tr\left\{n_F^i
(R_{++}^{ii})^{-1}\left[(r)^{\dagger}R_{--}r-(\bar{r})^{\dagger}
R_{--}\bar{r}\right] \right\},
\label{ascurppar}
\en
where $\bar{r}$ denotes the matrix of anti-particle reflection
coefficients, and $n_F^i$ is the Fermi-distribution for the incident
particle as in eq.  (\ref{J^Lgroup}). The same equation can be derived
for particles incident from the broken phase.

{\bf Phase space analysis.} We would like to determine in which regions of
phase space the
reflection of strange quarks can be substantial. To find this, one
must determine the region of energies, $\omega$, and momenta parallel
to the surface, $p_{||}$, for which the $s$-quark excitations exist in
the unbroken phase but not in the broken phase, or vice versa.

Consider first the case when
the parallel momenta of the incident particles in the unbroken phase
are large, $p_{||} \gg \omega_0$. For this region of phase space the
dispersion relations for normal quasiparticles have the same form as
the familiar Lorentz-invariant dispersion relations for massive
particles but with a modified effective mass.  In the unbroken phase
we have (section \ref{ss:unbroken})
\eq
\omega^2 = |k|^2 + 2{\omega^u_0}^2,
\label{reldisprel}
\en
while in the broken phase we have
\eq
\omega^2 = |k|^2 + 2{\omega^b_0}^2 + m_s^2.
\label{reldisprelb}
\en
One can make a Lorentz boost in the direction parallel to the wall and
remove $p_{||}$ from the problem, as done above. In this frame the
dispersion relations are again (\ref{reldisprel}, \ref{reldisprelb})
with the substitution $|k| \rightarrow k_t$ and $\omega^2 \rightarrow
\omega^2/\gamma^2$. Since $\omega^u_L > \omega^b_L > \omega_R$, there
are additional forbidden regions, depending on which chiralities are
under consideration, which are not present for the $p_{||}=0$ case
analyzed previously, making it difficult of estimate the contribution
from this region\footnote{In the first preprint version of this paper
we did not distinguish between broken and unbroken $\omega$'s and thus
this complication was overlooked.}

Let us turn now to consideration of the case $p_{||} \ll \omega_0$.
We can ask what fraction of $s$-quarks are totally reflected due to
level crossing as was discussed for $p_{||} = 0$ in section
\ref{sec:scattering}.  The $L-R$ and $R-L$ level crossings occur for
\eq
|p| = \frac{3}{2}(\omega_L - \omega_R) \sim 6 ~GeV.
\label{modp}
\en
Simple geometry gives the flux coming from particles satisfying this
condition.  If the reflection coefficients depend only
weakly on $p_{||}$, then the contributions from $L-R$
and $R-L$ s-quark total reflection give $ J_{3d} = \frac{1}{6 \pi}
|p|^2 J_{1d}\sim 5 \cdot 10^{-4}T^2 J_{1d}.$
This provides a rough estimate of the minimal contribution of this
region of phase space, however we should not place much confidence in
it as a real estimate.  Studying the differential
equation (\ref{ppar}), it is far from clear that we are justified is
assuming a weak $p_{||}$ dependence of the $\Delta$'s.  Moreover
neglect of the $L-L$ and $R-R$ scattering amplitudes cannot be
justified, but their contribution cannot be determined without going
much farther toward the solution of these equations than we have.

{\bf Range of validity of the Dirac Operator.}  The actual values
of $\omega_0$ and $T$ of interest are $\omega_0 \sim 50$GeV and
$T=100$GeV.  From the dispersion relation (\ref{reldisprel}), one sees
that the condition for the validity of the high-temperature expansion
($\omega^2 - |p|^2 \ll T^2$) is only marginally satisfied for the
large $p_{||}$ relativistic kinematics.  Thus the dispersion relation
needs to be obtained in a more suitable approximation before one can
accurately describe the physics of the $3d$ problem.

\mysection{Mixing in the broken phase}
\label{app:mixing}
\hspace*{2em}
Our purpose in the present appendix is to give the reader a feeling
for the qualitative differences between the eigenstates in the broken
and unbroken phases, and especially for the effects of mixing.  We
will therefore restrict the discussion here to the case of $p
\rightarrow 0$, and make a perturbative expansion in the scalar field
vev, $\phi$, dropping quadratic terms.  The mixing in the broken phase
for the charge $-\frac{1}{3}$ sector is then described by the matrix
$O_D$ ($D$ for down), where
\eq
O_D(\Omega_L^2 + \frac{16\alpha_s\phi^2}{9\alpha_W\sigma^2}
KM_d^2K^{\dagger}){O_D}^{\dagger} = {{\bar{\omega}}_L}^{D^2};
\en
where $\omega_L$ and $\omega_R$ are given in eqs. \ref{omega_L} and
\ref{omega_R} above, and it is henceforth understood that when the
notation $R$ is used, one must substitute, e.g., $\omega_U$ or
$\omega_D$ for $\omega_R$, as appropriate. $\sigma = 246$ GeV is the
zero temperature vev.  The particle eigenstates\footnote{Note that
physical states in the broken phase do not carry any fixed chirality,
so that subscripts $L$ and $R$ have nothing to do with actual
chirality and are merely labels to distinguish the two distinct
states.} are related to the initial fields by:
\eq
D_L = {O_D}^{\dagger}{\bf D}_L ,~D_R =
-\frac{32g_s\phi}{3\sqrt{6}g_W^2\sigma}M_d K^{\dagger}{O_D}^{\dagger}
{\bf D}_L,
\en
and
\eq
D_L= \frac{32g_s\phi}{3\sqrt{6}g_W^2\sigma}M_d {\bf D}_R,~D_R = {\bf
D}_R.
\en
The mass gap in the broken phase is ${{\bar{\omega}}_L}^{D^2}$ for
${\bf D}_L$ and
\eq
{\bar{\omega}_D}^2 = \Omega_D^2 -
\frac{16\alpha_s\phi^2}{9\alpha_W\sigma^2}M_d^2
\en
for ${\bf D}_R$.

The equations for the up quarks can be written in the same way and we
present them for completeness.  The mixing in the up sector is
described by the matrix $O^U$, where
\eq
 O_U(\Omega_L^2 + \frac{16\alpha_s\phi^2}{9\alpha_W\sigma^2}
M_u^2){O_U}^{\dagger} = {\bar{\omega}_{LU}}^2.
\en
The particle eigenstates are related to the initial fields by
\eq
U_L = {O_U}^{\dagger}{\bf U}_L ,~U_R =
-\frac{32g_s\phi}{3\sqrt{6}g_W^2\sigma}M_u {O_U}^{\dagger} {\bf U}_L,
\en
and
\eq
U_L= \frac{32g_s\phi}{3\sqrt{6}g_W^2\sigma}M_u {\bf U}_R,~U_R = {\bf
U}_R.
\en
The mass gap in the broken phase is ${\bar{\omega}_{LU}}^2$ for ${\bf
U}_L$ and
\eq
{\bar{\omega}_U}^2 = \Omega_U^2 -
\frac{16\alpha_s\phi^2}{9\alpha_W\sigma^2}M_u^2
\en
for ${\bf U}_R$.

It is important to notice that the mixing matrices of quarks in the
broken phase differ from the corresponding matrices in the unbroken
phase.  Note also that the change of mixing between unbroken and
broken phases in the down sector is considerably larger than in the up
sector, due to the fact that masses of up quarks are larger than
masses of down quarks.  This reveals the mechanism by which degeneracy
in one sector inhibits CP violation in the other, independently of the
convention of in which sector CKM phases occur.  It is the reason that
CP-violating effects are most profound for scattering in the down
quark sector.

\mysection{Equations in terms of observables}
\label{app:observables}
\hspace*{2em}
As we discussed in Appendix \ref{app:dirac}, the equation for the
scattering matrix does not contain exponentially rising terms only if
the imaginary parts of the particle momenta in the broken phase are
small enough: $|Im p_i| < a/2$. While this inequality holds for the
charge -1/3 quarks for all energies, for zero wall velocity, it breaks
down for the top quark, since its mass is quite large in the broken
phase, and also for the change -1/3 sector when the wall velocity
becomes larger than $\sim$0.4.  Therefore, under these circumstances
the equations should be modified. The idea is quite simple. It is
obvious that all physical reflection and transmission amplitudes must
be perfectly finite independently of the top quark mass or wall
velocity. In other words, if we would write equations for the
scattering amplitudes themselves, there would be no exponentially
large terms floating around.

Let us concentrate on the problem of left quark reflection and choose
the contour in the complex plane for which at $x \rightarrow +\infty$
the first three eigenvalues of the matrix $DR$ correspond to the
transmitted wave. In particular, $Im \; p_i(+\infty) > 0$ for $i =
1,2,3$. We denote the eigenvectors by
\eq
p_L = p_i, ~i=1,2,3, ~~ p_R = p_i,~ i=4,5,6.
\en
In order to simplify the notation we write an equation for the
scattering matrix $V$ in the form
\eq
\frac{\partial V}{\partial z} =\Omega V
\en
\eq
\Omega =  -(RE)^{-1}fR\frac{\partial e}{\partial z}E
\en
\eq
\Omega =
\left( \begin{array}{cc}
\Omega_{LL} & \Omega_{LR}\\
\Omega_{RL} & \Omega_{RR}  \end{array} \right).
\en
Some of the elements of the matrix $\Omega$ can be exponentially
large. Using the equation for $V$ and expressions for transmission and
reflection coefficients through $V$ (see eqs.(\ref{ru},\ref{rb})) we
get a set of non-linear equations:
\eq
\frac{\partial r^u}{\partial z} = - t^b\Omega_{RL}t^u
\en
\eq
\frac{\partial t^u}{\partial z} = \Omega_{LL}t^u - r^b \Omega_{RL}t^u
\en
\eq
\frac{\partial r^b}{\partial z} = \Omega_{LL}r^b - r^b \Omega_{RR}
+\Omega_{LR} - r^b \Omega_{RL}r^b
\en
\eq
\frac{\partial t^b}{\partial z} = - t^b \Omega_{RL}r^b - t^b
\Omega_{RR}
\en
with initial conditions at $x \rightarrow \infty$
\eq
r^u = r^b = 0,~~t^u = t^b = 1.
\en
The important point is that not all of the reflection and transmission
coefficients are observable. If, for instance $p_3$ and $p_6$ are
complex, we cannot have the third flavor in the broken phase since it
cannot propagate there. In other words, only the quantities
$(r^b)_{ij},~~i,j = 1,2$, $(t^b)_{ij},~~j \neq 3$ and $(t^u)_{ij},~~i
\neq 3$ have direct physical meaning. We will not change physical
quantities if we go to another set of matrices defined by
\eq
R_2 = \exp(-Im \int_{z_0}^z p_L dz)r^b\exp(Im \int_{z_0}^z p_R
dz),~~R_1 = r^u
\en
\eq
T_2 = t^b\exp(Im \int_{z_0}^z p_R dz),~~T_1 = \exp(- Im \int_{z_0}^z
p_L dz)T_2.
\en
Now, the equations for these new variables do not contain any
exponentially large terms at all due to our sign convention,
\eq
\frac{\partial R_1}{\partial z} = - T_2\bar{\Omega}_{RL}T_1
\label{noexp1}
\en
\eq
\frac{\partial T_1}{\partial z} = -Imp_L T_1 + \bar{\Omega}_{LL}T_1 -
R_2
\bar{\Omega}_{RL}T_1
\en
\eq
\frac{\partial R_2}{\partial z} = -Im p_L R_2 + R_2 Im p_R +
\bar{\Omega}_{LL}R_2 - R_2 \bar{\Omega}_{RR}
+\bar{\Omega}_{LR} - R_2 \bar{\Omega}_{RL}R_2
\en
\eq
\frac{\partial T_2}{\partial z} = + T_2 Im p_R - T_2
\bar{\Omega}_{RL}R_2 - T_2
\bar{\Omega}_{RR}
\label{noexp4}
\en
where
\eq
\bar{\Omega} =  -(RE\prime)^{-1}fR\frac{\partial e}{\partial
z}E\prime,~~ E\prime = \exp(iRe\int_{z_0}^z p(z)dz).
\en

We have checked that these equations give precisely the same values
for the reflection coefficients as the scattering matrix formalism,
which is a good check of the correctness of the numerical integration
schemes.  We also used them for scanning the up-quark case, but no
significant asymmetry was found there. This was expected since for the
up quark sector the change in mixing angles in going from the unbroken
to the broken phase is substantially smaller than it is for the down
sector.

The equations in this form are quite convenient and allow one to
integrate out particles with momenta large compared with some typical
scale. For example, for up-sector reflection one expects that an
interesting effect can appear only near the $c$-reflection threshold,
since at higher energies the momenta of the $c$ and $u$ quarks are
nearly degenerate.  Near the $c$-threshold the momentum of the
$t$-quark is huge, and the approximate solution to equations
(\ref{noexp1}- \ref{noexp4}) is
\eq
(R_2)_{3i} = (R_2)_{i3} = (T_1)_{3i} = (T_2)_{i3} = 0,
\label{int_out}
\en
with other equations unchanged. In this approximation the equations
for physical reflection amplitudes do not contain either the momentum
of the $t$-quark nor the eigenvectors corresponding to it.
Nevertheless, the $t$-quark does not decouple completely since it
influences the structure of other particle eigenstates through the
mixing.  Formally, the $t$-quark must be taken into account at the
stage at which one solves the equation to determine the eigenstates
and eigenfunctions.

The same kind of procedure can also be used to integrate out the
$b$-quark.  In the region of $s$-quark reflection, the momentum of the
$b$-quark is large compared with other particle momenta. In a sense,
rapid oscillations of the wave function are equivalent to exponential
suppression. This observation actually works quite well, as can be
checked by solving the equations for reflection coefficients in terms
of observables, throwing away the $b$-quark as in eq. (\ref{int_out})
but keeping it in the equations for $e,~f,$ and $p$.  This procedure
gives essentially the same result for the asymmetry as obtained by our
standard procedure.

We can understand the strong dependence of the asymmetry on the top
quark mass as follows.  The mass of the top quark contributes to the
mass gap of the left chiral $b$-quark.  Changing the top mass
therefore changes the mixing among the charge $-1/3$ left chiral
quarks, and thus influences the properties of the light flavors
indirectly, through the mixing in the physical eigenstates.

\mysection{Quadratic approximation}
\label{app:quadratic}
\hspace*{2em} In the low momentum approximation to which we confined ourselves
in other sections, the relation between energy and momentum of the
quasiparticles was unambiguous. For any fixed energy we had just one
momentum corresponding to it. This fact is lost when one treats the
dispersion relations exactly. For example, at second order in the
momentum, the dispersion relation looks like
\eq
\omega = \omega_0 + \frac{1}{3}p + \frac{p^2}{3\omega_0},
\en
so that if $\omega > \frac{11}{12}\omega_0$, there are two solutions
for the particle momenta. Physically, this means that there are two
quasiparticle excitations degenerate in energy but not in momentum,
one of which corresponds to the normal branch and the other to the
abnormal one. If the energy is close to the mass gap, then the
momentum of one of the excitations is small, and it is this excitation
whose scattering off the domain wall we have considered. The other
value of the momentum corresponds to the abnormal branch and is not
small: $p \sim \omega_0$. If we leave aside the question of the
stability of the abnormal excitations at such high momentum (in ref.
\cite{smilga,brpis} they are argued to be unstable), then the scattering
problem is much more complicated than we had before.  Instead of 3
possible initial and 3 possible final states, we have now 6 initial
and 6 final states.  The three additional states are the high momenta
excitations corresponding to the abnormal branch.  On the other hand,
if $\omega < \frac{11}{12}\omega_0$, the momenta are complex so that
particles with these nergies cannot propagate even in the unbroken
phase and their wave function must be zero everywhere.  Actually, in
the region of energies corresponding to $s$-quark total reflection,
$\omega$ lies well below $\omega_L$ for the $b$-quark, so that the
$b$-quark actually is absent in the initial states.  (Note that this
is not the case in linear approximation.)

As we discussed in Appendix \ref{app:observables}, particles with
large momenta decouple from the low-momentum $s$ and $d$ quarks. This
justifies the use of the linear approximation in the paper: in spite
of the fact that we treat $b$-quark excitations in the wrong way, as
having very short wavelength rather than zero amplitude, the $b$-quark
decouples from the equations for the scattering matrix in either case,
so that it is immaterial which method we use.  The influence of
$b$-quark properties on the asymmetry is due to its effect in fixing
the initial physical eigenstates in the light sector, which we do
correctly no matter what approximation is used for the $b$-quark in
the differential equation.  The linear approximation is much faster to
numerically integrate, so that is our method of choice.  We present,
however, for the sake of completeness, the method which one can use
for more complete analysis of the system.  In particular it would be
useful if one wished to study an intermediate range of momenta: large
enough that linear approximation is inadequate, but small enough that
the decoupling is not complete.

Let us expand the fermionic mass operators in the momentum, up to
second order terms in $p$. The change $p \rightarrow i
\frac{\partial}{\partial x}$ gives then an equation
\eq
\left( \begin{array}{cc}
a_L (\frac{\partial}{\partial x})^2 + b_L i\frac{\partial}{\partial
x}+ c_L &{\cal M}\\ {\cal M}^{\dagger}& a_R (\frac{\partial}{\partial
x})^2 + b_R i\frac{\partial}{\partial x}+ c_R
\end{array} \right)
\left(\begin{array}{c}L\\ R \end{array}\right) =0
\en
where
\eq
a_L = \frac{\omega_L^2}{3\omega^3},~~b_L =
1-\frac{\omega_L^2}{3\omega^2},~~ c_L =
\omega(1-\frac{\omega_L^2}{\omega^2}),
\en
and corresponding equations for the right quarks.  Let $A_{L,R}$ and
$B_{L,R}$ be the momenta of the left and right particles in the
unbroken phase. They are solutions to the quadratic equation
\eq
a_{L,R} p^2 - b_{L,R} p -c_{L,R} = 0.
\en
Let us introduce new variables, which correspond to particle
excitations in the unbroken phase with momenta $A_{L,R}, B_{L,R}$:
\eq
\Psi_1 = (i\frac{\partial}{\partial x}-B_L)\sqrt{a_L}\Psi_L,~
\Psi_2 = (i\frac{\partial}{\partial x}-A_L)\sqrt{a_L}\Psi_L,~
\en
\eq
\Psi_3 = (i\frac{\partial}{\partial x}-B_R)\sqrt{a_R}\Psi_R,~
\Psi_4 = (i\frac{\partial}{\partial x}-A_R)\sqrt{a_R}\Psi_R.
\en
Now, the equation for these variables have the same form as in the
usual case,
\eq
i\frac{\partial}{\partial x}\Psi = DR \Psi,
\en
where the operator $D$ and matrix $R$ are hermitean and defined by
\eq
R = \left( \begin{array}{cccc}
\frac{Re d_L}{|d_L|^2}&\frac{iIm d_L}{|d_L|^2}&0&0 \\
\frac{-iIm d_L}{|d_L|^2}&\frac{-Re d_L}{|d_L|^2}&0&0 \\
0&0&\frac{Re d_R}{|d_R|^2}&\frac{iIm d_R}{|d_R|^2}\\ 0&0&\frac{-iIm
d_R}{|d_R|^2}&\frac{-Re d_R}{|d_R|^2}
\end{array} \right) ,
\en
\eq
D = \left( \begin{array}{cccc} A_L Re d_L & iA_L Im d_L
&-\frac{1}{\sqrt{a_L}}{\cal M}\frac{1}{\sqrt{a_R}}
&-\frac{1}{\sqrt{a_L}}{\cal M}\frac{1}{\sqrt{a_R}} \\ -iA_L^* Im d_L &
-B_L Re d_L &-\frac{1}{\sqrt{a_L}}{\cal M}\frac{1}{\sqrt{a_R}}
&-\frac{1}{\sqrt{a_L}}{\cal M}\frac{1}{\sqrt{a_R}} \\
-\frac{1}{\sqrt{a_R}}{\cal M}^{\dagger}\frac{1}{\sqrt{a_L}}
&-\frac{1}{\sqrt{a_R}}{\cal M}^{\dagger}\frac{1}{\sqrt{a_L}}&A_R Re
d_R & iA_R Im d_R \\ -\frac{1}{\sqrt{a_R}}{\cal
M}^{\dagger}\frac{1}{\sqrt{a_L}} &-\frac{1}{\sqrt{a_R}}{\cal
M}^{\dagger}\frac{1}{\sqrt{a_L}}&-iA_R^* Im d_R & -B_R Re d_R
\end{array} \right),
\en
\eq
d_{L,R} = \frac{1}{a_L}\sqrt{b_{L,R}^2 + 4 a_{L,R}c_{L,R}}.
\en
The conserved baryonic current is given by the usual expression
\eq
J^B = \Psi^{\dagger}R\Psi
\en
Now these equations can be solved with the help of formalism described
in Appendices \ref{app:dirac} and \ref{app:observables}.

\mysection{$D_B$ and Scattering Length of the Quasiparticles}
\label{app:D_B}
\hspace*{2em}
One might think that with such a degree of concurrence among the
estimates for the scattering length listed in section
\ref{sec:prediction} that we can be rather confident
that $\lambda \sim (4-5)/T$.
However when one considers the calculations leading to these
estimates, one sees that the scattering length in those discussions is
that of a particle with the typical momentum $\sim T$, while we need
the scattering length of a {\em quasi-particle}, with relatively low
momentum.  The quasi-particles are gauge-singlets when viewed at a
scale longer than some screening length, given by the inverse Debye or
magnetic mass.  Moreover at least in the 1-dimensional problem, the
quarks which produce the asymmetry, and thus the quarks whose
diffusion length is relevant to determining $f_{sph}(\rho)$, have energies
considerably lower than the thermal average.  Let us therefore
investigate whether the scattering length obtained in the references
quoted above is a good approximation to the actual diffusion length
which we require.

The diffusion of the quasiparticles is clearly a 3-dimensional
phenomenon, so we must consider the general problem in which the
quarks incident on the domain wall have non-zero $p_{||}$.  We first
must decide whether, in the full 3-dimensional problem, the quarks
carrying the asymmetry current will have relatively low energies, as
is the case for the 1-d problem.  Since we know that the $p_t$ of
$s$-quarks which reflect is in all cases low (see Appendix
\ref{app:parallel}), the only way these quarks could have an energy
typical of the thermal medium would be if they have $p_{||} \gg
\omega_0$.  However quarks with this large value of $p_{||}$ see a
bubble thickness of $\sim (p_{||}/p_t) a^{-1}$ which may be much
larger than the expected mean free path $\sim 5/T$ of quasiparticles
having typical thermal energies.  If this is the case, these particles will
experience many inelastic collisions in the scattering process and the
purely quantum mechanical description of their scattering would be
inapplicable.  In fact, on account of the loss of quantum coherence,
the contribution of such particles to the asymmetry should be very
much reduced, since CP violation arises because of the difference in
phases between $\delta_{CP}$ and the quantum mechanical scattering
phase shift, which changes sign in going from particles to
antiparticles.

Thus the particles whose scattering contributes to the asymmetry
current have $p_{||}$ and $p_t \lsi \frac{3}{2}(\omega_L - \omega_R)
\sim 6$GeV (see Appendix \ref{app:parallel}), so the mean free path
which is relevant to estimating the $D_B$ which enters $\rho$, is the
mean free path of a quasiparticle with physical (energy, momentum)
$\sim (50,6)$ GeV.\footnote{See section
\ref{sec:thermal} for a discussion of the physical momenta of the
quasiparticles.  On account of averaging over all directions of
incidence of its collision-mate, the magnitude of the quasiparticle
momentum is in any event irrelevant to estimating the average $s$ of
its collisions.} Denoting the typical energy of a collision-mate by
$\bar{E} \sim T$ and doing the angular average, one finds $<s> = 2
\omega \bar{E}$.

Now we need the effective Lagrangian governing the quasiparticle
interaction.  On account of Debye and magnetic screening of the color
of the quasi-quark,
the propagator appearing in the matrix element is $\sim (q^2-
M_{screening}^2)^{-1}$.  Taking the Debye screening length as a lower
limit to this scale, $M_{screening}^2 > M_{D}^2 =
\frac{(N_f + 2 N_c)}{6}g^2 T^2$, and assuming the coupling constant to
be just the ordinary QCD coupling appropriate for this energy scale,
leads to\cite{bmpr:viscosity}:\footnote{Their calculation included all
tree level diagrams.}
\eq
\sigma_{qq} \lsi \frac{\alpha_s^2 <s>}{M_D^4} = \frac{<s>}{(8 \pi T^2)^2},
\en
where we have made use of the fact that the squared momentum transfer
is less than $<s>/2$ and thus is small compared with $M_D^2$.  Now
summing over quark and gluon collision-mates, and noting that
$\sigma_{qg} = \frac{9}{4}\sigma_{qq}$, we find
\eq
n \sigma = \sigma_{qq} ( n_{q,\bar{q}} + \frac{9}{4}n_g ) \lsi 0.013
T.
\en
The standard estimate of the mean free path in a gas is
\eq
\lambda = \frac{\bar{v}}{\bar{V}}\frac{1}{ n \sigma},
\en
where $\bar{v}$ denotes the average speed of the particle whose mean
free path is being computed and $\bar{V}$ denotes its average speed
relative to other particles in the gas.  The diffusion coefficient for
a gas is
\eq
D = \frac{ \bar{v} \lambda }{3}.
\en
Taking the average speed of quasiparticles to be $\sim 1$ while taking
$\bar{v}$, the average speed of the the low-momentum quasiparticle
which is diffusing to be $\sim 1/3$, gives a low and thus conservative
estimate\footnote{Another reason that the true $D_B$ and $\lambda$ may
be larger than this estimate is that in modeling the effective
Lagrangian we reduced just the range of interaction but not the
effective coupling, as compared to the fundamental Lagrangian.} $D_B
\gsi 3/T$ and $\lambda \gsi 25/T$. Hopefully this discussion has
revealed that the problem of estimating $\lambda$ and $D_B$ require
real understanding of the nature of the quasiparticles and their
short- and long-distance interactions.  Thus estimates of these
quantities must be regarded as highly provisional.  In the following
we use the ranges $D_B \sim (3-5)/T$ and $\lambda \sim (4-25)/T$.

\newpage


\begin{figure}
\epsfxsize=\hsize
\epsffile{dr_s_UB.ps}
\caption{Dispersion relation for $s$-quarks in the unbroken phase,
neglecting mixing. The figure is essentially identical for the broken
phase, except for the neighborhood of the crossing.}
\label{dr_s_UB}
\end{figure}

\begin{figure}
\epsfxsize=\hsize
\epsffile{dr_b_UB.ps}
\caption{Dispersion relation for $b$-quarks in the unbroken phase,
neglecting mixing.}
\label{dr_b_UB}
\end{figure}

\begin{figure}
\epsfxsize=\hsize
\epsffile{dr_b_B.ps}
\caption{Dispersion relation for $b$-quarks in the broken phase,
for zero CKM angles.}
\label{dr_b_B}
\end{figure}

\begin{figure}
\epsfxsize=\hsize
\epsffile {dr_dsb_B.ps}
\caption{Dispersion relation for $d$ (short-dashed), $s$
(long-dashed), and $b$ (solid) -quarks in the broken phase, focusing
on the region of total reflection of the $d$ and $s$. It is the $R_n$
branch of the $b$ which is visible in this region.  CKM angles have
been set to zero.}
\label{dr_dsb_B}
\end{figure}

\begin{figure}
\epsfxsize=\hsize
\epsffile {std.ps}
\caption{Dependence of $\Delta$ on $\omega$, in GeV, for the ``canonical''
choices of masses and mixing angles, and wall velocity $v=0$.}
\label{std}
\end{figure}

\begin{figure}
\epsfxsize=\hsize
\epsffile {mt90.ps}
\caption{Dependence of $\Delta$ on $\omega$, in GeV, for $m_t=90$ GeV
and the ``canonical'' choices for the other masses and the mixing
angles, and $v=0$.}
\label{mt90}
\end{figure}

\begin{figure}
\epsfxsize=\hsize
\epsffile {mt210.ps}
\caption{Dependence of $\Delta$ on $\omega$, in GeV, for $m_t=210$ GeV
and the ``canonical'' choices for the other masses and the mixing
angles, and $v=0$.}
\label{mt210}
\end{figure}

\begin{figure}
\epsfxsize=\hsize
\epsffile {th2301.ps}
\caption{Dependence of $\Delta$ on $\omega$, in GeV, for $\theta_{23}=0.01$
and the ``canonical'' choices for the masses and other mixing angles,
and $v=0$.}
\label{th2301}
\end{figure}

\begin{figure}
\epsfxsize=\hsize
\epsffile {mtdep.ps}
\caption{Dependence of $\Delta_{int}$ (in GeV) on $m_t(T=0)$, in GeV}
\label{mtdep}
\end{figure}

\begin{figure}
\epsfxsize=\hsize
\epsffile {th23dep.ps}
\caption{Dependence of $\Delta_{int}$ (in GeV) on $sin(\theta_{23})$}
\label{th23dep}
\end{figure}

\begin{figure}
\epsfxsize=\hsize
\epsffile {mbdep.ps}
\caption{Dependence of $\Delta_{int}$ (in GeV) on $m_b(T=0)$, in GeV}
\label{mbdep}
\end{figure}

\begin{figure}
\epsfxsize=\hsize
\epsffile {adep.ps}
\caption{Dependence of $\Delta_{int}$ (in GeV) on inverse wall-thickness $a$,
in
units of $T=100$ GeV, other parameters having their canonical values.}
\label{adep}
\end{figure}

\begin{figure}
\epsfxsize=\hsize
\epsffile {vdep.ps}
\caption{Dependence of $\Delta_{int}$ (in GeV) on wall velocity, other
parameters having their canonical values. }
\label{vdep}
\end{figure}

\begin{figure}
\epsfxsize=\hsize
\epsffile {uR01.ps}
\caption{Dependence of $\Delta$ on $\omega$, in GeV, for the
individual reflection $s_R \rightarrow d_L$.  Also shown is the sum of
this asymmetry and that for $s_R \rightarrow s_L$, which is practially
equal and opposite, so that the sum is 2-3 orders of magnitude less
than the individual asymmetries.}
\label{std_i}
\end{figure}

\end{document}